\documentclass[%
 reprint,
superscriptaddress,
showpacs,preprintnumbers,
nofootinbib,
 amsmath,amssymb, 
 aps,
 prd,
 longbibliography,
]{revtex4-1}

\usepackage{cancel}
\usepackage{accents} 
\usepackage{mciteplus,slashed}
\usepackage{amssymb,cancel,amsmath}
\usepackage{dcolumn}
\usepackage{bm}
\usepackage[caption=false]{subfig}
\usepackage{appendix}
\usepackage{physics}
\usepackage{enumerate}
\usepackage[T1]{fontenc}	
\usepackage{csvsimple}
\usepackage{hyperref}
\usepackage[section]{placeins}
\usepackage[capitalise]{cleveref}
\usepackage{booktabs}
\usepackage{graphicx}
\usepackage{mathrsfs}
\graphicspath{{Figures/}}
\usepackage[utf8]{inputenc}

\usepackage[dvipsnames]{xcolor}
\usepackage[normalem]{ulem}

\newcommand{\be}{\begin{equation}}
\newcommand{\ee}{\end{equation}}
\newcommand{\ba}{\begin{align*}}
\newcommand{\ea}{\end{align*}}
\newcommand{\bpm}{\begin{pmatrix}}
\newcommand{\epm}{\end{pmatrix}}
\newcommand{\bea}{\begin{eqnarray}}
\newcommand{\eea}{\end{eqnarray}}
\newcommand{\Gev}{\,\,\mathrm{GeV}}

\newcommand{\Lag}{\mathcal{L}}

\newcommand{\benum}{\begin{enumerate}}
\newcommand{\eenum}{\end{enumerate}}
\newcommand{\bi}{\begin{itemize}}
\newcommand{\ei}{\end{itemize}}

\newcommand{\Ugroup}{\text{U}}
\newcommand{\SUgroup}{\text{SU}}
\newcommand{\KeV}{~\mathrm{keV}}
\newcommand{\MeV}{~\mathrm{MeV}}
\newcommand{\GeV}{~\mathrm{GeV}}
\newcommand{\eV}{~\mathrm{eV}}
\newcommand{\TeV}{~\mathrm{TeV}}
\newcommand{\cm}{~\mathrm{cm}}

\newcommand{\gsim}{\lower.7ex\hbox{$\;\stackrel{\textstyle>}{\sim}\;$}}
\newcommand{\lsim}{\lower.7ex\hbox{$\;\stackrel{\textstyle<}{\sim}\;$}}

\newcommand{\citer}[1]{Ref. \cite{#1}}

\def\EL{E} 
\def\ER{\mathcal{E}} 
\def\PL{P}
\def\PR{\mathcal{P}}

\begin{document}

\title{Dipole portal to heavy neutral leptons}

\author{Gabriel Magill}
\email{gmagill@perimeterinstitute.ca}
\affiliation{Perimeter Institute for Theoretical Physics, 31 Caroline St. N., Waterloo, Ontario N2L 2Y5, Canada}
\affiliation{\mbox{Department of Physics \& Astronomy, McMaster University, 1280 Main St. W., Hamilton, Ontario L8S 4M1, Canada}}
\author{Ryan Plestid}
\email{plestird@mcmaster.ca}
\affiliation{Perimeter Institute for Theoretical Physics, 31 Caroline St. N., Waterloo, Ontario N2L 2Y5, Canada}
\affiliation{\mbox{Department of Physics \& Astronomy, McMaster University, 1280 Main St. W., Hamilton, Ontario L8S 4M1, Canada}}
\author{Maxim Pospelov}
\email{mpospelov@perimeterinstitute.ca}
\affiliation{Perimeter Institute for Theoretical Physics, 31 Caroline St. N., Waterloo, Ontario N2L 2Y5, Canada}
\affiliation{Department of Physics and Astronomy, University of Victoria, Victoria, BC V8P 5C2, Canada}
\affiliation{Theoretical Physics Department, CERN, 1211 Geneva, Switzerland}
\author{Yu-Dai Tsai}
\email{yt444@cornell.edu}
\affiliation{Perimeter Institute for Theoretical Physics, 31 Caroline St. N., Waterloo, Ontario N2L 2Y5, Canada}
\affiliation{Laboratory for Elementary Particle Physics, Cornell University, Ithaca, NY 14850, USA}

\date{\today}

\begin{abstract}
We consider generic neutrino dipole portals between left-handed neutrinos, photons, and right-handed heavy neutral leptons (HNL) with Dirac masses. 
The dominance of this portal significantly alters the conventional phenomenology of HNLs. 
We derive a comprehensive set of constraints on the dipole portal to HNLs by utilizing data from LEP, LHC, MiniBooNE, LSND as well as observations of Supernova 1987A and consistency of the standard Big Bang Nucleosynthesis. 
We calculate projected sensitivities from the proposed high-intensity SHiP beam dump experiment, and the ongoing experiments at the Short-Baseline Neutrino facility at Fermilab. Dipole mediated Primakoff neutrino upscattering and Dalitz-like meson decays are found to be the main production mechanisms in most of the parametric regime under consideration. 
Proposed explanations of LSND and MiniBooNE anomalies based on HNLs with dipole-induced decays are found to be severely constrained, or to be tested in the future experiments.

\end{abstract}

 \maketitle 
\section{Introduction \label{sec:1-intro}}

\begin{figure*}[!ht]
\begin{center}
\includegraphics[width=0.9\linewidth]{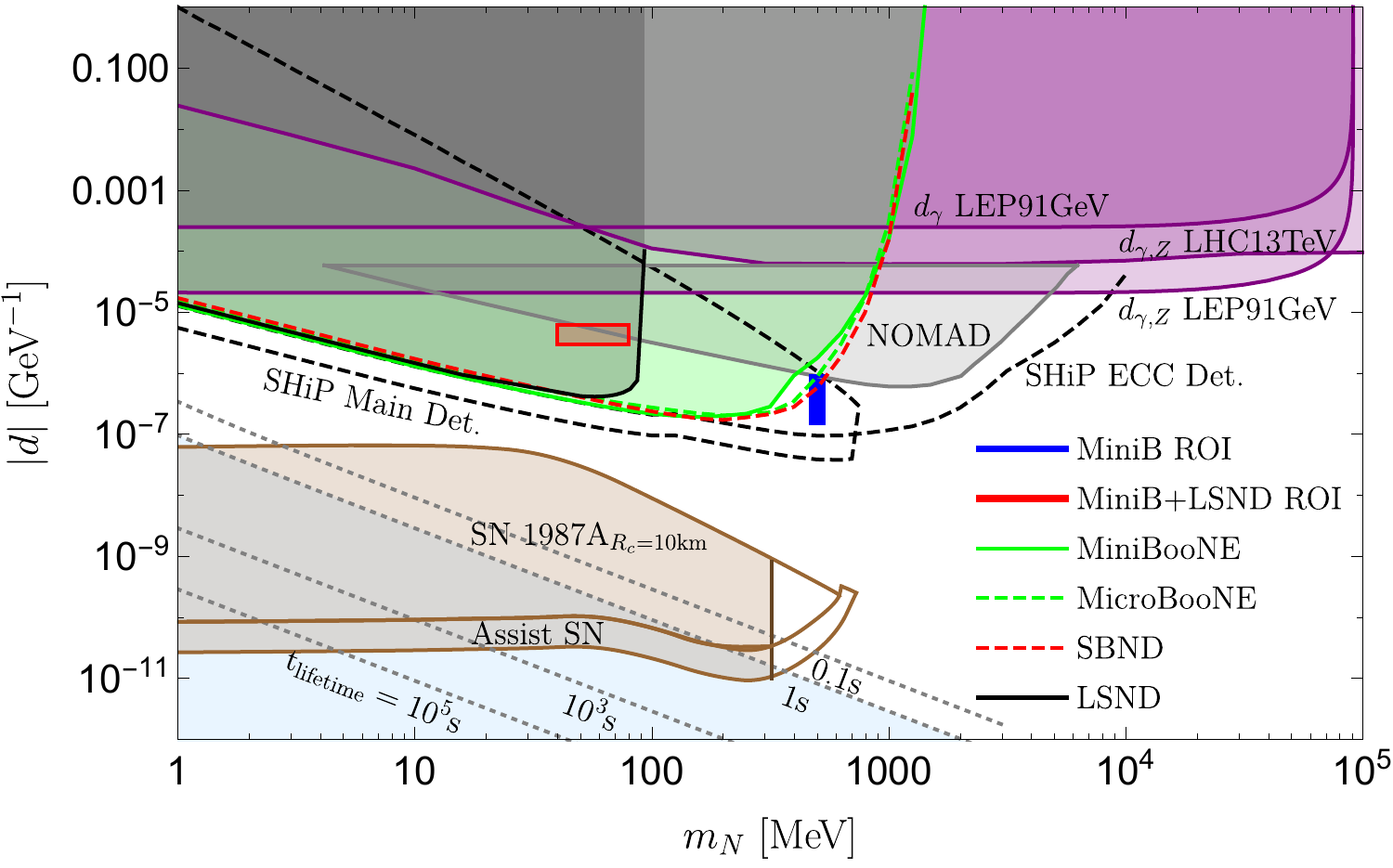}
\end{center}
\caption{
Overview of projected sensitivities ($95\%$ CL) and constraints obtained from SHiP, LHC, LEP, Supernova 1987A and experiments at the Short-Baseline Neutrino facility at Fermilab. 
We also show previously calculated favored regions of interest (ROI) in parameter space for MiniBooNE and LSND, and constraints from NOMAD. Limits are shown for the dimension 5 ($\gamma$ mediator) and dimension 6 ($\gamma+Z$ mediators) extensions. See \cref{tab:HEPcouplingsLabel} for an explanation of the labels. Each curve is discussed and presented in the paper. \label{fig:masterPlot}}
\end{figure*}


The Standard Model of particles and fields (SM) shows remarkable resilience under the scrutiny of 
numerous particle physics experiments. In particular, the LHC experiments have put significant constraints 
on new hypothetical colored states, pushing their masses to a TeV scale and beyond. At the same time, 
owing to its smaller production cross sections, the electroweak extensions of the SM are far less constrained, 
and a plethora of new models may be hiding at energies of a few hundred GeV and beyond. If such sectors 
are considered to be heavy, their impact on the SM physics can be encoded in the higher-dimensional extensions of the 
SM. Moreover, the electroweak singlet components of such sectors can be light, and still coupled to the SM states. 
In the last few years, significant attention has been paid to the models containing new singlet fermionic states $N$
(often referred to as heavy neutral leptons) that can couple to the 
SM leptons $L$ and Higgs field $H$ via the so-called neutrino portal coupling, $NLH$ (see {\em e.g.} \cite{Asaka:2005pn,Gorbunov:2007ak}). Owing to the neutrality of $N$, its mass 
$m_N$ is a free parameter with a wide range of possibilities from the sub-eV scale and up, all the way to the Planck scale. 
This range is somewhat narrower if $N$ is indeed taking part in generating masses for the light active neutrino species. 
A great deal of experimental activity is devoted to searches of $N$ particles, that may show up in cosmological 
data, in neutrino oscillation experiments, in meson decays, beam dump experiments and at high energy colliders. (For a recent 
overview of neutrino portal see {\em e.g.} \cite{Alekhin:2015byh}.)

Given large interests in searches of heavy neutral leptons, 
in this work we will analyze a less conventional case of $N$ particles coupled to the SM via the so-called dipole portal
encoded in the following effective Lagrangian, 
\be
\mathcal{L}\supset \bar{N}(i\slashed{\partial}- m_N) N+ (d \bar{\nu}_L\sigma_{\mu\nu}F^{\mu\nu}N + h.c).
\label{eq:simpledipole}
\ee
Here $F^{\mu\nu}$ is the electromagnetic field strength tensor, and $\nu_L$ is a SM neutrino field. This is an effective Lagrangian that 
needs to be UV completed at energy scales not much larger than $\Lambda \sim d^{-1}$. We are going to stay on the effective field 
theory grounds, noting that since our results show the sensitivity to $d$ to be much better than TeV$^{-1}$, the UV completion scale can be
raised above the electroweak scale. For now, \cref{eq:simpledipole} is also applicable only at energies 
below the weak scale, as it does not respect the full SM gauge invariance. 
Indeed, $F^{\mu\nu}$ should be a part of the $\Ugroup(1)$ and/or $\SUgroup(2)$ field strength, and the insertion of the Higgs field $H$ is also 
required, so that $d \propto \langle H \rangle \Lambda^{-2}$. For most of our analyses we will be interested in values of $m_N$ in the 
interval from 1\,MeV to 100\,GeV, and at relatively small energies, so that a treatment using \cref{eq:simpledipole} is indeed sufficient.

The main assumption made in \cref{eq:simpledipole} is the absence, or subdominance, of the mass mixing operator $NLH$.
When the mass mixing operator is dominant, the production and decay of $N$ particles is mostly governed by its interaction 
with the SM particles via weak bosons. The phenomenological consequences of these minimally coupled particles $N$ is
well understood. In contrast, if the leading order operator is suppressed, the dipole operator offers novel signatures and 
features in the production and decay of $N$, such as a much enhanced role of electromagnetic 
interactions in the production and decay of $N$. This case has so far being addressed only in a handful of works
\cite{Gninenko:2009ks,Gninenko:2010pr,McKeen:2010rx,Masip:2012ke,Masip:2011qb,Gninenko:2012rw}, and here we would 
like to present a comprehensive analysis of the dipole $N$ portal, and derive constrains on $d$ that result from a variety of different experiments, both at high and medium energies.

Previously dipole interactions of neutrinos have been studied in several specific contexts (that we are aware of). If the SM neutrinos 
have a large flavor off-diagonal EM dipole moment, the interaction of solar and reactor neutrinos may get enhanced.
This provides stringent limits on dipole moments of SM neutrinos \cite{Giunti:2014ixa}. 
Some theoretical and phenomenological aspects of the Dirac HNL dipole operator were discussed in Refs. \cite{Aparici:2009fh,Aparici:2013xga} (see also a more recent general discussion of dimension 5 effective operators in the neutrino sector \cite{Caputo:2017pit}). Another prominent place where the transitional $\nu-N$ dipole appears is the literature on searches of sterile neutrino dark matter
via a dipole-induced decay $N\to \nu\gamma$ (\cite{Abazajian:2017tcc} and references therein). 
A more closely related case to the topic of our study has arisen as a consequence of trying to accommodate MiniBoone and LSND anomalies,
that we would like to discuss now in more detail. 

While there is an overall theoretical/experimental consistency for the three-neutrino oscillation picture, there are several experimental 
results that do not fit in. Two notable exceptions are the anomalies observed at the intensity frontier experiments LSND and MiniBooNE \cite{AguilarArevalo:2007it,Athanassopoulos:1996jb}. 
In these experiments, an excess of low energy electron (anti-)neutrinos have been observed, the source of which is currently unknown. Conceivably, there are two possibilities: new physics or some unaccounted SM processes. 
Thus, for example, single photons produced via poorly understood SM neutrino interactions 
with nuclei \cite{Hill:2010zy} might lead to some partial 
explanation of the anomalies. (At the signal level, a single photon
cannot be distinguished from charged-current quasi-elastic events by MiniBooNE's Cherenkov detector.) 

The most popular proposal is the existence of a light ($m\sim\text{eV}$) sterile neutrino 
(\cite{Abazajian:2012ys} and references therein), which mediates the anomalous oscillation required to explain the observed excess signal. 
A possibility of eV sterile neutrinos being at the origin of the MiniBooNE and LSND oscillation results is strongly challenged 
by cosmological data. Indeed, the required parameters for mass splitting and mixing angle will lead to a complete thermalization of a new sterile species via oscillation mechanism. This stands in sharp disagreement with cosmological data (in particular, cosmic microwave background (CMB), Big Bang Nucleosynthesis (BBN) and late-time cosmology) that constrain not only the total number of thermally populated 
relativistic degrees of freedom in the early Universe, but also limits the total neutrino mass $\sum m_\nu \leq 0.17\eV$ at 95\%CL \cite{Couchot:2017pvz}. Consequently, a single eV sterile neutrino is not consistent with cosmology in the absence of new physics. 
At the very least, the minimal model would need to be modified to suppress the oscillations in the early Universe, which is usually achieved
at the expense of significantly enlarging the sterile neutrino sector {\em e.g.} by new types of interactions with dark matter and/or baryons
\cite{Dasgupta:2013zpn,Hannestad:2013ana}. Thus, the sterile neutrino solution to the MiniBooNE and LSND anomalies naturally leads to the idea of a {\em dark sector}, with new matter 
and interaction states.
\\

An alternative attempt to accommodate the anomalies without using eV-scale sterile neutrinos requires 
some dark sector states comparable in mass to the lightest mesons. 
Thus, it has been noted that the presence of a new sub-GeV neutral fermion $N$ may mimic the signals observed at
MiniBooNE and LSND \cite{Gninenko:2009ks,Gninenko:2010pr}. 
The necessary ingredient of this proposal is a new fermionic state $N$ 
in the 10-to-few-100\,MeV mass range and the dipole coupling in \cref{eq:simpledipole}. This coupling mediates a relatively prompt decay of $N$ 
to a normal neutrino and a photon, a signature that can be confused with the ``normal'' electron or positron final state in charged current 
events \cite{Gninenko:2009ks,Gninenko:2010pr}.
Whether this model can simultaneously account for both anomalies without running into problems with other constraints remain an 
open issue (see the discussions in Refs. \cite{Gninenko:2009ks,Gninenko:2010pr,McKeen:2010rx,Masip:2012ke,Masip:2011qb,Gninenko:2012rw}). At the same time the model has a clear advantage over the eV sterile neutrino model, as it creates no problems with cosmology, as $N$ states will decay to the SM at early times before the neutrino decoupling. 

Continuing investment in neutrino physics will eventually lead to better understanding of the origin of these two anomalies. 
The Short-Baseline Neutrino program (SBN) \cite{Antonello:2015lea} is going to be instrumental in testing the 
MiniBooNE anomaly. The design consists of three Liquid Argon time projection chamber (LAr-TPC) detectors that overcome the difficulties present at MiniBooNE by providing excellent photon/electron discrimination. Furthermore, the SBN program will use a near detector (SBND) to control systematic errors related to the neutrino beam content. Being close to the proton target, SBND will see a much larger neutrino flux than the mid-range detectors and will allow a more accurate measurement of the neutrinos before oscillation. In addition, a further increase in sensitivity may result from a proposed new experiment at CERN, Search for Hidden Particles (SHIP) \cite{Alekhin:2015byh}, that will be able to significantly advance the probes to $N$ states, and should also test their dipole interactions. For an analysis of a more conventional CC-dominated model of HNLs in application to Fermilab experiments we refer the reader to a recent paper \cite{Ballett:2016opr}.
\\ 

Motivated by the relative simplicity of the neutrino dipole portal model and its potential applicability to neutrino anomalies, it is very useful to have a comprehensive survey of the model over a large region of parameter space. We therefore consider the energy, intensity and astrophysics frontiers, 
where this portal can be probed. A plot summarizing our results is shown in \cref{fig:masterPlot}, and the rest of the paper considers each probe individually. 
The existing constraints from previous dark matter experiments can be improved by the SBN and SHiP. 
From astrophysics, $\text{MeV}$ HNLs could contribute to the supernova cooling, in particular that of Supernova 1987A (SN 1987A). This happens when the coupling $d$ is large enough so that the star can produce $N$ in sufficient quantity, but small enough so that $N$ can escape and cool down the star without being significantly impeded. For lifetimes longer than $0.1\text{s}-1\text{s}$, $N$ is relevant for, and can modify predictions of, BBN. The late decays of HNLs would modify the proton to neutron ratio, and with some reasonable 
assumptions about the initial cosmological temperatures being high, this puts an upper bound on the lifetime of $N$. We find that there is significant overlap of this region with SN constraints. Lastly, for above$\GeV$ masses, we turn to particle colliders and recast existing searches from the LHC and LEP. Going to particle colliders allows us to probe simple completions of the model which preserve the $\SUgroup(2)\times \Ugroup(1)$ structure of the SM. In these extended models, we have additional production channels stemming from $Z$ and $W$ bosons.

The paper is organized as follows. In Section 2, we provide more details on the model including the possible SM gauge invariant completions and the connections to neutrino masses. In section 3, 4 and 5, we consider the intensity, energy and astrophysical frontiers respectively. Finally, we conclude in section 6 with general remarks. 

\section{Generic features of neutrino dipole portals}\label{sec:2-dipole-portal}
\subsection{Main qualitative features of dipole portal}
\begin{figure}[!ht]
\centering
\begin{subfloat}[Weak meson decays]%
{\includegraphics[clip,width=0.4\linewidth]{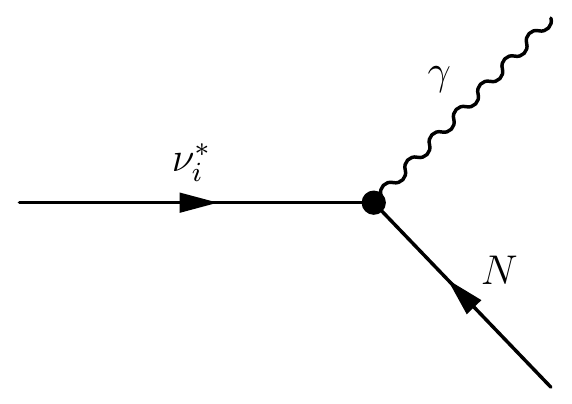}
\label{fig:v-1} }%
\end{subfloat}%
\begin{subfloat}[Dalitz-like decay]%
{\includegraphics[clip,width=0.4\linewidth]{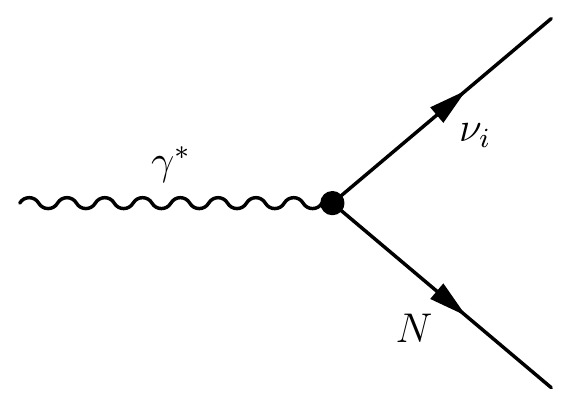}
\label{fig:v-2}}%
\end{subfloat}

 \begin{subfloat}[Primakoff upscattering]%
{\includegraphics[clip,width=0.4\linewidth]{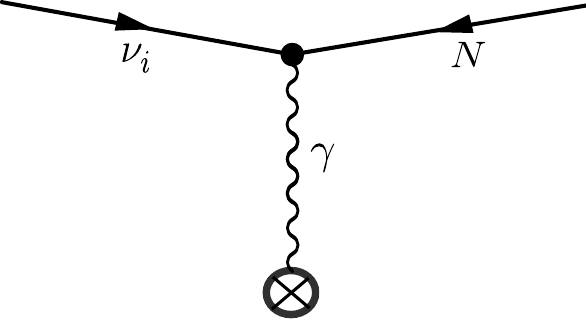}
\label{fig:v-3}}%
\end{subfloat}%
\begin{subfloat}[$N\rightarrow \gamma\nu$ (signal)]%
{ \includegraphics[clip,width=0.4\linewidth]{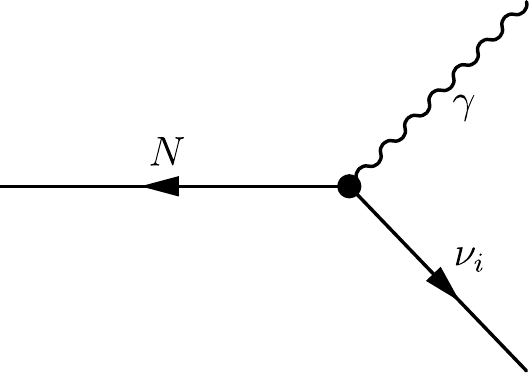}
\label{fig:v-4}}%
\end{subfloat}%
\caption{Dipole portal processes for $N$: \protect\subref{fig:v-1} Production of $N$ from off-shell neutrinos arising from weak meson decays ({\em e.g.} from $\pi,K\to \mu^+ \nu^*$); \protect\subref{fig:v-2} Production of $N$ from off-shell photons arising from Dalitz-like meson decays ({\em e.g.} from $\pi^0,\eta \to \gamma^* \gamma$); \protect\subref{fig:v-3} Production of $N$ from on-shell neutrinos via Primakoff-type upscattering (via photon exchange with the nucleus); \protect\subref{fig:v-4} Decays of $N$ to single photon final states (the main signal studied in this paper). Processes \protect\subref{fig:v-1} and \protect\subref{fig:v-2} are important for production of low mass $N$ at neutrino experiments. Process \protect\subref{fig:v-3} dominates production in supernovas at lower $N$ masses, and at neutrino experiments. Process \protect\subref{fig:v-4} is relevant for energy injection at BBN, for neutrino beam dump experiments, and controls the escape probabilities in supernova for large $N$ masses.\label{fig:v}}
\end{figure}

The consequences of the dipole portal in \cref{eq:simpledipole} can be easily understood by considering the four vertex alignments presented in \cref{fig:v}. The presence of an electromagnetic coupling to neutrinos allows for mesons to decay in two novel ways: Dalitz-like decays mediated by off-shell photons and neutrinoless weak decays with a single photon in the final state. In terms of producing $N$, incident neutrinos can upscatter via the dipole portal, which can be a more efficient production process than mass mixing mechanisms that have been traditionally considered. 
The decay of an HNL in our model will be dominated by single photon production, and for the values of $d$'s we consider in this paper, will occur much more rapidly than in mechanisms that are mediated by the weak force. This single photon signature was identified in Refs. \cite{Gninenko:2009ks,Gninenko:2010pr} as a promising signal, however the production mechanisms outlined above were not included.

We now here our discussion to beam dump experiments. There, production of $N$ will dominantly proceed via neutrino upscattering, wherein an incoming neutrino scatters via a photon to produce $N$. If the incoming neutrino scatters off the whole nucleus and the process happens coherently (i.e. $\sigma\propto Z^2$), we can get a crude estimate for the sensitivity one can achieve. In the limit of infinite mass of the nucleus, the problem reduces to the scattering in the external EM field $A_\mu=(A_0(\vec{q}),0)$ created by the nucleus. Calculating the cross section to logarithmic accuracy for $-\frac{1}{R_\text{nuc}^2}\le t\le -\frac{m_N^4}{4E_\nu^2}$, we find
\be
\sigma_{\nu\rightarrow N}=4\alpha Z^2 |d|^2\times \log\left(\frac{4E_\nu^2}{m_N^4R_\text{nuc}^2}\right).
\ee
Therefore for masses $m_N=50\MeV$, an incoming neutrino energy of $1\GeV$ and $R_\text{nuc}^{-2} \sim 0.3\GeV^2$, we can expect a production cross section per nucleus of roughly 
\begin{equation}
\sigma=4.5\times\left(\frac{Z}{18}\right)^2\left(\frac{d}{10^{-6}\GeV^{-1}}\right)^2 \times 10^{-38}\cm^2.
\label{eq:ballpark}
\end{equation}
It is worth noting that the $d^2$ scaling of the cross section makes it also a relatively mild, logarithmic function of energy, provided 
that $N$ is kinematically accessible. 


For characteristic values of $d$ suggested in \cref{eq:ballpark} and small masses, we can also expect $N$ to be long-lived. This opens up the possibility of HNLs being produced outside of the detector. For example, they could be produced in the dirt or line of sight leading up to the detector, and/or via mesons from the protons-on-target via Dalitz-like decays. Meson production via the dipole portal is an interesting new production mechanism we will discuss, and from dimensional arguments it is clear that the scaling of the meson decay branching to $N$ will occur via
$Br_{M\to N}\propto d^2m_M^2$, where $m_M$ is the mass of the decaying meson.

The decay length associated with the $N\to \nu\gamma$ process is another very important quantity. Given a decay rate of
\be
\Gamma_{N\to \nu\gamma}=\frac{|d|^2m_N^3}{4\pi},
\ee
and an HNL energy of $E_N=1\GeV\gg m_N$, the decay length and lifetime of $N$ scale as
\begin{align}\begin{split}
t_\text{dec}&=\tau\gamma=1.3\times 10^{-6}\text{s}\left(\frac{50\MeV}{m_N}\right)^4\left(\frac{10^{-6}\GeV^{-1}}{d}\right)^2\\
L_\text{dec}&= c\tau\beta\gamma\approx 400\text{m}\left(\frac{50\MeV}{m_N}\right)^4\left(\frac{10^{-6}\GeV^{-1}}{d}\right)^2.
\label{eq:ldecaytdecay}
\end{split}\end{align}
This turns out to be a very convenient length scale for beam dump experiments, if $m_N$ and $d$ have the fiducial values suggested above.
\subsection{Dirac vs Majorana masses and gauge invariant completions}

\begin{figure}[!ht]
\centering
\includegraphics[width=0.9\linewidth]{
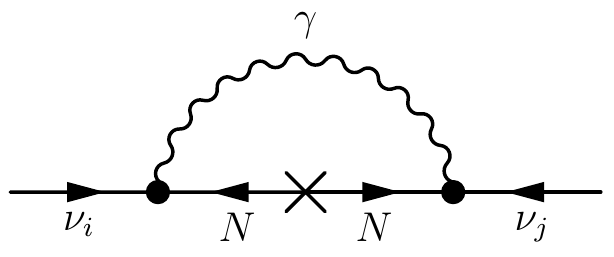}
 \caption{Loop level contribution to the $\nu$ mass mixing matrix in the presence of a Majorana mass term for the heavy neutral lepton $N$. With only Dirac masses, such diagrams will not be generated.}
\label{fig:majoranamassterm}
\end{figure}

If $N_D$ is a Dirac fermion, composed of two Weyl fields
\begin{equation}
N_D=\begin{pmatrix}
N \\
\scalebox{0.9}{$N^{c\dagger}$}
\end{pmatrix},
\end{equation}
one of which is completely decoupled from the SM, then the HNL is decoupled from the mechanism that generates active neutrino masses. Thus, we assume both the absence of mass mixing between $\nu$ and $N$, and a vanishing Majorana mass for $N$. This choice is technically natural and can be achieved by---for example---assigning $N$ the same lepton number as the SM leptons. If such a symmetry is not imposed, and a sizeable Majorana mass term, $\mathfrak{m}_N$, is present then the process shown in \cref{fig:majoranamassterm} can take place. Naive counting of divergences shows that the induced Majorana mass for the neutrinos, $\mathfrak{m}_\nu$ will scale as $\mathfrak{m}_{\nu} \sim d^2\Lambda^2\mathfrak{m}_{N}/16\pi^2$, where $\Lambda$ is the cutoff scale associated with the UV completion of the model, which can be as high as $d^{-1}$. This contribution, despite all the uncertainties, will be much larger than the required mass scale for the neutrinos, unless $N$ is Dirac, or quasi-Dirac with a small Majorana-type mass splitting satisfying $\mathfrak{m}_{N} \ll m_N$.
%
%
%
Quasi-Dirac $N$ would typically lead to {\em larger} values of $d$ than otherwise would be suggested by a simple application of 
the see-saw relation. Consider a model where the SM neutrinos couple to $N$ via a mass mixing interaction of the form $m_{\nu N}~\nu N$. This naturally generates dipole couplings between the SM neutrinos, sterile neutrino and the photon via a loop diagram. The dipole coupling generated is given in \cite{Pal1982,Shrock1982} as 
\be\begin{split}
d&=\frac{3m_{\nu N}}{32\pi^2}\frac{eG_F}{\sqrt{2}}\\
&=1.2\times 10^{-9}\Gev^{-1}\left(\frac{m_{\nu N}}{50\MeV}\right).
\end{split}\ee

The strength of this radiatively generated dipole portal is dictated by the mass mixing with the active neutrinos, and therefore constrained by patterns of the neutrino mass matrices. In particular, in the case of a type-I see-saw mechanism with the Majorana mass of $m_N=50\MeV$, observed neutrino masses would imply $m_{\nu N}\sim \KeV$ and consequently $d\sim 10^{-13} \GeV^{-1}$. We do not impose such a stringent constraint and consider $d$ to be an independent parameter. In fact, 
the size of $d$ can be much larger if the effective mixing angle between $\nu$ and $N$ is 
much larger than the naive see-saw relation implies. This may happen, for example, within an inverse see-saw model \cite{Mohapatra:1986aw,Mohapatra:1986bd}, where a mostly
Dirac fermion $N$ is supplemented with a small Majorana mass, so that the mass mixing parameter $m_{\nu N}$ is much larger than naively implied. 


Above the electroweak scale an $\SUgroup(2)\times \Ugroup(1)$ interpretation of $d$ would require a Higgs insertion, so that the dipole interaction is really a dimension 6 operator. Therefore, in the limit of large $\Lambda$ the maximum expected $d$ is
\be
d_\text{max}\sim \frac{e\mathrm{v}}{\Lambda^2}\sim\frac{100\GeV}{\Lambda^2}
\ee
where strong dynamics at the scale $\Lambda$ is presumed, and $\mathrm{v}$ is the Higgs field vacuum expectation value. Otherwise, if the new sector is perturbative, we would expect a loop factor, and $d_\text{max, pert}\sim\GeV/\Lambda^2$. 
To consider neutrino dipole couplings which respect the full gauge symmetries of the Standard Model, we write down the Lagrangian
\be
\Lag \supset \bar{L}\left(d_\mathcal{W}\mathcal{W}_{\mu\nu}^a\tau^a+d_BB_{\mu\nu}\right)\tilde{H}\sigma_{\mu\nu}N_D+h.c.
\label{eq:fullgenerallagrangian}
\ee
where $\tilde{H}=i\sigma_2H^*$ and $\tau^a=\sigma^a/2$. After spontaneous symmetry breaking of the Higgs, one obtains
\begin{align}\begin{split}
\Lag \supset & d_W\left(\bar{\ell}_L W^{-}_{\mu\nu}\sigma^{\mu\nu}N_D\right)\\
+ &\bar{\nu}_L[d_\gamma F_{\mu \nu}-d_Z Z_{\mu\nu}]\sigma^{\mu\nu}N_D
+h.c.
\label{eq:brokenphaseUVcompletion}
\end{split}\end{align}
where $W^{-}_{\mu\nu}\equiv \partial_\mu W^-_\nu-\partial_\nu W^-_\mu$. The dipole couplings in the broken phase are related to those in the unbroken phase via 
\begin{align}\begin{split}
d_\gamma&=\frac{\mathrm{v}}{\sqrt{2}}\qty(d_B \cos\theta_w +\frac{d_\mathcal{W} }{2} \sin\theta_w) \\
d_W &=\frac{\textrm{v}}{\sqrt{2}} \frac{d_\mathcal{W}}{2} \times \sqrt{2}\\
d_Z&= \frac{\mathrm{v}}{\sqrt{2}}\qty(\frac{d_\mathcal{W}}{2}\cos\theta_w-d_B \sin\theta_w)
\label{eq:brokenUnbrokenCouplings}
\end{split}\end{align}
where the additional factor of $\sqrt{2}$ in the expression for $d_W$ is a consequence of the normalization of $W^{-}=(\mathcal{W}^{1}+i\mathcal{W}^{2})/\sqrt{2}$. 
Note that the three ``dipole moments'' in the broken phase $d_\gamma$, $d_Z$ and $d_W$ are determined by only two parameters in the unbroken phase $d_\mathcal{W}$ and $d_B$; they are linearly dependent. 
 Notice that the normalization of the photon field strength term in \cref{eq:brokenphaseUVcompletion} matches that of \cref{eq:simpledipole}.\\ 

Although we have suppressed the relevant indices, the dipole coupling can be flavor dependent. Experiments at SBN will constrain $d^e_B$ and $d^\mu_B$. SHiP in addition will be sensitive to $\nu_\tau$, and thus an ideal setting to study all ``dipole couplings''. 
For both LHC and LEP, we turn on only the $d^\mu_{\gamma,B,W}$ coupling for simplicity. One can also turn on $d^e_{\gamma,B,W}$ and $d^\tau_{\gamma,B,W}$ that have an O(1) effect on the result.\\

Having established that a neutrino dipole portal is ultimately a dimension 6 operator, one might wonder if there are any non renormalizable SM only operators that are phenomenologically equivalent to our new physics signal. If so, one would need to perform a global fit on the whole basis of Wilson coefficients instead of focusing on just one operator. The case of SM only operators after electroweak symmetry breaking is considered in \cref{sec:backgrounds}. \citer{Grzadkowski:2010es} on the other hand provides a classification of all dimension 5 and 6 SM only operators above the electroweak scale (i.e. invariant under $\SUgroup(3)\times \SUgroup(2)\times \Ugroup(1)$). In order to replicate our signature, we need at least one photon, one neutrino and an additional gauge boson. If we assume that no particles except neutrinos escape detection, and furthermore that the interactions are $2\rightarrow\text{N}$, then none of the dimension 5 or 6 operators in \citer{Grzadkowski:2010es} contribute to single photon processes at beam dump experiments, LEP or the LHC. 


\section{Intensity Frontier}\label{sec:4-intensity-frontier}
We consider probing HNLs at beam dump experiments and our analysis focuses on neutrino experiments hosted at CERN, Los Alamos and Fermilab. Fermilab is building a substantial Short-Baseline Neutrino oscillation program \cite{Antonello:2015lea} that among other physics 
goals will settle the question of sterile neutrinos at $\Delta m^2\sim 1\eV^2$. It will consist of 3 LAr-TPC detectors called SBND, MicroBooNE and ICARUS, which will be spread out over a 600m range from the proton target.
The SBN program is designed to achieve a $5\sigma$ sensitivity in the parameter space of $(3+1)$ sterile neutrino models consistent with LSND at $99\%\text{CL}$. These detectors can resolve photons from electrons with a $94\%$ photon rejection rate. \\

At CERN, we will be interested in the past experiment NOMAD and future proposal SHiP. The proposed SHiP experiment is unique among beam dump experiments in that it features very large neutrino energies and a sizeable flux of electron, muon and tau neutrinos. Furthermore, the use of lead inside the neutrino detector, $Z=82$, will provide an ideal setting to take advantage of coherent production, which scales as $Z^2$. At Los Alamos, we consider the LSND experiment which will prove to be useful at low HNL masses. In what follows, we discuss the various production mechanisms at beam dumps, the main backgrounds involved in the search, and our results.

\subsection{Production mechanisms}
At neutrino beam dump experiments, HNL production can happen in three principle ways. The first---and most familiar--- mechanism is mass mixing, however this is subdominant in our analysis by assumption. The two dominant production mechanisms are therefore meson decays 
and Primakoff upscattering, both of which are explained in greater detail below. In principle DIS production via Drell-Yan like processes is also possible, but we found this to be subdominant. 

\subsubsection{Primakoff upscattering}
\begin{figure}[!ht]
\centering
\begin{center}
\includegraphics[width=1\linewidth]{
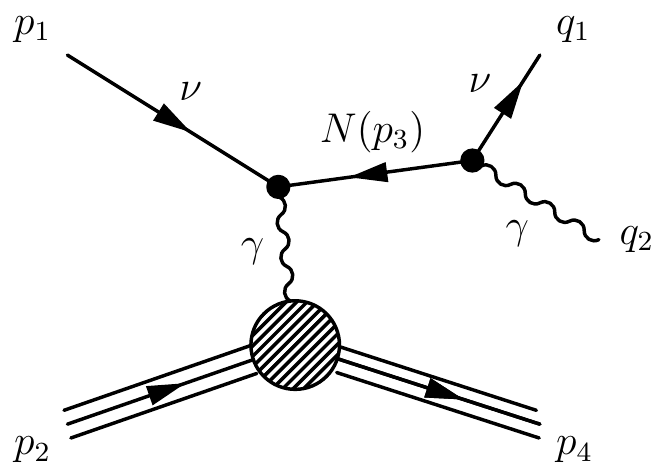}
\end{center}
\caption{Tree level neutrino scattering process with a final state photon, arising from dipole portal to HNL. We work in the narrow width approximation, and assume the above diagram factorizes. }
\label{fig:signalfeynmandiagram}
\end{figure}

Neutrino upscattering is the dominant production mechanism for $N$ across a wide range of masses for the experiments we consider. It happens when an incoming neutrino interacts with matter and upscatters into a long-lived HNL state $N$. The HNL subsequently decays into a neutrino and a photon; an explicit example is provided in \cref{fig:signalfeynmandiagram}.
This process can either happen inside the fiducial volume of the detector or in the line of sight separating the proton target from the detector. In all our results, we employ the narrow width approximation, since $N$ is usually produced on-shell and travels some distance before decaying. Having an HNL lifetime and energy consistent with the necessary flight distance is enforced by 
\be
P_\text{dec}(L_1,L_2)=\exp[-L_1/L_\text{dec}]-\exp[-L_2/L_\text{dec}].
\label{eq:decayprobability}
\ee
In \cref{sec:neutrinoupscattering}, we present the details of how the cross section is obtained for coherent and diffractive scattering. We apply the cuts described in \cref{sec:analyticcuts} to ensure proper kinematics of the photon. There, it is also fully described how the region of integration of $t$ is determined. Once we have obtained the cross section, cuts, photon detection efficiency and luminosity, we can set limits following the discussion in \cref{sec:sensitivity}.


\subsubsection{Meson decays}

At low mass, HNLs are long lived and represent a kinematically allowed decay channel for light mesons. Unlike mass mixing induced decays, the dipole portal allows for electromagnetically mediated Dalitz-like pathways in addition to weak decays mediated by an off-shell neutrino. The qualitative features that can allow for significant production of HNLs are
\begin{enumerate}[(i)]
\item High meson multiplicity per proton ({\em e.g.} pions).
\item $\textrm{BR}(\text{Meson}\rightarrow X+\gamma)=\order{1}$ ({\em e.g.} $\pi^0$, $\eta$) or $\textrm{BR}(\text{Meson}\rightarrow X+\nu)=\order{1} $ ({\em e.g.} $\pi^\pm$, $K$).
\end{enumerate}
In terms of meson production at the experiments we consider, the largest difference between them is that immediately following the proton target, SBN features a $50\text{m}$ meson decay chamber, whereas SHiP has a hadron stopper. This divides our discussion into prompt ($\tau^\text{rest}\lesssim 10^{-12}s$) and long-lived ($\tau^\text{rest}\gtrsim 10^{-12}s$) mesons. Only the former will contribute to HNL production at SHiP, whereas both will be relevant at SBN due to its long decay chamber. To obtain rates, we calculate the differential cross section of HNL production from mesons in the meson rest frame, which we combine with the meson fluxes in the lab frame. The details of these calculations are outlined in \cref{sec:mesonphasespaceappendix}. \\ \\
The species we have included in our analysis are shown in \cref{tab:mesonmultiplicityperPOT}, from which it is clear that the prompt mesons are $\pi^0$ and $\eta$. For both of these, the dominant channel for HNL production is 
\be
\pi^0,\eta\rightarrow \gamma (\gamma^*\rightarrow \nu_a N)
\label{eq:dalitzdecay}
\ee
We immediately see that these radiative Dalitz-like decays will be useful for improving the sensitivity to $d_e$ and $d_\tau$ flavored couplings, since the process in \cref{eq:dalitzdecay} is universal in the flavor $a$. By contrast, neutrino upscattering at beam dump experiments is limited by smaller incident fluxes of $\nu_e$ and $\nu_\tau$ neutrinos, as compared to $\nu_\mu$ neutrinos. The long-lived mesons we consider are $\pi^\pm$ and $K^\pm$. They can produce HNLs via an off-shell neutrino decay
\be
\pi^\pm,K^\pm\rightarrow \mu^\pm\left(\accentset{(-)}{\nu}_\mu{\tiny ~}^*\rightarrow \gamma\accentset{(-)}{N}\right).
\label{eq:longlivedmesondecays}
\ee
When considering decays to electron flavor, such as $K,\pi \to e\nu_e$, one typically expects a chiral suppression of $\order{m_e^2/m_\mu^2}$ in the branching ratio relative to the muon channel. While we 
concentrate on \cref{eq:longlivedmesondecays} for muon flavors at SBN, we note that 
$K,\pi \to eN\gamma$ will avoid chiral suppression 
due to the chirality-flipping nature of the dipole portal. 
The $K^+$ states, whose rates are about a tenth of those of pions, are important because they allow production of heavier HNLs. \\ \\
To get a handle on which mesons are expected to contribute most, we calculated the average multiplicities of each meson per proton on target at SBN. Our results are shown in \cref{tab:mesonmultiplicityperPOT}. The $\pi^-$ multiplicity has been calibrated to match that of Table X in \cite{PhysRevD.79.072002}, and we find very good agreement for the other meson multiplicities. No distribution parameters for $K^-$ and $\eta$ were available, and so we rescaled those of $K^+$ to match expectations. Both $K^-$ and $\eta$ contributions are very small, so the discrepancy in average momentum and angle as compared with Table X has a negligible effect on our results. We conclude that pions will be the most important mesons for sourcing low mass HNL particles.

\begin{table}
\begin{tabular}{ccccl}
\toprule
\multicolumn{1}{c}{\shortstack{Meson\\ Species}}	& \multicolumn{1}{c}{\shortstack{Multiplicity\\per POT}} 	& \multicolumn{1}{c}{\shortstack{$\langle p \rangle$\\$[\text{GeV}/c]$}}	& \multicolumn{1}{c}{\shortstack{$\langle \theta \rangle$\\$[\text{mrad}]$}} & \multicolumn{1}{c}{\shortstack{$\langle\tau\rangle$\\$[\text{sec}]$}} 	\\
\midrule
 $\pi^-$ & 0.9004 & 0.83 & 527 & $~2.6\cdot 10^{-8}$\\
 $\pi^+$ & 0.9784 & 1.07 & 423 & $~2.6\cdot 10^{-8}$\\
 $\pi^0$ & 0.9098 & 0.89 & 483 & $~8.4\cdot 10^{-17}$ \\
 $K^+$ & 0.0689 & 1.33 & 410 & $~1.2\cdot 10^{-8}$ \\
 $K^-$ & 0.0024 & 1.29 & 409 & $~1.2\cdot 10^{-8}$ \\
 $\eta$ & 0.0295 & 1.35 & 403 & $~5.0\cdot 10^{-19}$ \\
\bottomrule
\end{tabular}
\caption{Meson multiplicities, average momentum and average angle at the SBN facility. Pions are assumed to follow a Sanford-Wang distribution, while kaons and etas are calculated based on the Feynman Scaling distribution.}
\label{tab:mesonmultiplicityperPOT}
\end{table}

\subsection{Backgrounds \label{sec:backgrounds}}
The main backgrounds for HNLs will be single photon signatures, arising from mis-reconstructed $\pi^0$ or radiative resonance decays such as $\Delta\rightarrow N\gamma$. At SHiP, there is not much publicly available information, and therefore we consider various benchmark estimates for these backgrounds. We guide our estimate by considering the observed single photon backgrounds at NOMAD, rescaled to account for differences in the target mass and number of protons on target. \\

On the other hand, the SBN collaboration has estimated the number of single photon events that can fake a $\nu_e$ CC signature in each of its detectors. We can estimate the total single photon background by taking this number and dividing it by $6\%$ to factor out the photon rejection rate.
%
We then impose a $200\MeV$ threshold in our results since the single photon backgrounds grow with decreasing energy. To account for signal photons that may have been lost, we apply a $20\%$ signal efficiency cut. 

The backgrounds at LSND are similar in spirit to those at MiniBooNE, in that electron-like events arise from both electron and photon sources. In order to obtain constraints, we base our analysis at LSND on an electron-neutrino elastic scattering search \cite{Auerbach:2001wg}, respecting the fiducial geometry and energy cuts described in that paper. Examining Fig. 1 and 10 of \cite{Auerbach:2001wg}, we note that the incident neutrino flux favors energy values between 30-50$\MeV$, whereas the collected electron-like sample peaks at energies around 22$\MeV$. Single photons from HNL decays on the other hand tend to be much harder and closer in energy to their parent SM neutrino. We therefore explore two different recast strategies. In the first case, we impose a lower threshold on the incident neutrino energy of 18$\MeV$. This corresponds to the full dataset collected by LSND, comprising of roughly 300 predicted background and data events. In the second strategy, we impose a lower energy cut of 40$\MeV$ in an attempt to better discriminate our new physics signal from SM backgrounds. This cut amounts to keeping roughly 27 predicted background and data events. We find that the latter strategy provides slightly better sensitivities to HNLs, and these are the LSND results that feature in all of our plots. 

Lastly, diagrams containing loops of charged leptons and either a $W$ or $Z$ boson, can induce an effective $\gamma \gamma \nu \nu$ vertex in the SM and provide a potential source of single photon backgrounds. We have explicitly estimated the size of this background in \cref{sec:LoopSMBackgrounds} and it is many orders of magnitude lower than the HNL production cross section estimated in the previous section, and can therefore safely be ignored.

\subsection{Experimental results and prospects}
\label{sec:experiments}
In what follows we describe and summarize the implications of existing measurements at LSND and MiniBooNE. We also comment on the projected reach of ongoing and future experiments such as MicroBooNE and SHiP.
\subsubsection{LSND}
The LSND oscillation anomaly, which consists of an excess of $\bar{\nu}_\mu\rightarrow \bar{\nu}_e$ events \cite{Athanassopoulos:1996jb}, has historically motivated interest in sterile neutrinos. While common interpretations of the excess typically involve very light sterile states, more recently it has been proposed that a dipole portal coupled with HNLs with $m_N\approx 50~\text{MeV}$ could explain the excess \cite{Gninenko:2009ks,Gninenko:2010pr,Masip:2012ke}. It is therefore of great importance to consider the observations at LSND and their implications for dipole portals to HNLs.

The setup at LSND involves a neutrino flux coming primarily from $\mu^+$ and $\pi^+$ decays at rest \cite{Auerbach:2001wg}. Consequently the dominant production channel of HNLs is through neutrino upscattering. In modelling the production of HNLs at LSND we include Primakoff upscattering of neutrinos, as well as decays in flight for $\pi^0$, decays at rest for $\mu^+$ and decays both at rest and in flight for $\pi^+$. We account for the change in LSND's source of neutrinos, LAMPF, and include two years of data assuming a water based target and three years of operation using a high-Z target (mostly tungsten) \cite{Auerbach:2001wg}. For our purposes the primary effect of the target material is to modify the incident flux of neutrinos, and mesons.

The decays in flight of $\pi^+$ and $\pi^0$ are modelled assuming a Burman-Smith distribution with appropriate parameters for both water and tungsten \cite{BURMAN1990621,Burman_Smith_1989}. Additionally, the decay at rest of $\mu^+$ and $\pi^+$ contribute to the production of HNLs. The decay mode of interest for $\pi^0$ is a Dalitz-like decay, while for $\mu^+$ and $\pi^+$ an off-shell neutrino mediates the production of HNLs. This off-shell neutrino can be either $\overline{\nu}_\mu$, or $\nu_e$ and we include both of these processes in our analysis. Summing all of these processes, and appropriately boosting the HNLs from decays in flight, leads to an incident flux of HNLs which may enter the detector and decay leaving a single photon signature. 

On top of a flux of HNLs due to pion and muon decays, Primakoff upscattering of neutrinos in transit on their way to the detector can provide an additional source of HNLs. Alternatively, upscattering can occur within the detector itself. These processes have to be considered separately since much longer decays are possible in the case of the former, the target material upon which the neutrino upscatters is different, and angular cuts will be dictated by the different geometries.

When upscattering in transit to the detector, the medium of interest is the dirt---and other terrestrial material---along the line of sight between the source and the detector. In our analysis this is modelled as SiO$_2$ and we include both coherent and diffractive scattering. The produced HNL must be directed in a range of solid angle so as to guarantee that it passes through the detector. The range of angles for which this occurs is different depending on how far away the HNL is produced from the detector. To account for this effect, we analyze ten evenly spaced points between the source and the detector. At each of these points, given a flux of neutrinos, we calculate the number of HNLs that would both be produced \emph{and} enter the fiducial volume of the detector. The LSND detector is off-axis from the neutrino source, and is roughly cylindrical in shape, and so we define the angular cuts such that the HNL would pass through the bottom-near and top-far corners (relative to the neutrino source) of the detector; the angular cuts are implemented as described in \cref{sec:analyticcuts} and account for fiducial cuts at the bottom of the detector. In addition to passing through the detector, the HNL's subsequent decay must occur within the fiducial volume for a signal to be observed. We account for this effect by including the probability that the HNL decays in the fiducial volume \cref{eq:decayprobability}. Angular cuts within the detector are, as before, described in \cref{sec:analyticcuts}.

It is also possible that upscattering occurs within the fiducial volume of the detector. For LSND this implies a target composed of CH$_2$ (mineral oil) for the incident neutrinos, and implies furthermore that neutrinos can be produced and subsequently decay along the entire line of sight. We account for this effect at leading order in the limit of $L_\text{dec}\gg L_\text{fid}$, which is the relevant regime when considering the minimal bound on the dipole-coupling of the HNL. We restrict the production of HNLs to the forward pointing hemisphere (\emph{i.e.} an angular cut of $\theta\leq \pi/2$), due to experimental cuts. Additionally, we only include the effects of coherent scattering due to the presence of a hadronic veto within the detector.

\subsubsection{Fermilab's SBN program}\label{ssec:miniboone}
At Fermilab, we are interested in the past experiment MiniBooNE, as well as ongoing experiments involving the SBND and MicroBooNE detectors.
At MiniBooNE, we consider the existing search for $\nu_\mu\rightarrow \nu_e$ quasi-elastic scattering events \cite{AguilarArevalo:2007it}. When limited to reconstructed neutrino energies of $475<E_\nu^{QE}<1250\MeV$, they find very good agreement between background and data.
However, for energies between $300$ and $475\MeV$, MiniBooNE sees a persistent excess. MiniBooNE, being an oil based Cherenkov detector, cannot distinguish electrons from photons. A possible explanation for the excess \cite{Hill:2010zy} is from the $\Delta\rightarrow N\gamma$ process faking a $\nu_e$ signal. A direct chiral perturbation theory calculation finds these rates to be twice as big as data driven estimates from MiniBooNE. 

\begin{figure}[!ht]
\begin{center}
\includegraphics[width=0.9\linewidth]{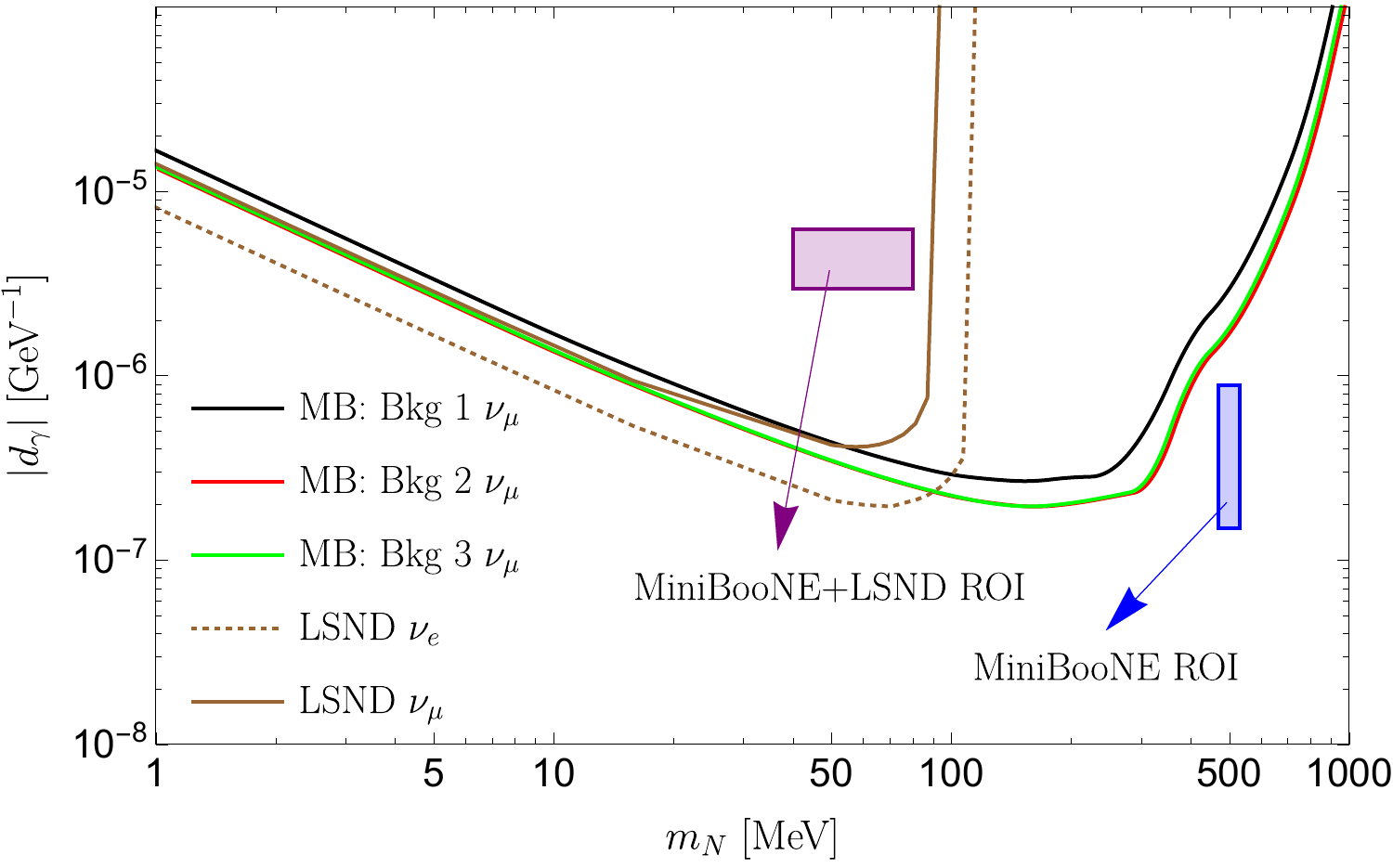}
\end{center}
\caption{
$95\%$ CL limits for HNL particles using MiniBooNE and LSND $\nu_e$CC measurements. In light of the experimental anomaly, background option 1 (Bkg 1) uses the data and backgrounds as is, option 2 includes an alternative stronger $\Delta\rightarrow \gamma N$ background estimate \cite{Hill:2010zy}, and option 3 includes only neutrino energies in the anomaly-free region ($E_{\nu_e}>470\MeV$). We also overlay regions of interest (ROI) from the MiniBooNE and LSND anomalies (see text).}
\label{fig:MiniBooNEExclusionPlot}
\end{figure}

The more exotic interpretations of the MiniBooNE and LSND anomalies \cite{Gninenko:2009ks,Gninenko:2010pr} involve additional single photons from new physics coming from an HNL model with a large dipole coupling $d$ and an active neutrino mass mixing term in the range $|U_{N\nu}|^2\simeq 10^{-3}-10^{-2}$. In that case, production of HNL arises from neutral current $\nu$ scattering that leads to the production of HNL. In \cref{fig:MiniBooNEExclusionPlot}, we revisit the constraints from MiniBooNE by considering both production and decay stemming only from the dipole portal. Since it is difficult to reconstruct HNL energies (due to energy being carried away by outgoing neutrinos), we take an inclusive approach and sum over all the backgrounds and data bins. We calculate the allowed 95\%CL HNL limits following the procedure in \cref{sec:sensitivity} for three different assumptions, which we denote by Bkg 1, 2 and 3 in \cref{fig:MiniBooNEExclusionPlot}. Firstly, we use the data and backgrounds as given in \cite{AguilarArevalo:2007it}. Secondly, we repeat the analysis after including the additional sources of backgrounds identified in \cite{Hill:2010zy}. And lastly, we compute constraints taking into account only the $E_\nu>470\MeV$ region. Based on \cite{Hill:2010zy}, we assume a $25\%$ photon identification efficiency to account for resolution and smearing effects. The photon energy detection threshold is $140\MeV$. Comparing our results to \cite{Gninenko:2010pr} where dipole portal production mechanisms are ignored, we see that around $50\MeV$ masses production from dipole portal is actually dominant. An explicit calculation reveals that for the best fit parameters in \cite{Gninenko:2010pr}, the dipole production cross section is roughly 20 times larger than production from mixing, and so this explanation appears to be excluded. This point is discussed in \cite{Masip:2012ke}, and in the same work, the authors attempt to accommodate the constraint from the muon capture with photon emission at TRIUMF \cite{Bernard:2000et, McKeen:2010rx} by introducing an additional heavy neutrino $\nu_{h'}$. In this way $N$ can decay to $N\rightarrow \nu_{h'}\gamma$ as a main decay channel, and the branching ratio to $\nu_\mu$ can be adjusted to accommodate the LSND/MiniBooNE anomalies while evading muon capture bounds. This same model was recently considered in the context of coherent and diffractive scattering at both MiniBooNE and MicroBooNE \cite{Alvarez-Ruso:2017hdm}. In contrast, we make no attempt to go beyond the minimal dipole coupling and we therefore exclude the favored regions of \cite{Gninenko:2010pr}. 

\begin{figure}[!ht]
\begin{center}
\includegraphics[width=0.9\linewidth]{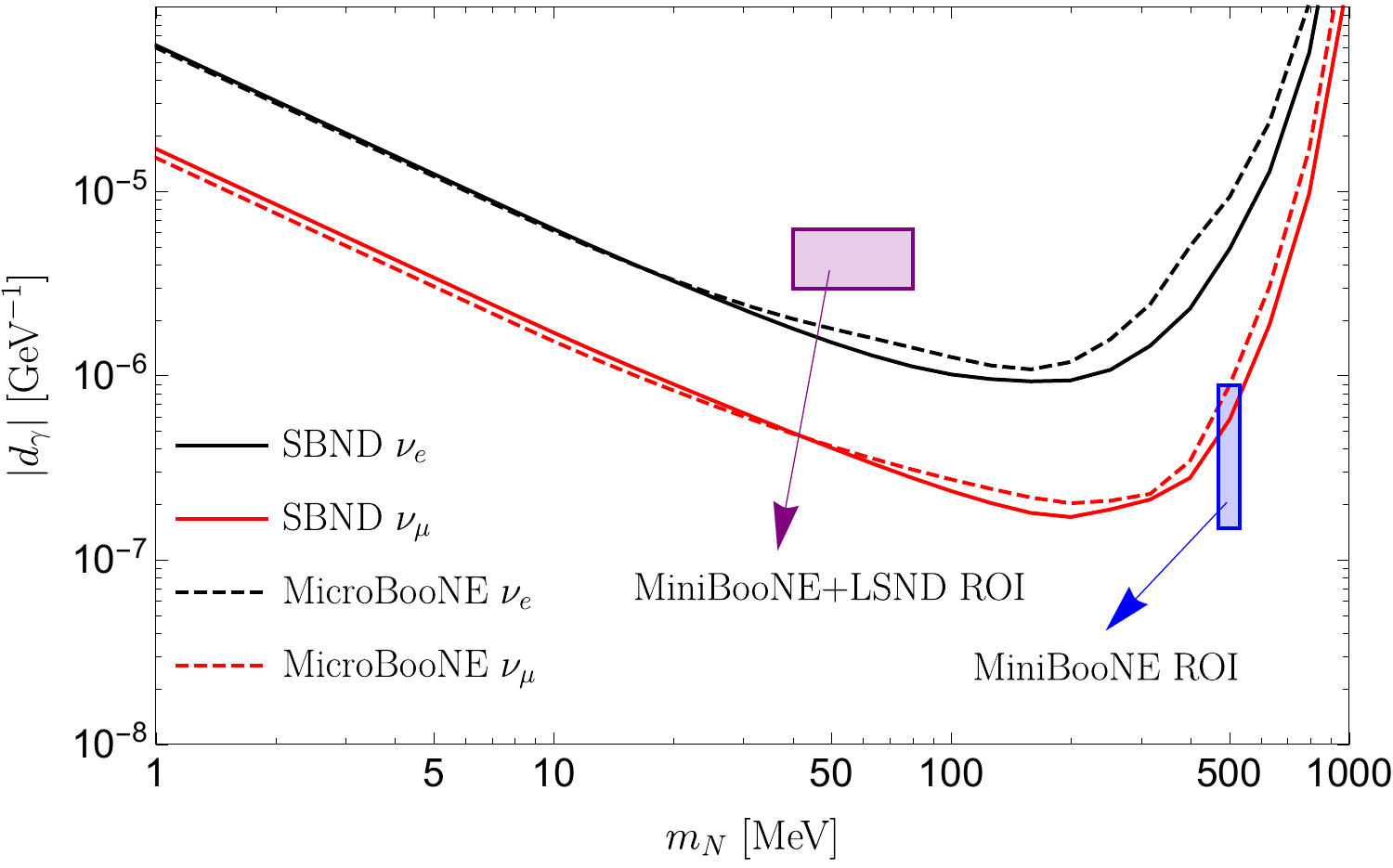}
\end{center}
\caption{Projected $95\%$ CL sensitivities at Fermilab's upcoming Short-Baseline Neutrino program \cite{Antonello:2015lea}. Results for electron (black) and muon (red) dipole couplings are shown for the SBND near detector (solid) and the MicroBooNE middle detector (dotted). Backgrounds are calculated based on expected lifetime single photons (see text).}
\label{fig:figureMicroLAr1dedmu}
\end{figure}

For $500\MeV$ HNL masses explaining MiniBooNE data, we find that production from mixing dominates. Therefore in order to obtain stronger dipole-only constraints, we turn to ongoing and future experiments. Our results for SBND and MicroBooNE are shown in \cref{fig:figureMicroLAr1dedmu}. They assume $6.6 \times 10^{20}$ POT of data in SBND and $13.2 \times 10^{20}$ POT of data in MicroBooNE. As we see, after only 3 years of data taking, they can start cutting into favored parameter space, provided photon data is collected in this duration.

\subsubsection{SHiP and NOMAD \label{ssec:3d-NOMAD}}
\begin{figure}[!ht]
\begin{center}
\includegraphics[width=0.9\linewidth]{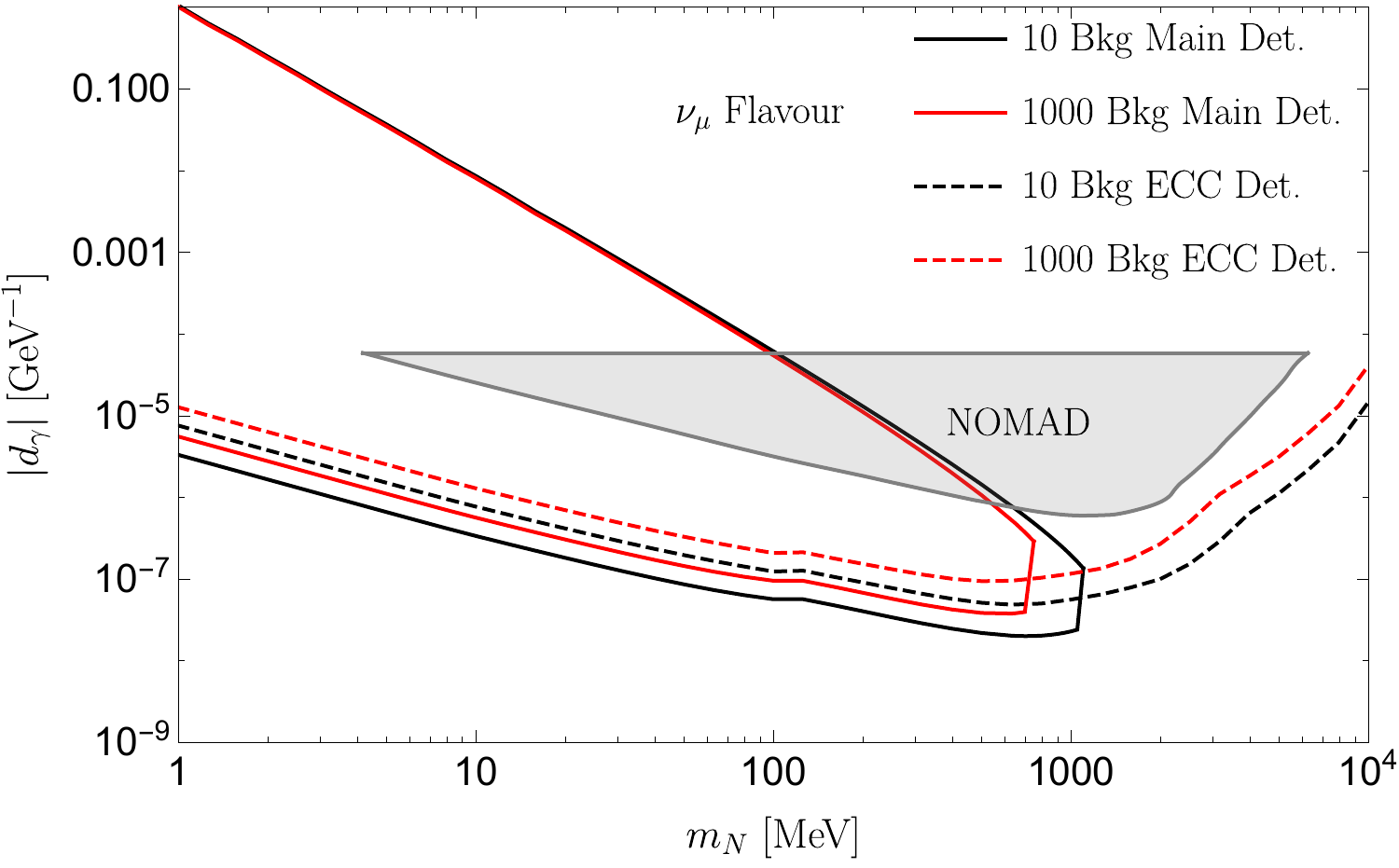}
\end{center}
\caption{Projected $95\%$ CL sensitivities at SHiP for muon neutrino dipole moments. Solid (dotted) lines indicate the main (ECC) detector, and black (red) lines represent 10 (1000) background events during the lifetime of the experiment. We also overlay existing constraints \cite{Gninenko1998b,Gninenko1998a} from NOMAD.}
\label{fig:captionfigureCERNdmu}
\end{figure}

At CERN, we will consider the NOMAD experiment, which ran from 1995-1998 \cite{Vannucci:2014wna,Altegoer1998a,Altegoer1998b}, and the proposed SHiP experiment. Both of these experiments are based on CERN's Super Proton Synchrotron, and consequently have neutrino fluxes extending to larger energies as compared to Fermilab. NOMAD has already performed a search for single photon production. Using this data corresponding to $1.45\times 10^{18}$ POT, Monte Carlo simulations of HNL signals (with no mass mixing) where performed \cite{Gninenko1998a,Gninenko1998b} to simulate the Primakoff process $\nu_\mu Z\to N Z$. The signature of interest was an isolated electromagnetic shower corresponding to a single photon with energy distributed from 0 to $E_\nu$, with $ E_{\nu_\mu}/2$ as an average. The backgrounds, estimated to be roughly 10 events, come mainly from $\pi^0$ production, as well as $\nu_e$ CC interactions. The full results\footnote{The dipole coupling in their paper, ($\mu_\text{trans}$), differs from ours by a factor of $2$ ($\mu_\text{trans}=2d$).} from their simulation are shown in \cref{fig:captionfigureCERNdmu}.

CERN has also proposed a future high energy facility called SHiP \cite{Anelli:2015pba}. If indeed funded and built, it would provide some of the strongest probes of heavy neutral leptons to date \cite{Alekhin:2015byh}. At SHiP, neutrinos are produced by $400\GeV$ protons impinging on a molybdenum and tungsten target. A hadron stopper immediately after the target allows only prompt meson decays, and a magnetized iron shield deflects muons. Following this is an emulsion cloud chamber near detector (which we will refer to as ``ECC detector'') containing lead bricks, a vacuum decay chamber followed by the main detector (which we will refer to as ``main detector''). 
The length of the whole experiment would be on the order of $100\text{m}$. It is advantageous to consider HNL production from prompt mesons, the line of sight, and lead bricks in order to maximize our sensitivity to a large range of HNL lifetimes. We apply a photon detection efficiency of $80\%$ and an energy threshold of $0.1\GeV$. \\

\begin{figure}[!ht]
\begin{center}
\includegraphics[width=0.9\linewidth]{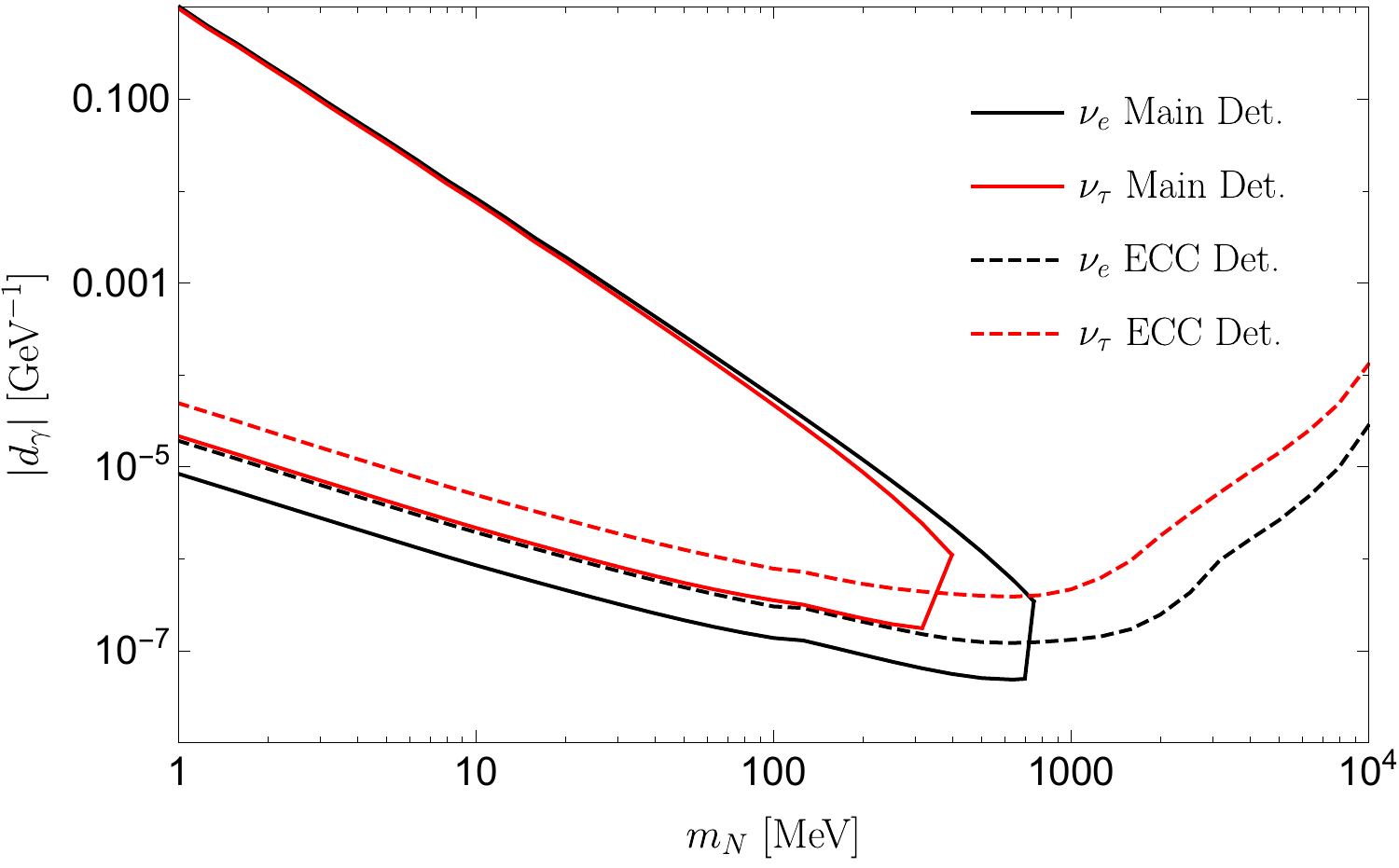}
\end{center}
\caption{Projected $95\%$ CL sensitivities at SHiP for electron (black curve) and tau (red curve) neutrino dipole moments. Solid (dotted) lines indicate the main (ECC) detector. In this plot, we assume 100 background events.}
\label{fig:ExclusionPlotSHiPflavors}
\end{figure}

A unique feature of SHiP is that it is expected to have a sizeable flux of $\nu_e$ and $\nu_\tau$ neutrinos. Therefore, we can interpret the results of the single photon search as constraints on $d^f_\gamma$, for a given flavor $f$. Recall that flavor indices in \cref{eq:simpledipole,eq:fullgenerallagrangian} are suppressed and a priori general. The projected sensitivities achievable at SHiP are shown in \cref{fig:captionfigureCERNdmu} for muon flavors assuming 2 different benchmark choices for the number of background events (10 and 1000 background events). In \cref{fig:ExclusionPlotSHiPflavors}, we show the sensitivity for electron and tau dipole moments assuming 100 background events. At SHiP, single photon rates have not yet been studied. We can obtain a naive estimate by comparing to NOMAD, which had about 10 background events with 100 times less protons-on-target than SHiP. Therefore, with higher luminosities coupled to improved detector capabilities, it is reasonable to estimate around 100-1000 background events in the SHiP ECC detector. This detector will probably have more background events than the main detector, since the latter is surrounded by veto structures designed to reduce backgrounds as much as possible. For the SHiP curves appearing in \cref{fig:masterPlot}, we assume 1000 backgrounds events in both detectors in order to provide a conservative estimate. 

\section{Energy Frontier \label{ssec:3c-LEP}}

\subsection{Production mechanisms}
Beam dump experiments feature very large luminosities, however, the masses of $N$ which are accessible are limited by the incoming neutrino energy spectrum, typically peaked around 1$\GeV$, or between $10-20\GeV$ in the case of SHiP. In contrast, particle colliders can probe much larger masses at the expense of smaller luminosities \cite{Aparici:2009fh}. Additionally, since dipole operators must couple to either $B_{\mu\nu}$ or $W_{\mu\nu}$ above the electroweak scale there is the added possibility of on-shell production of the $Z$ and $W$ mediators. The HNL couplings appearing in all of the high energy plots for LEP and the LHC are defined as follows. We take the relations in \cref{eq:brokenUnbrokenCouplings} and rescale $d_{B,\mathcal{W}}\equiv \sqrt{2}~\overline{d_{B,\mathcal{W}}}/(\mathrm{v}\cos\theta_w)$ to obtain
\begin{align}
\begin{split}
d_\gamma&=\overline{d_B}+\frac{\tan\theta_w}{2}\overline{d_\mathcal{W}}\\
d_W&=\frac{\overline{d_\mathcal{W}}}{\cos\theta\sqrt{2}}\\
d_Z&=\frac{\overline{d_\mathcal{W}}}{2}-\tan\theta_w\overline{d_B}.
\end{split}\label{eq:reduced-dipoles}
\end{align}
\cref{tab:HEPcouplingsLabel} illustrates the assumptions made in each of the exclusion curves for LEP and the LHC.
\begin{center}
\begin{table}[h!]
\begin{tabular}{ | c | c | c | c | }
 \hline
 Exp. & Plot Label & Assumptions & Probed $d$ \\ \hline
 LEP & $d_\gamma$ & $\overline{d_\mathcal{W}}=0$, $d_Z=0$ & $\overline{d_B}$ \\
 & $d_{\gamma,Z}$ & $\overline{d_\mathcal{W}}=0$ & $\overline{d_B}$ \\\hline
 LHC & $d_{\gamma,Z}$ & $\overline{d_\mathcal{W}}=0$ & $\overline{d_B}$ \\
 & $d^a_{\gamma,W}$ & $d_\gamma=a\times \overline{d_\mathcal{W}}$ & $\overline{d_\mathcal{W}}$ \\
 \hline
\end{tabular}
\caption{Assumptions and conventions used in obtaining constraints at LEP and the LHC for the minimal HNL models and the HNL extensions respecting the $\SUgroup(2)\times \Ugroup(1)$ symmetry of the Standard Model. 
\label{tab:HEPcouplingsLabel}}
\end{table}
\end{center}
We now discuss the mechanisms for producing HNLs at LEP and the LHC, and then discuss the details of the analyses and our results.

\subsubsection{LEP}
At LEP, production will proceed via $e^+e^-\to (N\to\gamma\nu)\bar{\nu}+h.c.$. The signature to look for is thus a single photon final state with missing energy. This channel can proceed via either $Z$ or $\gamma$ mediators depending on the dipole coupling in the unbroken phase (see \cref{eq:brokenphaseUVcompletion,eq:reduced-dipoles}). Therefore the total production cross section at $s=m_Z^2$ for $e^-e^+\to N\bar{\nu}$ integrated over all angles is
\begin{align}\begin{split}
&\sigma_{N\nu}=\frac{\alpha \left|\overline{d_B}\right|{}^2 \left(m_N^2-m_Z^2\right){}^2 \left(2 m_N^2+m_Z^2\right)}{6 \text{cos$^2\theta $}_w \text{sin$^2\theta
 $}_w m_Z^6 \Gamma _Z^2}\times \\
&\left(\text{tan$^2\theta $}_w m_Z^2
 \left(C_A^2+C_V^2\right)+4 \text{cos$^2\theta $}_w \text{sin$^2\theta $}_w \Gamma _Z^2\right),
\end{split}\end{align}
where we treat the electron as massless and assume that $\overline{d_\mathcal{W}}=0$. The axial and vector couplings are defined as $C_A=-1/2$ and $C_V=-1/2+2\sin\theta_W$. In practice, we apply the experimental angular photon and energy cuts described in \cref{sec:expresultsandprospects,sec:analyticcuts} and do not make approximations on the masses of electrons. 

\subsubsection{LHC}
At the LHC, there are two main production channels we can consider. The first channel is analogous to LEP, and consists of oppositely charged quarks and anti-quarks interacting via an s-channel photon or $Z$ boson: $q_i\bar{q}_i\to (N\to\gamma\nu)\bar{\nu}+h.c.$. This gives the same signature as LEP, up to subtleties that will be discussed in \cref{sec:expresultsandprospects}. In addition to neutral currents, the LHC provides us with the opportunity to study interactions proceeding via charged currents. The charged current couplings appeared as one of two possible couplings above the electroweak scale in \cref{eq:brokenphaseUVcompletion}, and leads to a final state consisting of a single photon, charged lepton and missing energy---for example: $u_i\bar{d}_j\to (N\to\gamma\nu)\ell^+$. For the LHC, the rate of production of HNLs is calculated using MadGraph5\_aMC$@$NLO v2.5.5 \cite{Alwall:2014hca}, making use of FeynRules2.3 \cite{Alloul:2013bka,Christensen:2008py} to load our implementation of the HNL model.

\subsection{Experimental results and prospects }
\label{sec:expresultsandprospects}
\subsubsection{LEP}
There have been many analyses dedicated to the $\gamma+E_\text{miss}$ final state \cite{Adriani1992,Akers1994,Abreu1996,Lopez1996}. We choose to focus on the results of LEP1, which ran at a center of mass (COM) energy corresponding to the $Z$ pole and accumulated about $200\text{pb}^{-1}$ of data, and LEP161 which ran at a COM energy of $161\GeV$ and accumulated $25\text{pb}^{-1}$\cite{Assmann2002}. Using partial luminosity and combining many analyses, LEP1 was able to set an upper bound of $0.1\text{pb}$ on the cross section of new physics contributing to the $\gamma+E_\text{miss}$ final state, within the angular acceptance range of $|\cos\theta_\gamma|\le 0.7$ and requiring the outgoing photon to have a minimal energy of $0.7\GeV$. We also enforce that the HNL decays within $1\text{m}$ of the interaction point using \cref{eq:decayprobability}. To set constraints using LEP data that extend to slightly larger HNL masses, we point out that LEP's $161\GeV$ run also set an upper bound of $1\text{pb}$ on the single photon cross section from new physics. 
%
%
%
\subsubsection{LHC}
To probe the coupling $d_Z$, we recast a recent dark matter search at $\sqrt{s}=13\TeV$ by ATLAS \cite{Aaboud:2017dor} involving final states containing at least one photon with $E^\gamma_T>150\GeV$, missing energy greater than $150\GeV$, and 0 or 1 jets. Events in our MadGraph simulation were generated with 0 or 1 photon, and no jets. Owing to the systematic uncertainties in the modelling of initial state radiation, only background predictions with 1 jets are shown in the ATLAS paper. We use a data-driven method to estimate the background events with 0 jets by looking at the ratio of data events reported to contain either 0 or 1 jet.
%
%
Following this, we see a deficit of data events in both the 0 and 1 jet channels as compared to the background predictions, which will motivate us to adopt the $CL_s$ method for estimating the sensitivity at the LHC, which we describe in \cref{sec:sensitivity}. The dominant background for this search was the irreducible $Z(\rightarrow\nu\nu)\gamma$ process, followed by $W(\rightarrow \ell\nu)\gamma$ in which the final state lepton was not detected. In addition to all the cuts described in the paper, we also impose a probability function requiring the HNL to decay before the closest distance to the ECAL barrel, namely $r=1.5\text{m}$ from the beamline. We take the photon ID efficiency to be 92\% \cite{Aaboud:2016yuq}. \\

The LHC also provides us with the opportunity to probe the charged current HNL extension. We make use of a $\sqrt{s}=8\TeV$ CMS search for supersymmetric models with gauge-mediated breaking \cite{CMS:2015loa}. In its analysis, the collaboration searched for 1 electron/muon with transverse momentum greater than $25\GeV$, 1 or more photons, and missing energy greater than $120\GeV$. The dominant backgrounds in this search were misidentified photons, misidentified leptons,
and electroweak backgrounds. In the case of CMS, the transverse distance from the beamline to the ECAL barrel is $1.29\text{m}$, and the detection efficiency for electrons and muons are 80\% and 90\% respectively. There are no requirements on the number of jets, however they show results consistent with low jet activity by requiring that the scalar $p_T$ sum of jets ($H_T$) be smaller than $100\GeV$. In our event generation, we do not consider associated jet production, which provides us with a conservative estimate. We simulate production of $N$ and $\ell$ from a $W$ boson via the $d^a_{\gamma W}$ coupling, and decays of $N$ to a neutrino and photon via the $d_\gamma$ coupling. We do this for various relative magnitudes between $d^a_{\gamma W}$ and $d_\gamma$. 

In both the CMS and ATLAS searches, results are shown in terms of several signal regions defined by an additional requirement on the missing energy. We cycle through each of these signal regions and independently calculate the sensitivity in order to find the most constraining missing energy requirement.
%
%
We now briefly comment on ways in which one could extend the reach of this analysis. Access to longer HNL lifetimes could be achieved by using the location of the photons hitting the ECAL barrel and endcaps, and statistically mapping these back to the original direction of the HNL. Then, on an event-by-event basis, we could select different maximal distances in the probability of decay cut. Currently, we only used the distance of closest approach between the IP and the ECAL barrel. An additional possibility is to allow the HNL to decay somewhere inside of the ECAL as opposed to before reaching the surface. To avoid potential difficulties with triggering however, this might have to be done in association with jets or leptons. Lastly, tau flavored couplings could be explicitly probed in the $\ell+\gamma+\slashed{E}_T$ analysis by tagging tau leptons. This would be a nice complement to neutrino beam dump experiments, whose characteristic energies and neutrino flavors often prohibit tau production. We do not include tau leptons in our simulations.

\subsubsection{Results}
\begin{figure}[!ht]
\begin{center}
\includegraphics[width=0.9\linewidth]{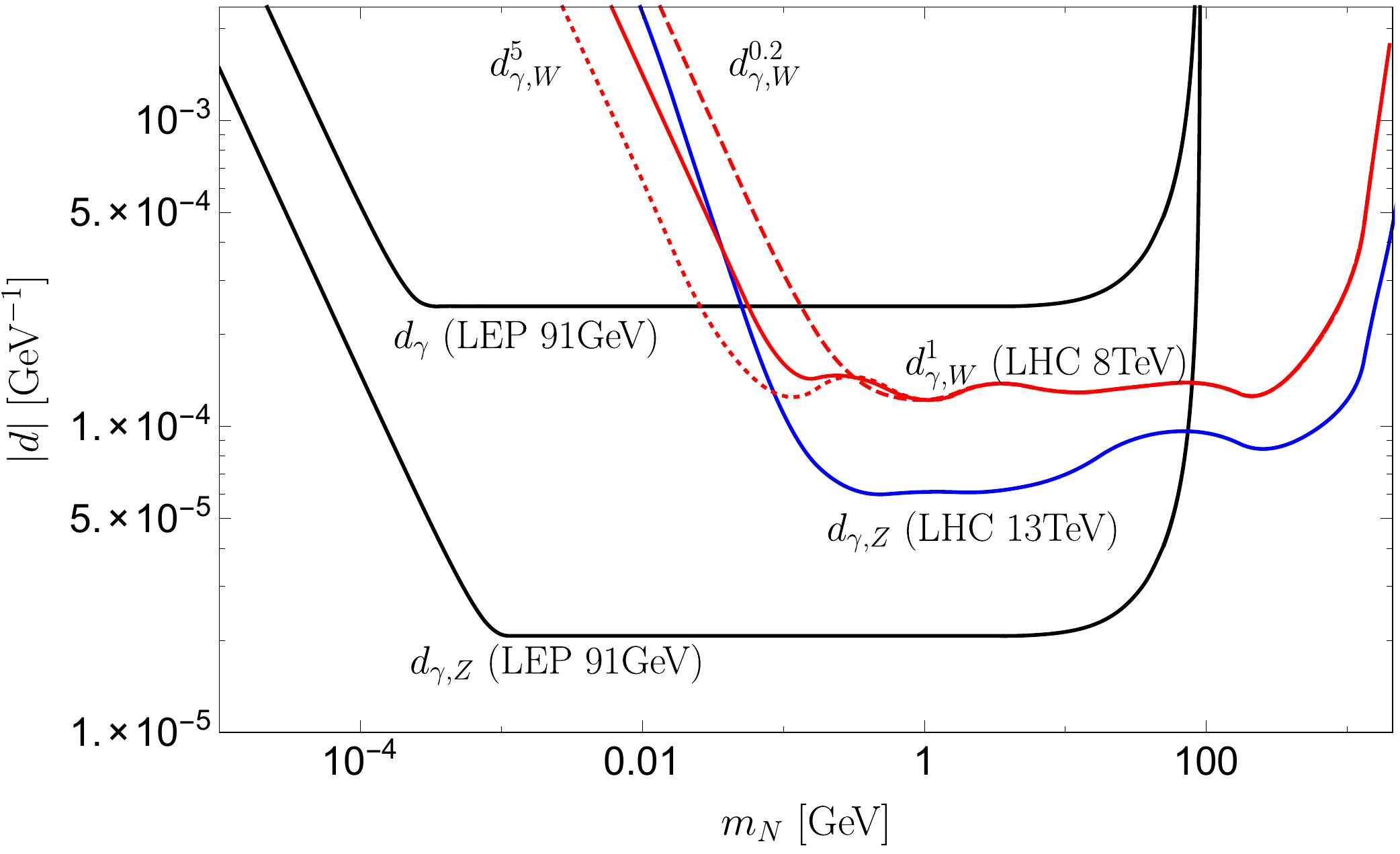}
\end{center}
\caption{$95\%$ CL sensitivities at LHC and LEP. Limits are shown for the dimension 5 ($\gamma$ mediator) and dimension 6 ($\gamma$, $Z$ and $W^\pm$ mediators) extensions. For the LHC $8\TeV$ results involving a photon and charged lepton final state, we consider various relations between the production ($d_{\gamma W}^a$) and decay ($d_\gamma$) couplings. See \cref{tab:HEPcouplingsLabel} for an explanation of the plot labels.}
\label{fig:ExclusionPlotLHCLEP}
\end{figure}

The compilation of the high energy limits on the dipole couplings is presented in \cref{fig:ExclusionPlotLHCLEP}.
All constraints have a characteristic ``U'' shape. The right boundary of the excluded region is controlled by the kinematic reach, 
and in the case of the LHC extends beyond a TeV. The left boundary (small $m_N$) is controlled by the lifetime of $N$, as smaller $m_N$ 
leads to the longer lifetime of $N$ and the loss of the $\gamma$ signal in the detector. The bottom part of the constraints is controlled by the 
rates and backgrounds, and is approximately independent on $m_N$ as in this region the production cross section is $m_N$ independent, and 
its decay is relatively prompt. It is interesting that below $m_Z/2$ the LEP experiments are still capable of providing better sensitivity 
to the neutrino dipole portal. 

%

\section{Cosmology and Astrophysics}\label{sec:CosAstro}

\subsection{Big Bang Nucleosynthesis}\label{sec:3b-big-bang-nucleosynthesis}

Cosmology provides a very sharp tool in limiting the coupling constants of metastable heavy particles. In particular, consistency of BBN-predicted $^4$He and deuterium yields with observations shows that the Universe was dominated by electrons, photons and SM neutrinos at very early epochs with temperature $T\sim 1$~MeV. Any massive relic surviving in large abundances down to these temperatures, or conversely having a lifetime in excess of $0.1$ seconds, will distort this balance, and contribute to the Hubble rate during the proton-neutron freeze-out. Since most of the neutrons end up in $^4$He, this possibility constrains the lifetime of heavy metastable relics {\em if} they are populated to large thermal abundances. 

Therefore, we are led to investigate the mechanisms that populate HNLs in the early Universe. The analysis of the conventional mass-mixed case in its impact on BBN was performed in Ref. \cite{Ruchayskiy:2012si}, and the mechanisms for thermal population of HNLs through neutrino oscillations is quite established \cite{Dodelson:1993je}. Here we notice that the processes that populate $N$'s through a dipole portal can be divided into two categories. 
\begin{enumerate}[(i)]
\item Inverse decays\footnote{The importance of inverse decays in astrophysical constraints of the neutrino dipole portals, including BBN and supernova bounds, was first discussed in \cite{Aparici:2009fh}.}, $\nu +\gamma \to N$. These processes are important at $T\sim m_N$, and can be derived from the 
width of $N$.
\item $2\to 2$ processes, such as $f^+f^- \to N\bar\nu$\,or\,$\bar N \nu$, where $f$ is a SM fermion, as well as all crossing-related processes. While higher order in the coupling constant, these rates are enhanced in the UV. 
\end{enumerate}
At any given temperature in the early Universe, the abundance of $N$ particles is set either by equilibrium, if their interaction rates are faster than the Hubble rate, or by the approach to equilibrium regulated by
\begin{equation}
\frac{n_N}{n_f} \sim \frac{\langle \sigma v \rangle n_{\bar f}}{H(T)},
\end{equation}
where $n_f$ are $n_N$ is the number density of charged species and HNLs, $H(T)$ is the Hubble rate, and 
$\langle \sigma v \rangle n_{\bar f}$ is the temperature-dependent rate for creating an HNL per unit of time. 
The most important for us is the scaling of the above expression with temperature and parameters of our model. 
Making a simple parametric estimate we arrive to 
\begin{equation}
\frac{\langle \sigma v \rangle n_{\bar f}}{H(T)} \propto \alpha g_*^{-1/2} M_{\rm Pl} d^2 T,
\label{eq:rate}
\end{equation}
where $ M_{\rm Pl} $ is the Planck mass and $g_*$ is the effective number of degrees of freedom appearing from the definition of the Hubble rate, $H(T) \simeq 1.66 g_*^{1/2} T^2 M_{\rm Pl}^{-1}$. The most important feature of \cref{eq:rate}, besides the self-explanatory dependence on $ M_{\rm Pl}$ and $d$, is its scaling with temperature. The rate is enhanced in the UV, and therefore, it is {\em the highest temperatures} in the system that determine 
the initial abundance of $N$. Therefore, strictly speaking, one cannot determine the initial abundance of $N$ without ever specifying the initial temperature relative to $d^{-1}$. On the other hand, assuming that the Universe at some point had temperature $T\sim d^{-1}$, the ratio in \cref{eq:rate} is then larger than one for all values of $d$ covered by our master plot, Fig. 1. Therefore, with this assumption, one can be sure that $N$ was in fact thermalized in the early Universe.

Once $N$ is thermally populated, it will last until the lifetime of the Universe is comparable to $\tau_N$. To predict how much energy the thermally-created reservoir of $N$ stores, one would need to understand at what temperatures HNLs decouple, which can be estimated parametrically by equating the r.h.s. of \cref{eq:rate} to one. This gives the decoupling temperature of
\begin{equation}
T_{\rm decouple} \sim {\rm 1\,GeV} \times \frac{\tau_N}{0.1\,{\rm s}}\times \left(\frac{m_N}{10\,{\rm MeV}}\right)^3,
\end{equation}
where we re-expressed $d$ in terms of the lifetime formula for $N$. The decoupling of $N$ means that at temperatures $T< T_{\rm decouple}$ the decays of heavy SM particles heat up the SM bath but not $N$, and its relative energy density is somewhat diluted as $g_*$ at decoupling will be larger than at the time of decay. At the same time, for $N$ heavier than an MeV, there is a possibility for a significant enhancement of the $N$ energy density at decay due to them becoming nonrelativistic. The ratio $\rho_{N}/\rho_{\rm SM}$ will gain an enhancement factor $m_N/T_{\rm dec}$, where $T_{\rm dec}$ is the temperature corresponding to the time of the decay of $N$, 
$H(T_{\rm dec}) \sim \tau_N^{-1}$ (in the assumption that $T_{\rm dec}<m_N$). Consequently, our estimate becomes 
\begin{equation}
\label{eq:estimate}
\frac{\rho_{N}}{\rho_{\rm SM}} \propto \frac{g_N}{g_*(T_{\rm decouple})}\times \frac{m_N}{T_{\rm dec}},
\end{equation}
where $g_N = 7/8\times 4$ as $N$ caries four fermionic degrees of freedom. 
This estimate can be used to constrain the lifetime of HNLs as $\rho_{N}/\rho_{\rm SM}$ is constrained at $T\sim 1$\,MeV through the $n/p$ freeze-out. If $\tau_N\sim 0.1\,{\rm s}$, the ratio in \cref{eq:estimate} is $O(1)$, while only less than 10\% variations are allowed
(see, {\em e.g.} Ref. \cite{Fields:2014uja}). 



\subsection{Supernova SN 1987A \label{sec:3a-supernova}}
The modification of energy generation and transfer in stars can also serve to limit the viable parameter space for a dipole neutrino portal. In particular, SN 1987A has proved to be a useful probe of weakly coupled particles below the GeV scale \cite{Raffelt1996,Dreiner2003,Dreiner:2008tw,Dreiner:2013mua,Fischer:2016cyd,Chang:2016ntp,Hardy:2016kme}. The typical consideration is as follows: weakly coupled particles may serve to substantially enhance the rate of cooling of a supernova, and if this cooling proceeds too quickly and the energy is able to escape without being reabsorbed, then nuclear processes at the core of the supernova can rapidly stop. This in turn leads to significant deviations between the predicted and observed neutrino pulses observed at terrestrial neutrino observatories \cite{Hirata1987,Alekseev:1988gp,Bionta:1987qt}. Therefore it is the rate of cooling, rather than the rate of production itself that is important. 

There are two considerations in determining whether HNLs (or any new weakly coupled particle) can spoil supernova predictions. First, for sufficiently weak coupling very few HNLs will be produced, and consequently they will not be able to efficiently cool the interior of the supernova. This naively suggests strong couplings can be excluded, however, if the coupling is sufficiently large, then any HNLs that are produced will be trapped. Provided this trapping occurs within the ``neutrinosphere'' (defined as $r<R_{\nu}$ where $T(R_\nu)=3~\text{MeV}$) \cite{Chang:2016ntp}, then the energy stored in the HNLs can be efficiently recycled and re-emitted in the form of neutrinos, ultimately having no impact on the observations at terrestrial detectors. A full treatment that captures this competition between production and absorption would involve a detailed study\footnote{\Cref{eq:prob-esc} assumes an outward radial path for the HNL and does not account for passage through the core of the supernova. Neglecting this $\order{1}$ effect is already an approximation \cite{Chang:2016ntp}.} of the following integrals \cite{Chang:2016ntp}
\begin{subequations}
\begin{equation}
 \dv{E}{t} = \int_{0}^{R_\nu} P_\textrm{esc}(r_0)\times \left\langle E_N \dv{\Gamma_\text{prod}}{r}\right\rangle(r_0)~\dd r_0
 \label{eq:emissivity-exact}
\end{equation}
\begin{equation}
 P_\text{esc}(r_0) =\exp\qty[- \int_{r_0}^{R_{\text{far}}} \frac{1}{\lambda_\text{MFP}} \dd r].
 \label{eq:prob-esc}
\end{equation}
\end{subequations}
where $\dd \Gamma /\dd r$ is the local rate of production of HNLs, $E_N$ denotes the HNL energy, $R_{\text{far}}$ is a large radius to which the escape probability is insensitive, and the average is taken with respect to the local thermal bath at $r_0$. The probability of escape $P_\text{esc}$ is found by exponentiating the line-of-sight integral of the mean free path, which in the case of the dipole portal will always be inversely proportional to the square of the dipole coupling $\lambda_\text{MFP}\propto 1/d^2$. 

For each HNL mass $m_N$, there will exist a minimal dipole coupling $d_\text{prod}(m_N)$ for which too few HNLs are produced to significantly alter the observed neutrino signal. Likewise, there will also exist a maximum dipole coupling $d_\text{abs}(m_N)$ such that for any stronger couplings the HNLs will be efficiently reabsorbed and will not cool the interior of the supernova appreciably. 
The region of excluded parameter space lies between theses two curves in the $d-m_N$ plane \emph{i.e.} $d_\text{prod}(m_N)<d_\text{excl}(m_N)<d_\text{abs}(m_N)$. Although \cref{eq:emissivity-exact,eq:prob-esc} are in general complicated, in the weak coupling regime ($d\lesssim d_\text{prod}$), and the strong coupling regime ($d\gtrsim d_\text{abs}$), the analysis simplifies. 

In trying to obtain the lower curve $d_\textrm{prod}(m_N)$ of \cref{fig:astrophysicsplot}, the coupling is small and so the probability of escape is nearly unity. We may therefore study the production of HNLs and neglect the absorptive properties of the bath. Furthermore, this may be done locally, as opposed to globally, at a characteristic radius. This approximation is often termed the ``Raffelt criterion'' \cite{Raffelt1996}, and is defined in terms of the energy carried by HNLs per unit volume, per unit time, $\dd \mathcal{E}_N/\dd t$ (being referred to as emissivity throughout this paper), at a fixed radius $r_0$ 
\begin{equation}
\dv{\mathcal{E}_N}{t}(r_0,t)\leq 10\% \dv{\mathcal{E}_\nu}{t}(r_0)\approx\qty[\frac{\rho(r_0)}{\textrm{g/cm}^3} ]\times 10^{19} \textrm{erg~cm}^{-3}\text{s}^{-1},
\label{eq:raffaelt}
\end{equation}
where $\dd \mathcal{E}_\nu/\dd t$ is the maximum energy per volume per time emitted via neutrinos. This criterion essentially requires that HNLs produced at some fixed radius $r_0$ carry no more than $10\%$ of the total energy lost to neutrinos per time. The emissivity constraints derived based on the Raffelt criterion and from the criterion with the integrated energy are compared explicitly in \cite{Dreiner:2013mua}. The difference is well within an order of magnitude as demonstrated for their scenario. 

In the limit of strong coupling, the relevant question is whether the produced HNLs can escape the supernova's neutrinosphere. Since $d_\text{abs}(m_N)\gg d_\text{prod}(m_N)$ we may assume a large flux of HNLs in the parameter space of interest, and so by \cref{eq:emissivity-exact}, it is the probability of escape that must inhibit cooling due to HNL production. As demonstrated by \cref{eq:prob-esc}, and the discussion thereafter, this quantity depends exponentially on the dipole coupling by way of the mean free path. Therefore, a reasonable criterion is that that $P_\text{esc}(d_\text{abs})=1/2$, since for $d\gtrsim d_\text{abs}$ this quantity will be exponentially suppressed. 
\begin{figure}
\includegraphics[width=\linewidth]{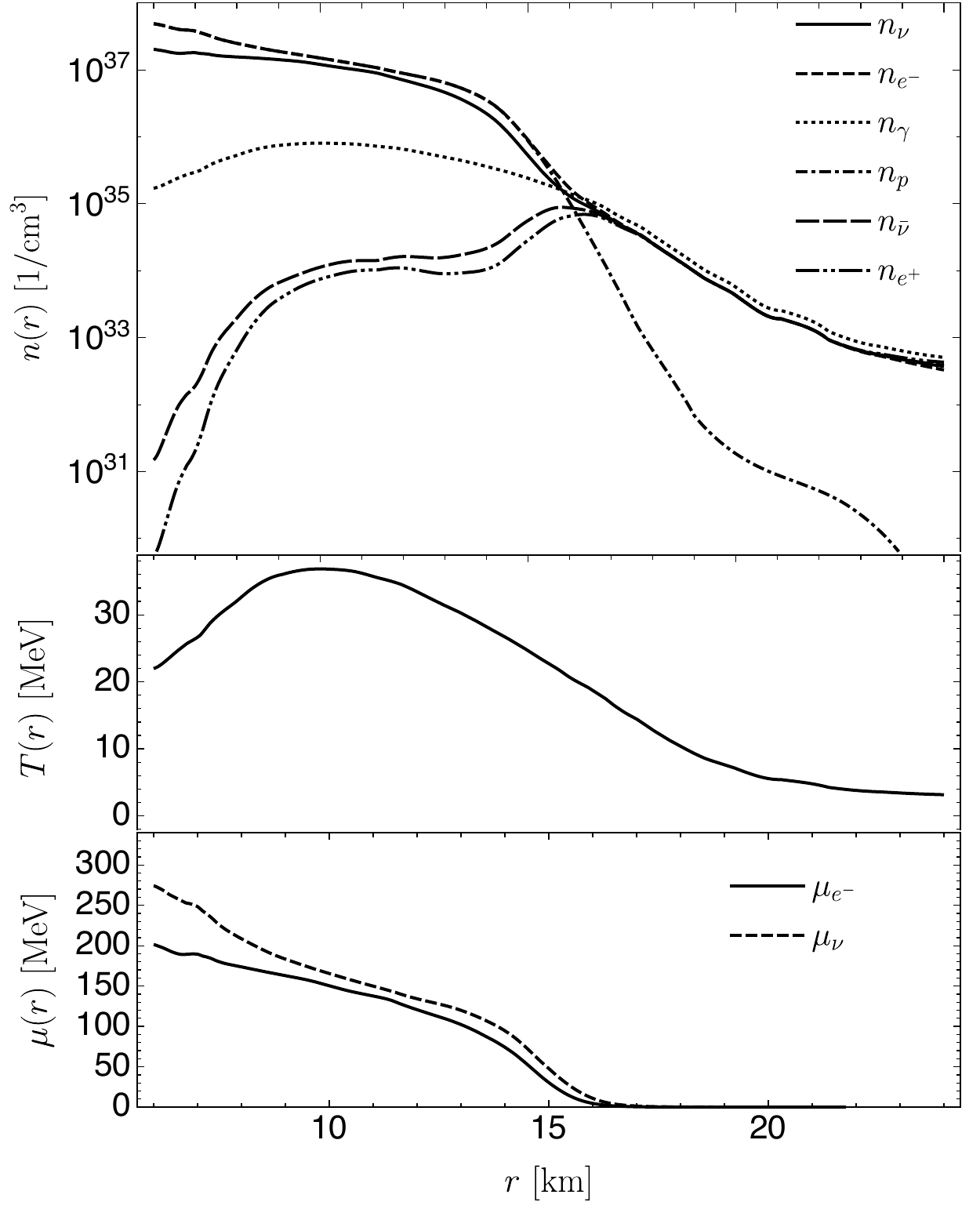}
\caption{Radial profiles of the number density, temperature, and chemical potentials at one second after the bounce from the simulation of an $18M_\odot$ progenitor \cite{Fischer:2016cyd}. \label{fig:SN-profiles}}
\end{figure}
Although the Raffelt criterion is most naturally imposed where the temperature is maximal, and densities are high, it is possible that this will lead to a rather conservative bound on $d_\text{abs}$. This is because, being produced in the hot and dense interior of the supernova, the HNLs must travel through several kilometres of absorptive material composed of electrons, protons, and neutrinos, all of which have number densities in excess of $10^{37}/~\text{cm}^3$. This feature is mitigated to some extent due to Pauli-blocking, however which effect is dominant is hard to determine. With this in mind, we perform our analysis at two radii $r_0^{(a)}=10~\textrm{km}$ and $r_0^{(b)}=14~\textrm{km}$. The former corresponds to the conventional choice \cite{Dreiner2003,Dreiner:2013mua,Raffelt1996,Chang:2016ntp} of the hottest ($T\approx 30~\text{MeV}$) and most dense ($n_e,n_\nu, n_p\approx 10^{37}/~\textrm{cm}^3$) region of the supernova. The latter choice, by contrast, includes slightly lower temperatures ($T\approx 20~\textrm{MeV}$) number densities ($n_e,n_\nu, n_p\approx 10^{36}/~\textrm{cm}^3$) but does not require transit through the most dense regions of the supernova due to the sharp decline in number density in the outward radial direction. 

Before turning to the details of the calculation of the emission rates and escape probabilities, we first summarize the physics that is included in our calculations. We use radial profiles corresponding to a supernova with an $18 M_\odot$ progenitor, which are obtained by digitizing the reference runs shown in Fig.\ 5 of \cite{Fischer:2016cyd}. In calculating the optical depth, the full radial dependence is accounted for, but as discussed above, we apply the Raffelt criterion at two fixed radii. We include all species present except for neutrons as they do not couple to HNLs via the dipole portal. In computing the optical depth, and emissivities, we account for the effects of quantum degeneracy including Pauli-blocking, which is found to modify the rate of production and to have a dramatic effect on the escape probabilities of HNLs. 

\subsubsection{Production\label{sec:prod}}
Supernovae typically have significant populations of protons, neutrons and photons, as well as electrons and neutrinos, and their associated anti-particles. Save the neutron, HNLs couple to all of these species at tree level via the dipole portal, and this allows for the following production mechanisms 
\begin{align}
\nu + e^\pm &\rightarrow N + e^\pm \quad &\text{(upscattering)}\\
\nu + p &\rightarrow N + p \quad &\text{(upscattering)}\\
e^+ + e^- &\rightarrow \overline{\nu} + N \quad &\text{(synthesis)}\\ 
\gamma + \nu &\rightarrow N \quad &\text{(inverse decay)}.
\end{align}
We point out that our analysis does not include thermal field theory effects, and so we omit the ``plasmon decay'' $\gamma\rightarrow \overline{\nu} N$ production mode. In general, ignoring the thermally acquired effective mass of photons in T channel scattering processes is only justified if the characteristic momentum flowing through the photon is much larger than its effective mass, which is on the order of $20-30\MeV$. Using vacuum propagators for the dominant HNL production process $e^-\nu\rightarrow e^-N$, we calculated the quantity $\sqrt{-\langle q^2\rangle}$ and found it to be greater than $70\MeV$ for all masses considered, eventually asymptoting to $m_N$ for heavy $N$. Furthermore, for all masses considered, ignoring the regime $\sqrt{-q^2}<30\MeV$ changes $\langle q^2\rangle$ by less than 4\%. In addition to thermal effects, we also neglect the influence of nucleon magnetic moments (because of the additional $\propto m_p^{-1}$ suppression), and for that reason neglect 
$\nu + n \rightarrow N + n$ production mode. Going back to the channels we consider, all of these have two incident species, and so the rate of production is controlled by the product of their densities (\emph{i.e.} $n_e n_\nu$ in the case of electron upscattering). In the case of upscattering, however, the chemical potential can be an order of magnitude larger than the temperature, and so Pauli-blocking of the outgoing SM product must also be taken into account. 

As discussed above, in considering the minimal dipole coupling that can spoil predictions from SN 1987A, we study the Raffelt criterion, \cref{eq:raffaelt}, at both $r_0=10~\textrm{km}$ and $r_0=14~\textrm{km}$.

The following integral equation defines the emissivity
\begin{equation}
\dv{\mathcal{E}_N}{t}=\int \frac{\dd^3p_1}{(2\pi)^3}\frac{\dd^3p_2}{(2\pi)^3} f_1 f_2 
\langle E_N \sigma \rangle_\mathfrak{F} ~v_\text{M\o{}l},
\label{eq:3a-emsisivity}
\end{equation}
where $f_a=1/(\exp[(E_a-\mu_a)/T(r_0)]+1)$ is the Fermi-Dirac distribution for species $a$, and $v_\text{M\o{}l}$ is the M\o{}ller velocity
\be	
v_\text{M\o{}l} = \sqrt{(\boldsymbol{v}_1-\boldsymbol{v}_2)^2-(\boldsymbol{v}_1\times \boldsymbol{v}_2)^2}.
\ee
The average, $\langle E_N \sigma\rangle_\mathfrak{F}$, is taken over phase space with the appropriate distribution functions included. For inverse decays, this is the trivial one-body phase space of the HNL, but for $2\rightarrow 2$ process the appropriate Pauli-blocking factor of the outgoing SM particle, $\mathfrak{F}(E_3):=1-f(E_3)$, is included, where $E_3$ is evaluated in the rest frame of the bath. Explicitly, for $2\rightarrow 2$ processes the average is defined as 
\begin{equation}
\langle E_N \sigma \rangle_\mathfrak{F} := \int\frac{\dd \Phi_2(p_3,p_N)}{4\mathcal{F}(s)} \mathfrak{F}(E_3) E_N \abs{\mathcal{M}}^2_{\text{prod}} 
\label{eq:prod-avg}
\end{equation}
where $\Phi_2(p_3,p_N)$ denotes the two-body Lorentz invariant phase space of the outgoing HNL and SM particles, $\mathcal{F}(s)$ the Lorentz-invariant flux factor, and $E_N$, like $E_3$, is evaluated in the rest-frame of the bath. The production matrix element $\mathcal{M}_\text{prod}$ is calculated at zero-temperature, and does not include---for example---the in-medium modification of the photon propagator.

Following \cite{Gondolo:1990dk,Dreiner2003,Cannoni:2013bza}, we can rewrite \cref{eq:3a-emsisivity} as
\begin{align}\begin{split}
\dv{\mathcal{E}_N}{t}=&\frac{1}{32 \pi^4}\int_{M^2}^\infty ds\int_{\sqrt{s}}^\infty dE_+\int dE_- \langle E_N \sigma \rangle_\mathfrak{F}\\
&\times F(s,m_1,m_2)~f\left(E_1, \mu_1\right) f\left(E_2, \mu_2\right)
\end{split}\end{align}
where
\begin{align}\begin{split}
& M^2=\text{Max}[(m_1 + m_2)^2, (m_N+m_\text{final})^2],\\
& E_1=\frac{E_+ + E_-}{2} \quad\text{and}\quad E_2=\frac{E_+-E_-}{2},\\ 
& F(s, m_1, m_2) = \sqrt{\frac{1}{4} \left(s-m_1^2-m_2^2\right)^2-m_1^2 m_2^2}.
\end{split}\end{align}
Using the Mandelstam variable $s$, we can show that $E_-$ depends on $s$, $E_+$, $m_1$, $m_2$, and $\cos\theta$, and that its associated bounds of integration are obtained by considering the limits $\cos\theta\rightarrow \pm 1$ with $E_+$ and $s$ held fixed.

\subsubsection{Escape\label{sec:escape}}
The escape probability \cref{eq:prob-esc} is dictated by the mean free path $\lambda_\text{MFP}$ of the HNL in the hot bath of the supernova. Demanding that the probability of escape is less than $50\%$ is equivalent to demanding that $-\ln {P_\textrm{esc}}\lesssim 2/3$. Since the dipole portal is the only coupling between the Standard Model and the HNL, all processes that contribute to $\lambda_\text{MFP}$ are proportional to $d^2$. It is therefore convenient to introduce a reduced mean free path $\lambdabar$, defined at a reference value $d=10^{-7}~\text{GeV}^{-1}$ via
\begin{equation}
\begin{split}
-\ln P_\textrm{esc}&= \int_{r_0}^{25~\textrm{km}} \frac{1}{\lambda_\textrm{MFP}(r)} \dd r\\
&=\qty(\frac{d}{10^{-7}~\text{GeV}^{-1}})^2 \int_{r_0}^{25~\textrm{km}} \frac{1}{\mathbf{\lambdabar}_\textrm{MFP}(r)} \dd r
\end{split}
\end{equation}
Implicit in the above analysis is the assumption that the path of the HNL is directed radially outwards. This underestimates the probability of absorption as it neglects paths that travel through the core and other overdense regions, however as discussed in Appendix B.\ of \cite{Chang:2016ntp} this effect is $\order{1}$ and can be captured by multiplying the optical depth by the substitution $\lambdabar_\text{MFP}\rightarrow \lambdabar_\text{MFP}/3$. We may then define the critical dipole moment where HNLs are efficiently trapped via the condition 
\begin{equation}
d_\textrm{abs} =\sqrt{\frac{2/3}{ 3\times \int_{r_0}^{25~\textrm{km}}\frac{\dd r}{\mathbf{\lambdabar}_\textrm{MFP}(r)}}} \times {10^{-7}~\text{GeV}^{-1}}.
\label{eq:d-abs-def}
\end{equation}
The above procedure does not take into account the flux of HNLs coming from the core of the supernova
can be exponentially large, and therefore some of amount of energy deposition can happen beyond $d_{\rm abs}$. The flux is a factor of $(d_\text{abs}/d_\text{prod})^2\sim 10^6$ larger than the lower bound, and so an even larger dipole coupling is required to efficiently absorb this large flux of HNLs, given roughly by $d_\text{abs}\rightarrow d_\text{abs} \times \log\qty(\frac{d_\text{abs}}{d_\text{prod}})^2$, which is approximately an order of magnitude larger, and consequently more stringent. Since we neglect this effect, our analysis can be considered conservative in this regard.  

Both single body decay of the HNL, and $2\rightarrow 2$ scattering contribute to the mean free path. In no particular order, the relevant processes are 
\begin{align}
N + e^\pm &\rightarrow \nu + e^\pm\quad &\text{(downscattering)}\\
 N + p &\rightarrow\nu + p \quad &\text{(downscattering)}\\
\overline{\nu}+N &\rightarrow e^+ + e^{-} \quad &\text{(annihilation)}\\ 
N &\rightarrow \gamma + \nu \quad &\text{(decay)}\\
N+\text{SN}&\rightarrow N+\text{SN}&(\text{gravitational trapping}).
\end{align}
%
We have included the full radial dependence of the temperature and chemical potentials in our calculation of \cref{eq:d-abs-def}. As can be clearly seen in \cref{fig:SN-profiles}, the chemical potentials of the neutrinos and electrons are significantly higher than the temperature within the interior of the supernova, therefore for HNLs produced at $r_0\approx 10~\textrm{km}$, Pauli-blocking and the Fermi-Dirac distributions of the absorptive species can play an important role in determining the escape probability. As discussed above, we compute the reduced optical depth integral at a reference dipole coupling of $10^{-7}~\textrm{GeV}^{-1}$ and include the effects of Pauli-blocking via
\begin{align}
 \frac{1}{\lambdabar_\textrm{MFP}(r)} &= \sum_\alpha \langle n_\alpha\sigma_{\alpha N} \rangle(r) + \beta\gamma(r_0) \langle \Gamma_N\rangle (r) .
\end{align} 
Here $\alpha\in \{ e^-, e^+,\nu_{e,\mu,\tau}, \overline{\nu}_{e,\mu,\tau},\gamma, p\}$ labels the species that can absorb HNLs, and $n_\alpha$'s are their Fermi-Dirac distributions. The thermal averages $\langle n_\alpha \sigma_{\alpha N} \rangle$ and $\langle \Gamma \rangle$ includes the thermal distribution of the absorptive bath for the $2\rightarrow 2$ absorption, and the associated Pauli-blocking of outgoing SM particles for both decays and $2\rightarrow 2$ processes. 

We fix the incident HNL energy to be $\langle E_N\rangle (r_0, m_N)$, defined as the average energy per HNL produced at $r_0=10~\text{or}~14~\text{km}$, and this implies a boost factor for the HNL $\beta\gamma(r_0, m_N)$. In practice we compute the average energy numerically, however the qualitative behaviour can be understood as follows. The dominant production mechanism over most of the mass-range is Primakoff upscattering off of electrons which is Pauli-blocked on the outgoing electron. For $m_N \gtrsim \mu_e$, the momentum transfer required to create an HNL typically kicks electrons above the Fermi surface and imparts the HNL with three-momentum of order $P_N\sim \order{\mu_e}$. Therefore the momentum can be estimated using elementary kinematics. In contrast, for low masses the effects of Pauli-blocking must be accounted for by demanding a large momentum transfer, $q^2\approx -\mu_e^2$, so as to kick the electron above the Fermi-surface. Taking the average neutrino to be $\mu_\nu/2$ and averaging over angles then leads to the estimate 
\begin{equation}
\langle E_N\rangle \approx \begin{cases}
m_N + \frac{\mu_e^2}{2 m_N}& \text{for}~m_N\gg \mu_e\\
\frac{m_N^2}{\mu_\nu} + \mu_e & \text{for}~m_N\ll \mu_e 
\end{cases},
\end{equation}
where the chemical potentials are evaluated at $r_0$. We also assume the HNL's path is directed radially outward (and correct for the possibility of transit through overdense regions via a factor of $3$ as discussed above). 

The thermal averages $\langle n_\alpha \sigma_{\alpha N} \rangle$ and $\langle \Gamma_N \rangle$ take into account the radial profile of the supernova, as a consequence of the Pauli-blocking of outgoing SM particles and the thermal distributions of initial SM particles inheriting the radial dependence of the chemical potential and temperature profiles. As in the case of production, the matrix element $|\mathcal{M}_\text{abs}|^2$ is computed at zero temperature and we have checked that finite temperature corrections are under control.

Finally, the gravitational pull from the supernova could potentially trap the HNLs and prevent additional cooling of the supernova from happening. This is especially relevant for the high mass regime. Here we follow the simple energy argument introduced in \cite{Dreiner:2003wh} that determines the particle mass for which this effect becomes important.

The gravitational trapping has to be taken into account when 
\begin{equation}
\left<E_{\rm kin}\right>_{\rm HNL} \le 
\frac{G M_c m_{N}}{R_c},
\end{equation}
where $\left<E_{\rm kin}\right>_{\rm HNL}$ is the average kinematic energy of the HNLs, $G$ is the Newton constant, $M_c$ is the enclosed mass of the supernova within the radius $R_c$, at which the HNL of mass $m_N$ is produced. We take $M_c\approx M_{\rm SN}$ which is the mass of SN 1987A and calculate $\left<E_{\rm kin}\right>_{\rm HNL}$ at two radii $R_c = $10 km and 14 km, corresponding to the radii we choose for the emissivity and the optical depth considerations. We determine that for $m_{N}\gsim$ 320 MeV gravitational trapping is important at both $R_c = 10$ km and 14 km.

\subsection{Results}

\begin{figure}[!ht]
\centering
 \includegraphics[width=0.9\linewidth]{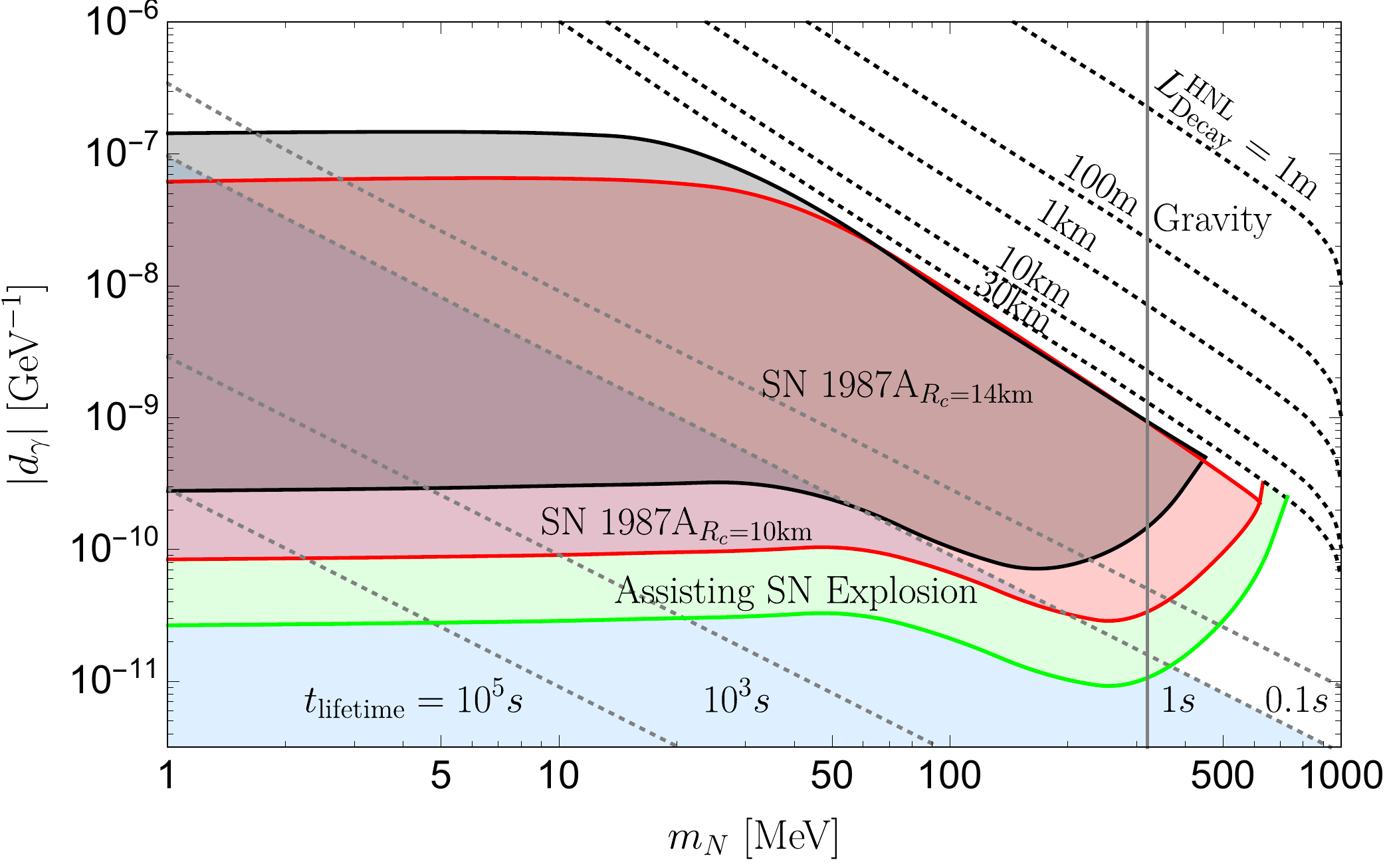}
 \caption{Emissivity and optical depth constraints (red) from supernovae SN 1987A, and parameter space facilitating its conversion to a neutron star (green). We also show lines of constant HNL lifetimes to gauge where BBN might be affected. Two radii of production $r_0$ are plotted for comparison, with one at the hottest densest radius $r_0=10~\text{km}$ and one closer to the edge of the high density region $r_0=14~\text{km}$. The gravitational trapping becomes significant for HNLs with mass above the vertical gray line, labeled ``Gravity''. 
 }
 \label{fig:astrophysicsplot}
 \end{figure}
 
We begin with the BBN limits, that rest on several assumptions. First, we assume that the temperatures in the early Universe were 
initially rather large, and as a consequence, HNLs got thermally populated. If the maximum ({\em i.e.} reheating after inflation) 
temperature was limited to a sub-GeV range (which is a rather extreme assumption), then domains of parameter space 
 with small $m_N $ and small $d$ will not be constrained by $n/p$ freeze-out, as the abundance of HNLs at 1 MeV can be much smaller. The
 second assumption is that we assume that the BBN proceeds along a standard scenario, and HNLs provide only a small perturbation. An alternative scenario, when the Universe is actually dominated by $N$, and its decay reheats the $\nu$ and $\gamma,e$ baths, might 
 not be excluded throughout the whole parameter space. Namely, the BBN provides only a handful of reliable predictions ($^4$He, D/H). It could be possible that for some ``islands'' on $\{m_N,d\}$ space, the outcome of the nuclear reaction network is similar to a standard BBN. 
 In this case, however, one would also have to make sure that the energy densities of neutrinos and photons are also consistent 
 with measurements of $N_{\rm eff}$. This may look as an additional fine tuning, and therefore we do not consider such
 an accidental possibility seriously.

 Thus, with the above caveats, if the $\rho_{N}/\rho_{\rm SM}$ ratio is larger than 0.1 at the time of n/p freeze-out, 
 the BBN is perturbed outside of its agreement with observations. Then it is possible to set the constraints on lifetime to be less
 than a fraction of a second (see Ref. \cite{Fradette:2017sdd} for a somewhat similar analysis of the Higgs portal relics). We choose to be on 
 a very conservative side, and set the limit for lifetime to be 1 sec, 
 shown by the diagonal line in \cref{fig:astrophysicsplot}. (At $m_N \sim $ 1 MeV, the decoupling temperature is 
 close to an MeV, and therefore $\rho_{N}/\rho_{\rm SM}> 0.2$ unless $N$ particles decay early. At $m_N> 10$ MeV, the decoupling 
 temperatures are in the GeV regime and larger, so that there can be a significant dilution by $g_*(T_{\rm decouple})$. However, 
 $m_N/T_{\rm dec}$ more than compensate for this dilution, along the $\tau_N = 1$ second line). We observe that on $\{m_N,d\}$ space
 the BBN constraints do not overlap with neutrino/beam dump or high energy experimental constraints. 
 
Our astrophysical results are collected in \cref{fig:astrophysicsplot}. As described in detail in the previous subsections, we have calculated present limits on heavy neutral lepton dipole moments stemming from supernovae cooling. The lower curve of the excluded region is found by requiring that the rate of energy produced by HNL (the emissivity) is larger than a tenth of that from neutrinos. The upper curve is obtained by enforcing that $\int \lambda_{MFP}^{-1}\dd r<2/3$, namely that the probability of an HNL interacting with something on its way outside the star (the optical depth) is small. 

Our analysis reveals that Pauli-blocking of electrons and neutrinos is an essential feature in determining both the emissivity and especially the optical depth. In the latter case, quantum degeneracy makes the hot and dense interior of the supernova nearly transparent to HNLs whose decay and downscattering is inhibited by a Fermi-sea extending up to momenta on the order of $\mu_\nu\approx 250~\text{MeV}$. Unintuitively, this means that the escape probability for an HNL produced at $r_0=10~\text{km}$ is nearly equal to that of one produced at the edge of the densest regions at $r_0=14~\text{km}$. Similarly, the production of HNLs is severely inhibited by the Fermi-sea of electrons. Naively, the high densities of electrons and neutrinos shown in \cref{fig:SN-profiles} favor HNL production, and this suggests that Primakoff upscattering is the dominant production mechanism. This is, in fact, the case at low masses (but only marginally so), however at higher masses ($m_N\gtrsim 50~\text{MeV}$) inverse decays actually come to dominate despite the number density of photons being two orders of magnitude smaller. This is because the inverse decay is not Pauli-blocked. The consequences of quantum degeneracy are that the HNL behaves as if it is much more weakly coupled than one would expect based on naive predictions. 

The qualitative features of our results can be described as follows. 
The upper curve is dominated at low masses by downscattering off of electrons and neutrinos, and the inclusion of Pauli blocking increases the bound on $d$ due to the large chemical potentials (i.e. a large number of already occupied states) of these leptons for $r\ge 10~\text{km}$.
Downscattering is relatively insensitive to the mass of the HNL, (\emph{i.e.} $\sigma\sim d^2$) and so is eventually overtaken by the decay of the HNL which scales as $\Gamma\propto d^2 m_N^3$ and benefits from the absence of Pauli-blocking on the outgoing photon; this crossover between $m_N$ independent downscattering, and power-law decay lengths can be clearly seen in \cref{fig:astrophysicsplot}. The bottom curve is dominated primarily by upscattering of neutrinos off of electrons. This process is only Pauli-blocked on the outgoing electron, and benefits from high number densities of both electrons and neutrinos. In direct parallel with the escape probabilities, this process is eventually overtaken at large masses by inverse decays. The inverse decays scale as $m_N^4 d^2$ and provide the dominant contribution for $m_N\gtrsim 50~\text{MeV}$. The maximal emission is reached when $m_N\simeq \sqrt{s}\approx (T+\mu_\nu)$, but this production channel ceases to be viable at masses much higher than the average center of mass energy $m_N\gg \langle \sqrt{s} \rangle \approx (T + \mu_\nu)\approx 250 ~\text{MeV}$ because the HNL cannot be efficiently produced. Upscattering has a slightly higher kinematic limit of $m_N\gg \langle \sqrt{s} \rangle \approx (\mu_e+ \mu_\nu)\approx 500 ~\text{MeV}$ due to the large chemical potential of the neutrinos.

Gravitational trapping of the HNLs becomes important for large mass HNLs. 
Above the mass $m_{N}$= 320 MeV, the average kinematic energies of the HNLs are smaller than gravitational potential they feel from SN 1987A, as indicated with a vertical line in \cref{fig:astrophysicsplot}. The effect can to some degree alleviate the cooling bound of the SN on the HNLs since these HNLs can be gravitationally trapped and never travel out of the supernova.
We leave a more refined determination of the gravitational effect on the SN cooling to future works. 


Also on the plot there is a region called ``Assisting SN Explosions''. The detailed mechanism of core-collapse supernova explosion is an active research topic, and with the most explored mechanism being driven by neutrinos \cite{1966ApJ143626C}. Simulation results such as \cite{Rampp:2000ws,Liebendoerfer:2000cq} have tended to find that the neutrino-driven explosion struggles to reproduce the revival of the shock-waves for a successful explosion, and requires additional shock energy to match the observation during the core collapse. It is worth noting, however, that the most recent simulation based on a 3D progenitor model \cite{Muller:2017hht}, suggests that the neutrino-driven mechanism itself could possibly provide enough shock revival and explain the observed explosion energies. It is likely that a larger range of progenitors and more refined simulations are still required to fully understand the issue of SN explosions.


With these details in mind, it is worth noting that new degrees of freedom, for example, HNLs, have long been proposed to power SN explosions \cite{Zatsepin:1978ac}, and were most recently proposed to assist neutrinos in reviving the shock waves and augment their energies \cite{Fuller:2009zz}. We briefly review the mechanism for the reader. The star begins by collapsing under its gravitational pull, causing a bounce off of the inner core. This radiates an outward shock. The shock gets stuck, because of dissociation of heavy nuclei, and gets revived by SM neutrino heating and hydrodynamic effects, producing an explosion. This depletes the star's core of leptons. The outward shock then encounters a matter envelope surrounding the star. At this point, previous simulations \cite{Rampp:2000ws,Liebendoerfer:2000cq} found that the shockwave is not able to expel the envelope, and the explosion is quenched. The matter in the envelope falls back into the core, possibly creating a black hole and preventing a neutron star final state from forming. If it was blown away, however, the core could live on as a neutron star, which is the observed remnant of the core-collapse supernovae. By adding HNLs (or any other metastable particles with right properties), they can escape to the envelope and decay into neutrinos and photons. This creates an additional outward radial pressure in the envelope and breaks up some of the heavy nuclei. The original shockwave then has an easier time expelling the envelope away, and wastes less energy dissociating the nuclei inside the envelope. Interestingly, even a small amount of additional energy injection could possibly result in a proper explosion \cite{Zatsepin:1978ac, Fuller:2009zz}. 

In \cref{fig:astrophysicsplot}, the upper bound of the `preferred' region for assisting supernova explosions is determined by consistency with SN 1987A limits. The lower bound, which is the main numerical result of \cite{Fuller:2009zz}, corresponds to having an energy emission from HNLs of $10^{51}$ergs. By contrast, the energy emitted by all Standard Model neutrino species in SN 1987A is $E_\nu\approx 3\times 10^{53}$erg \cite{Raffelt1996}. It is important to note that the simulation in \cite{Fuller:2009zz} assumes vacuum flavor mixing angles of $\sin^2\theta>10^{-8}$ for $\nu_\tau$ mixings and $10^{-8}<\sin^2\theta<10^{-7}$ for $\nu_\mu$ mixings, which are not present in our model. However, the main features of their analysis still hold in our case, since the HNLs in our scenario can also generate the required amount of energy injection given in \cite{Fuller:2009zz}. To obtain the favored region, we have effectively redone the emissivity analysis described in \cref{sec:prod} using an emitted energy of $10^{51}$erg. Recall that the emissivity constraint is done requiring a power loss through HNLs less than $10\%$ that from SM neutrinos.
We find that the ``Assisting SN Explosion'' regime is mostly covered by the BBN constraint on HNLs with lifetimes longer than 1 second.
\\

\section{Discussion and Conclusion}

In this paper we have considered a variety of phenomenological consequences of a massive Dirac particle, that has a
dipole portal $d$ to the SM neutrinos and the photon, as a main source of production and decay of HNLs. 
The Dirac nature of the mass of $N$ is dictated by the arguments of the neutrino mass generation. 
Different variants of such models have been proposed in the past, as a way of 
mimicking the excess of neutrino signals observed at  LSND and MiniBooNE. 
We have provided an attempt at a comprehensive analysis of this 
class of model, assuming the dominance of dipole couplings. 

We find that the high energy probes (LEP and LHC) of HNLs through a dipole portal are giving sensitivity to $d$ at a scale of 
$(10\,{\rm TeV})^{-1}$ and better, mostly through the mono-photon type signatures. In particular, the sensitivity of the 
LHC experiments extends to the TeV scale $m_N$. High intensity beam dump and neutrino experiments (``intensity frontier'' experiments)
cannot reach to such high masses, but instead are able to probe much lower values of couplings for the sub-GeV masses. 
We find that the inclusion of the dipole production of $N$ disfavors common explanation of the MiniBooNE and LSND anomalies 
by already existing data. Interestingly, LSND itself provides the most stringent constraints on the dipole coupling at low masses, while the MiniBooNE, MicroBooNE, and SBND detectors provide the leading constraints at slightly higher masses. At the peak sensitivity to the dipole coupling, for $m_N \sim {\rm few}\,100\,$MeV, the experiments probe scales of 
$d\sim (10^{-7}-10^{-6})\,{\rm GeV}^{-1}$, which is far beyond the weak scale. Future experimental facilities, including SBND, and in particular SHiP, will be able to help improve sensitivity to these couplings. For the SHiP main detector, the level of the single photon backgrounds is not currently well understood, and while we use our optimal estimates at this point, detailed simulations can help 
better evaluation of sensitivity to dipole portal. 
Astrophysics, in particular physics of SN explosions, further restricts the parameter space for the model, probing up to a few hundred MeV scale masses and 
a 
$d\sim (10^{-7}-10^{-10})\,{\rm GeV}^{-1}$ range of couplings. The cosmological bounds are somewhat model dependent as they are sensitive to the 
high-temperature regime of the early Universe for which we do not have the direct experimental data. In the most likely eventuality of 
high initial temperatures, the constraints on lifetime are in the 1 second range and better, disfavoring low-$d$, low-$m_N$ corner of the 
parameter space. Overall, the HNL coupled to the dipole portal adds to new physics models that can be studied both at high and medium energies, and in
astrophysical/cosmological settings. We conclude our paper with a few additional comments:
\begin{enumerate}[(i)]

\item One of the reasons the current model can be studied with such a variety of tools is the fact the dipole portal we explore,
below the electroweak scale, is a dimension 5 portal. It gives cross sections that scale as $\sigma \propto d^2$. This is  similar to the interactions
of axion-like particles $a$ ({\em e.g.} $g_{a\gamma\gamma} a F\tilde F$), which is also dimension 5. 
Indeed, one can observe broad numerical similarities between
sensitivity to $g_{a\gamma\gamma}$ and our derived sensitivity to $d$.

\item We have covered only a handful of the existing intensity frontier searches that we think to be the most sensitive. It is possible that some other experiments (such as {\em e.g.} CHARM, CCFR, and T2K) may also provide additional constraints on the model.
Among new planned facilities, some would involve unprecedented intensities (DUNE),
 and it is possible that new levels of sensitivity to $d$ can be derived there as well. 

\item There are several experimental setups proposed at the LHC to probe long-lived particles, including MATHUSLA \cite{Chou:2016lxi} and CODEX-b \cite{Gligorov:2017nwh}, and a small detector to probe weakly coupled states in the forward regime, FASER \cite{Feng:2017uoz}, which has already considered HNLs (but not their neutrino dipole interactions) \cite{Kling:2018wct}. These setups could potentially extend our reach at the energy frontier. However, since the lifetime of the HNLs in our scenario scales as $m_N^{-3}$ as seen in \cref{eq:ldecaytdecay}, the decay lengths may be too short in the near GeV mass range to significantly improve on the reach of existing probes.

\item 
We have provided a SM gauge invariant completions of the dipole portal operator. 
This should not be confused with a proper UV completion (which was briefly discussed in \cite{Aparici:2009fh,Aparici:2013xga}). Such UV completion may also point to a potential tuning issue that can arise in this model. Operators 
(\ref{eq:fullgenerallagrangian}) can radiatively induce significant mass mixing operator, $LNH$, which we have assumed to be small and/or absent. It will be important to find out whether tuning-free UV completions of this model exist. This task falls outside the scope of this paper. 

\end{enumerate}

\section*{Acknowledgements}
We thank Ornella Palamara, Roni Harnik, and Zarko Pavlovic for the correspondence and for the details of the experiments along the Booster Neutrino Beamline at Fermilab. We also thank Jae Hyeok Chang and Robert Lasenby for useful references in astrophysical and cosmological constraints, and Jonah Miller for discussions relating to supernova simulations. In addition, we thank Cliff Burgess, Pilar Hernandez, Richard Hill, Joseph Bramante and Christopher Brust for useful discussions. This research was supported in part by Perimeter Institute for Theoretical Physics. Research at Perimeter Institute is supported by the Government of Canada through the Department of Innovation, Science and Economic Development and by the Province of Ontario through the Ministry of Research and Innovation. This research was also supported by funds from the National Science and Engineering Research Council of Canada (NSERC), and the Ontario Graduate Scholarship (OGS) program. YT was supported by the Visiting Graduate Fellow program at Perimeter Institute.

\appendix
\crefalias{section}{appsec}
\section{Intensity Frontier}
\subsection{Neutrino upscattering \label{sec:neutrinoupscattering}}


We obtain an expression for $d\sigma/dt$. Consider first the matrix element for the production of $N$, which factorizes into a hadronic and a leptonic tensor, i.e.
\be
|\mathcal{M}|^2=\frac{|d|^2e^2}{q^4}L_{\mu\nu}W^{\mu\nu}.
\label{eq:lag_rp}
\ee
In terms of a right-handed projection operator, the leptonic tensor is
\begin{equation}\label{eq:4-lep-current}
L_{\mu\nu}=
4\operatorname{Tr}\biggl[\slashed{p}_1 P_R\:
\sigma_{\nu \alpha } q^{\alpha}
(\slashed{p}_3+m_N)
\sigma_{\mu\beta} q^{\beta}
\biggr].
\end{equation}
The hadronic current is given by
\be
\langle A| \Gamma^\mu |A'\rangle = F_1\gamma^\mu +F_2 \frac{i}{2 M_H} \sigma^{\mu\delta}q_\delta.
\label{eq:hadronmucurrent}
\ee
In the heavy nucleus limit, squaring \cref{eq:hadronmucurrent} gives
\be
W^{\mu\nu}=F_1^2\text{Tr}\biggr[\left(\slashed{p}_4+m_H\right)\gamma^\mu 
\left(\slashed{p}_2+m_H\right)\gamma^\nu\biggr].\\
\label{eq:4-had-current}
\ee
The representation of the form factors will depend on whether the scattering is coherent or inelastic. In the former case, the neutrino upscatters on the nucleus as a whole and the cross section scales as $Z^2$. Since $M_H=AM_\text{nucleon}$ and $|t|=|q^2|=Q^2$ is small, we retain only $F_1$ in \cref{eq:hadronmucurrent}, which we take to be the Woods-Saxon (WS) form factor. The WS form factor parameterizes the charge density of the nucleus as 
\be
\rho(r)=\frac{\rho_0}{1+\exp\left(\frac{r-r_0A^{1/3}}{a}\right)}
\ee
and takes its Fourier transformation with respect to the momentum exchange $q$ \cite{Magill2016,Jentschura2009}.
From \cref{eq:lag_rp}, we obtain 
\begin{align}\begin{split}
\frac{d\sigma}{dt}=-&\frac{2 \alpha d^2 Z^2 F_{\text{WS}}^2}{t^2 \left(s-m_H^2\right){}^2}\times \\
&\Big\{-t m_N^2 (2 s+t)+m_N^4 \left(2 m_H^2+t\right)+\\
&~~~~~~2 t \left(s-m_H^2\right)
 \left(-m_H^2+s+t\right)\Big\}
 \label{eq:dsigmadtSHiP}
\end{split}\end{align}
The $1/t^2$ pre-factor in the lab frame is proportional to $1/(E_N-E_\nu)^2$, meaning there is a phase space enhancement favoring $E_N=E_\nu$. \\

On the other hand, when the scattering is inelastic, the incoming neutrinos scatter off of the individual nucleons. When this happens, $|t|$ is of moderate size, $M_H=M_\text{nucleon}$ and we retain both form factors. $F_1$ and $F_2$ take on different values depending if they are for the neutron or proton. Their values are given \cite{Perdrisat2006,Beck:2001dz} by solving the system of equations
\begin{align}\begin{split}
G_{\gamma,E}^{p,n}&= F_{1,\gamma}^{p,n} - \frac{Q^2}{4 M_\text{Nucleon}^2} F_{2,\gamma}^{p,n} \\
 G_{\gamma,M}^{p,n} &= F_{1,\gamma}^{p,n} + F_{2,\gamma}^{p,n}
\end{split}\end{align} 
with 
\begin{align}\begin{split}
G_{\gamma,E}^{\{p,n\}} &= \{G_D,0\}				\\
G_{\gamma,M}^{\{p,n\}} &= \mu_{\{p,n\}} G_D 	\\
G_D &= \frac{1}{(1 + Q^2/0.71\GeV^2)^2}			\\
\mu_{p,n} &= \{2.793,-1.913\}.
\end{split}\end{align} 

We then obtain 
\be
\sigma_{total}=Z\times \sigma_p+(A-Z)\times\sigma_n.
\ee
In contrast to the coherent scattering case, the inelastic cross section depends only linearly on $Z$ and $A$. Furthermore, values of $t$ for which we have inelastic scattering generically avoid the $t\rightarrow 0$ enhancement. \\
\subsection{Meson decays}\label{sec:mesonphasespaceappendix}
In determining the number of HNLs present at intensity frontier experiments, it is important to consider both Primakoff upscattering and direct decays of mesons into HNLs. The decay in flight of mesons will lead to a distorted spectrum of HNLs that depends on the details of the decay at rest, and the spectrum of incoming mesons. In this appendix we outline how to obtain the spectrum of HNLs given a spectrum of incident mesons.

 We denote the rest frame energy and momentum $\mathcal{E}$ and $\mathcal{P}$, and the angle relative to the boost vector in the rest frame as $\phi$, while lab frame quantities are defined analogously as $E$, $P$, and $\theta$. We first compute the rest frame differential decay rate as a function of the energy of the HNL $\dd \Gamma/\dd \ER$. Normalizing by the overall decay rate of the meson defines the differential branching ratio in the rest frame $\left. \dd\textrm{BR}/\dd \mathcal{E}\right|_\text{rest}=\left.(1/\Gamma) \cdot \dd{\Gamma}/\dd\mathcal{E}\right|_\text{rest}$. The most important contribution to HNL production is from pions, and so we quote the result of $\dd \Gamma/\dd \mathcal{E}$ in the rest frame for the process $\pi^0\rightarrow N\nu \gamma$ 
\begin{widetext}
\be 
\frac{\dd \Gamma}{\dd \ER}=-\frac{1}{2\pi m_\pi}\alpha ^2 d^2 F_{\pi }^2 \left[\PR \left(4 \ER^2 m_{\pi }^2-3 \ER m_{\pi } m_N^2-2 \ER m_{\pi
   }^3+m_N^4+3 m_{\pi }^2 m_N^2\right)-m_{\pi }^2 \left(4 \ER-m_{\pi }\right) m_N^2 \tanh^{-1}\left(\frac{\PR}{\ER}\right)\right].
\ee
\end{widetext}

In our meson calculations, we have set the lepton masses in some of the integration bounds to 0 in order to make the integrals tractable. For most of the meson decay channels, this approximation was found to have a minor effect on the results. For heavy $m_N$, the $\pi\rightarrow \mu N\gamma$ channel was found to be underestimated by this approximation, yielding a conservative estimate.
Next, for a given energy $\mathcal{E}$, the resultant distribution in the lab-frame can be found by considering
\begin{equation}
\EL=\gamma \ER - \beta \gamma \PR \cos\phi
 \label{eq:mesint1}
\end{equation}
and noting that the decay of a pseudo-scalar is isotropic in the rest frame. Consequently the lab energies are sampled uniformly from $[\EL_-, \EL_+]$ where $\EL_\pm =\gamma \ER \pm \gamma \beta \PR$. The population of the interval of phase space in the lab frame must be the same as its corresponding interval in the rest frame. This implies that a delta-function distribution in the rest frame is transformed to a box distribution with a width of $(\EL_+ - \EL_-)=2\gamma\beta \PR$ in the lab frame.

The same argument can be applied to obtain the maximum and minimum rest frame energies that can be boosted into a given infinitesimal window centered about $\EL$. These are given by 
\begin{equation}
\ER_\pm=\gamma\EL\pm\beta\gamma\PL.
\end{equation} 
Using this information we can construct the spectrum of HNL energies generated by a meson traveling at velocity $\beta$ in the lab frame
\begin{equation}
 \qty[\frac{1}{\Gamma}\dv{\Gamma}{E}]_\text{lab}=
 \int_{\ER_A}^{\ER_B} \frac{1}{2\gamma \beta P(\ER)}
 \qty[ \frac{1}{\Gamma} \dv{\Gamma}{\ER}]_\text{rest}\dd \ER 
\end{equation}
where the factor of $2\gamma \beta \PR $ accounts for the normalization of the box distribution discussed above. The quantities $E_A$ and $E_B$ are defined via 
$\ER_A(\EL,\gamma)=\min(\ER_-,\ER_\text{min})$ and $\ER_B(\EL,\gamma)=\max(\ER_+,\ER_\text{max})$ where $\ER_\text{min}$ and $\ER_\text{max}$ are the minimum and maximum energies of the HNL that are kinematically allowed in the rest frame. Notice that the limits of integration on the right-hand side are functions of the lab energy $\EL$ and the velocity $\beta$, or equivalently $\gamma$. 

Finally, we consider a spectrum of parent mesons. In this case a spectrum (e.g. $N(\gamma)=N(E/m_\pi)$ in the case of pions) is assumed to be given and we weight the contribution of each value of $\beta$ by this spectrum finally giving 
\begin{equation}
 N_\text{lab}(E_N)=\int_{\gamma_\text{min} }^{\gamma_\text{max}}
 \qty[\frac{1}{\Gamma}\dv{\Gamma}{E} ]_\text{lab} N(\gamma)\dd \gamma 
\end{equation}
the spectrum of HNL's produced from a given flux of mesons. 

The meson energy lab spectrum used was adjusted to account for the magnitude of the beam energy, and the meson masses under consideration. When considering SBN the Sanford-Wang \cite{deNiverville:2016rqh,PhysRevD.79.072002} distribution was used to model the incident pions, while for kaons and eta mesons the Feynman scaling hypothesis \cite{PhysRevD.79.072002} was employed. At SHiP where the incident proton beam has an energy of $450\GeV$ the BMPT \cite{deNiverville:2016rqh,BMPT:Bonesini:2001iz} distribution was used instead for both pions and eta mesons.
The use of the Feynman scaling approach was inspired by \citer{Mariani:2011zd}, which argues that low energy proton beams and high meson masses exhibit special mass effects that are not well captured by Sanford-Wang. The Feynman scaling approach assumes that $\frac{d^2\sigma}{dpd\Omega}$ depends only on $p_T$ and $x_F=p_{||}^\text{COM}/p_{||}^{COM,max}$, and is proportional to $(1-|x_F|)$. Mass effects tend to give stronger weight in the data in the $x_F=0$ regime. This is reflected in the Feynman Scaling approach, whereas Sanford-Wang keeps increasing as $x_F$ crosses over to negative values. At even lower energies, such as at LSND where the POT energy is around 0.8$\GeV$, we employ the Burman-Smith distribution \cite{Burman_Smith_1989,BURMAN1990621}. By fitting to datasets spanning a wide range of pion kinetic energies ($30-553\MeV$) the Burman-Smith distribution attempts to model the pion spectrum down to zero kinetic energy. At LSND, as low kinetic energy protons interact with the beam stop, pions which are produced are slowed down. The negative pions are absorbed in matter while the positive pions decay. Most of these $\pi^+$ are at rest, while some (2\%) decay in flight. For $\mu^+$ and $\pi^+$ that decay at rest, we take their spectrum to be isotropic. For $\pi^0$ and $\pi^+$ that decay in flight, we use the Burman-Smith distribution.


\subsection{Perturbative electroweak backgrounds}
\label{sec:LoopSMBackgrounds}
As another source of background, we consider non-resonance induced single photons from perturbative electroweak processes. Although it is intuitive that the loop suppressed SM backgrounds from $A\nu \rightarrow A\nu \gamma $ will be low, it is important to quantify by how much, as this process could occur via neutrinos interacting in the walls of the SHiP experiment. Our goal here is to show that this potential source of background is very small and under control.
The cross section for $\gamma \nu\rightarrow \gamma \nu$ has been explicitly calculated using effective operators \cite{Dicus1997,Abbasabadi2000}, and this provides a convenient way to calculate the SM contribution to $A\nu\rightarrow A\nu \gamma$ by way of the equivalent photon approximation (EPA). The EPA treats the nucleus as a static charge distribution which sources a Coulomb field coherently (see \citer{Belusevic1988,Budnev1975} for a comprehensive review). As discussed in \citer{Magill2016, Magill:2017mps} the full $\sigma_{\nu A}$ cross section can be calculated from the $\sigma_{\gamma \nu}$ cross section via
\begin{equation}
\sigma_{\nu A}=\int_{{s_{min}}}^{s_{max}}\dd s~\sigma_{\gamma \nu }(s)\int _{\qty(\frac{s}{2E_\nu})^2}^\infty \dd Q^2 P(s,Q^2)
\label{eq:2-EPA-convolution}
\end{equation}
where $E_\nu$ is the energy of the neutrino in the lab frame. The function $P(s,Q^2)$ can be interpreted as the probability of the nucleus sourcing a quasi-real photon with ``mass'' $Q^2$ whose center of mass energy with the incident neutrino is $s$. 
Typically the EPA reveals an IR logarithmic enhancement, due to the effective measure of $ds/s$ induced by $P(s,Q^2)$. 
This IR enhancement is offset due to the steep $s$ dependence of $\sigma_{\gamma \nu}(s)$ \cite{Abbasabadi2000}, which scales as $\sigma_{\gamma\nu}(s)\propto s^{2.8}$ for $s\rightarrow0$. 
Using the EPA approximation to calculate production in the lead bricks of SHiP for a representative neutrino energy of $E_\nu=20~\textrm{GeV}$, we find a SM background estimate of 
\begin{equation}
\frac{\sigma_{\text{bkg}
} }{\text{Pb atom}}= 5.7 \cdot 10^{-10} \textrm{fb} =5.7 \cdot 10^{-49} \textrm{cm}^2.
\end{equation}
This is many orders of magnitude lower than the HNL production cross section estimated in the previous section and can safely be ignored. The smallness of this process follows physically from Yang's theorem \cite{Gell-Mann1961,Dicus1997,Yang1950}.

\section{Sensitivity \label{sec:sensitivity}}
We wish to briefly outline the general strategy for how all of the projected and real exclusion limits were calculated. The strategy is based on the 2009 PDG on statistics \cite{Amsler2009}. We consider a counting experiment where the experiment has seen $n$ events, whereas $b$ were predicted from the Standard Model and $s$ from new physics. In a Bayesian framework given a posterior probability and likelihood function, one can set an upper limit at credibility level $1-\alpha$ by solving
\be
1-\alpha=\int_0^{s_\text{up}}p(s|n)ds=\frac{\int_{-\infty}^{s_\text{up}}L(n|s)\pi(s)ds}{\int_{-\infty}^\infty L(n|s)\pi(s)ds}.
\ee
Using a flat prior in the new physics signal rate and the Poisson likelihood function
\be
L(n|s)=\frac{(s+b)^n}{n!}\mathrm{e}^{-(s+b)},
\ee
this can be rewritten as
\begin{align}\begin{split}
\alpha &= \mathrm{e}^{-s_{up}}\frac{\sum_{m=0}^n(s_\text{up}+b)^m/m!}{\sum_{m=0}^n b^m/m!}\\
&=\frac{\Gamma_\text{upper incomplete}(1+n,b+s_\text{up})}{\Gamma_\text{upper incomplete}(1+n,b)}.
\label{eq:incompleteGammaSensitivity}
\end{split}\end{align}

Solving for $s_{up}$ gives us the number of signal events consistent with the observation and background prediction at $(1-\alpha)$CL. Throughout this paper, we choose $1-\alpha=95\%$. To estimate projected sensitivities, we assume that $n=b$, namely that the data collected exactly matches the background prediction. For the LHC data, we implement the $CL_s$ method due to the presence of under-fluctuations of the data compared to the background predictions. This consists in defining 
\be
\alpha_b = \int_n^\infty L(n'|b) dn'
\ee
and solving for $s_{up}$ in
\be
\alpha'\equiv\frac{\alpha}{1-\alpha_b}=5\%,
\ee
with $\alpha$ defined in \cref{eq:incompleteGammaSensitivity}. This method overcovers in order to avoid setting bounds to signal rates which we are insensitive to, which can happen precisely when the data under-fluctuates. In all these cases, once we have obtained $s_{up}$, we can solve for the new physics coupling in the equation
\begin{align}\begin{split}
s_{up}=\mathcal{L}\sigma_\text{prod}\text{Br}(N\rightarrow\gamma\nu)\epsilon_\text{cuts} A_\text{geom} P_\text{dec}(L_1,L_2).
\end{split}\end{align}
In the equation above, $\mathcal{L}$ is the luminosity of the experiment. In the case of beam dump experiments, there is often an implicit sum over neutrino energies, and $\mathcal{L}$ is obtained by considering the rates and cross section of CC events in the experiment, as thoroughly described in \cite{Magill2016}.

\section{Analytic cuts \label{sec:analyticcuts}}
The calculations applicable for the neutrino experiments and for LEP are all done analytically. We generically proceed by calculating on-shell production of the HNL in the geometric region of interest and apply efficiency cuts to ensure that the recoiling nucleus/nucleon and outgoing photon from the decay of $N$ have the correct properties. It is thus important to devise handles that allow us to estimate these cuts as a function of the energy of $N$. \\ 

Consider the reaction $\nu(p_1) A(p_2)\rightarrow N(p_3)A'(p_4)$, followed by $N(p_3)\rightarrow\gamma(q_2)\nu(q_1)$. It is a relatively simple exercise in field theory to obtain $d\sigma/dt$. From here, we must determine the bounds on $t$. When working in the coherent scattering regime, we limit ourselves to the range $-0.5\GeV^{2}<t<0$. For inelastic scattering, we limit ourselves to $-2\GeV^{2}<t<-(0.217\GeV)^2/A^{\frac{2}{3}}$, and the $t<-2\GeV^2$ region applies for DIS. Within these regions, we need to pick bounds on $t$ such that when evaluated in the lab frame, the angle of the HNL overlaps with the detector. We can further restrict the range of $t$ by considering recoil cuts on the outgoing nucleus and nucleon respectively. Assuming $p_2$ initially starts at rest we have
\begin{align}\begin{split}
t&=(p_4-p_2)^2\\
&=2M_H^2-2M_HE_4\\
\Rightarrow t&\le 2M_H^2-2M_HE_4^\text{tot cut}
\end{split}\end{align}
where $M_H$ is the mass of the nucleus or nucleon depending on the context. For completeness, we also derive that
\be
E_N=E_\nu+\frac{t}{2M_H}.
\ee
This equation will be convenient for limiting ourselves to values of $t$ in which $E_N$ is 4 times above the photon energy threshold of the experiment, and to ensure that $E_N$ is sufficiently boosted for production in the line of sight. 
In addition, we derive cuts that require the photon from $N\rightarrow \gamma\nu$ to point in the right direction and be above the energy threshold of the experiment. We enforce this by imposing an efficiency factor defined as
\be
\epsilon=\frac{\Gamma \Bigm\lvert_{E_\gamma>E_\gamma^\text{cut},\theta<\theta^\text{cut}}}{\Gamma\Bigm\lvert_{E_\gamma>0,\theta<\pi}}.
\ee 
To derive a closed form for $\epsilon$, assume $N$ travels with momentum $\vec{p}_N$ in the $z$ direction. The width in the lab frame is given by
\be
\Gamma(N\rightarrow \gamma \nu)=\frac{d^2m_N^6}{8\pi E_N^3}\int \frac{d\cos\theta}{(1-\beta\cos\theta)^2}\theta(E_\gamma^\text{sol}-E_\gamma^\text{cut})
\ee
where
\be
E_\gamma^\text{sol}=\frac{m_N^2}{2(E_N-|\vec{p}_N|\cos\theta)}.
\ee
In the limit $E_\gamma^\text{cut}=0$ and integrating $\cos\theta$ between -1 and 1, we obtain as expected
\begin{align}\begin{split}
\Gamma&=\frac{1}{\gamma_\text{boost}}\frac{d^2m_N^3}{4\pi }.
\end{split}\end{align}
Requiring $E_\gamma^\text{sol}\ge E_\gamma^\text{cut}$ is equivalent to imposing
\be
\cos\theta\ge \frac{1}{\beta}\left(1-\frac{m_N^2}{2E_NE_\gamma^\text{cut}}\right)\equiv \cos\theta_\gamma^\text{ cut}.
\ee
Putting everything together, in terms of $x=\cos\theta$ and $\beta$ the velocity, we have
\begin{align}\begin{split}
\epsilon&=\frac{\int_{x_\text{cut}}^1 \frac{dx}{(1-\beta x)^2}}{\int_{-1}^1 \frac{dx}{(1-\beta x)^2}} \\
&=\frac{(\beta +1)}{2}\times\frac{1- x_\text{cut}}{1-\beta x_\text{cut}}.
\end{split}\end{align}
We calculate $x_\text{cut}$ as
\be
x_\text{cut}=\text{Max}\left[\cos\theta^\text{$\circ$ cut},\cos\theta^\text{$\gamma$ cut}\right]
\ee
where we have included a cut on the angle of the emitted photon with respect to the direction of $N$. We will typically choose $\theta^\text{$\circ$ cut}=\frac{\pi}{4}$ since we want to be emit photons in a cone centered along the initial direction of $N$.

\bibliography{ms.bib}

\begin{thebibliography}{93}%
\makeatletter
\providecommand \@ifxundefined [1]{%
 \@ifx{#1\undefined}
}%
\providecommand \@ifnum [1]{%
 \ifnum #1\expandafter \@firstoftwo
 \else \expandafter \@secondoftwo
 \fi
}%
\providecommand \@ifx [1]{%
 \ifx #1\expandafter \@firstoftwo
 \else \expandafter \@secondoftwo
 \fi
}%
\providecommand \natexlab [1]{#1}%
\providecommand \enquote  [1]{``#1''}%
\providecommand \bibnamefont  [1]{#1}%
\providecommand \bibfnamefont [1]{#1}%
\providecommand \citenamefont [1]{#1}%
\providecommand \href@noop [0]{\@secondoftwo}%
\providecommand \href [0]{\begingroup \@sanitize@url \@href}%
\providecommand \@href[1]{\@@startlink{#1}\@@href}%
\providecommand \@@href[1]{\endgroup#1\@@endlink}%
\providecommand \@sanitize@url [0]{\catcode `\\12\catcode `\$12\catcode
  `\&12\catcode `\#12\catcode `\^12\catcode `\_12\catcode `\%12\relax}%
\providecommand \@@startlink[1]{}%
\providecommand \@@endlink[0]{}%
\providecommand \url  [0]{\begingroup\@sanitize@url \@url }%
\providecommand \@url [1]{\endgroup\@href {#1}{\urlprefix }}%
\providecommand \urlprefix  [0]{URL }%
\providecommand \Eprint [0]{\href }%
\providecommand \doibase [0]{http://dx.doi.org/}%
\providecommand \selectlanguage [0]{\@gobble}%
\providecommand \bibinfo  [0]{\@secondoftwo}%
\providecommand \bibfield  [0]{\@secondoftwo}%
\providecommand \translation [1]{[#1]}%
\providecommand \BibitemOpen [0]{}%
\providecommand \bibitemStop [0]{}%
\providecommand \bibitemNoStop [0]{.\EOS\space}%
\providecommand \EOS [0]{\spacefactor3000\relax}%
\providecommand \BibitemShut  [1]{\csname bibitem#1\endcsname}%
\let\auto@bib@innerbib\@empty
\bibitem [{\citenamefont {Asaka}\ and\ \citenamefont
  {Shaposhnikov}(2005)}]{Asaka:2005pn}%
  \BibitemOpen
  \bibfield  {author} {\bibinfo {author} {\bibfnamefont {Takehiko}\
  \bibnamefont {Asaka}}\ and\ \bibinfo {author} {\bibfnamefont {Mikhail}\
  \bibnamefont {Shaposhnikov}},\ }\bibfield  {title} {\enquote {\bibinfo
  {title} {{The nuMSM, dark matter and baryon asymmetry of the universe}},}\
  }\href {\doibase 10.1016/j.physletb.2005.06.020} {\bibfield  {journal}
  {\bibinfo  {journal} {Phys. Lett.}\ }\textbf {\bibinfo {volume} {B620}},\
  \bibinfo {pages} {17--26} (\bibinfo {year} {2005})},\ \Eprint
  {http://arxiv.org/abs/hep-ph/0505013} {arXiv:hep-ph/0505013 [hep-ph]}
  \BibitemShut {NoStop}%
\bibitem [{\citenamefont {Gorbunov}\ and\ \citenamefont
  {Shaposhnikov}(2007)}]{Gorbunov:2007ak}%
  \BibitemOpen
  \bibfield  {author} {\bibinfo {author} {\bibfnamefont {Dmitry}\ \bibnamefont
  {Gorbunov}}\ and\ \bibinfo {author} {\bibfnamefont {Mikhail}\ \bibnamefont
  {Shaposhnikov}},\ }\bibfield  {title} {\enquote {\bibinfo {title} {{How to
  find neutral leptons of the $\nu$MSM?}}}\ }\href {\doibase
  10.1007/JHEP11(2013)101, 10.1088/1126-6708/2007/10/015} {\bibfield  {journal}
  {\bibinfo  {journal} {JHEP}\ }\textbf {\bibinfo {volume} {10}},\ \bibinfo
  {pages} {015} (\bibinfo {year} {2007})},\ \bibinfo {note} {[Erratum:
  JHEP11,101(2013)]},\ \Eprint {http://arxiv.org/abs/0705.1729}
  {arXiv:0705.1729 [hep-ph]} \BibitemShut {NoStop}%
\bibitem [{\citenamefont {Alekhin}\ \emph {et~al.}(2016)\citenamefont {Alekhin}
  \emph {et~al.}}]{Alekhin:2015byh}%
  \BibitemOpen
  \bibfield  {author} {\bibinfo {author} {\bibfnamefont {Sergey}\ \bibnamefont
  {Alekhin}} \emph {et~al.},\ }\bibfield  {title} {\enquote {\bibinfo {title}
  {{A facility to Search for Hidden Particles at the CERN SPS: the SHiP physics
  case}},}\ }\href {\doibase 10.1088/0034-4885/79/12/124201} {\bibfield
  {journal} {\bibinfo  {journal} {Rept. Prog. Phys.}\ }\textbf {\bibinfo
  {volume} {79}},\ \bibinfo {pages} {124201} (\bibinfo {year} {2016})},\
  \Eprint {http://arxiv.org/abs/1504.04855} {arXiv:1504.04855 [hep-ph]}
  \BibitemShut {NoStop}%
\bibitem [{\citenamefont {Gninenko}(2009)}]{Gninenko:2009ks}%
  \BibitemOpen
  \bibfield  {author} {\bibinfo {author} {\bibfnamefont {S.~N.}\ \bibnamefont
  {Gninenko}},\ }\bibfield  {title} {\enquote {\bibinfo {title} {{The MiniBooNE
  anomaly and heavy neutrino decay}},}\ }\href {\doibase
  10.1103/PhysRevLett.103.241802} {\bibfield  {journal} {\bibinfo  {journal}
  {Phys. Rev. Lett.}\ }\textbf {\bibinfo {volume} {103}},\ \bibinfo {pages}
  {241802} (\bibinfo {year} {2009})},\ \Eprint {http://arxiv.org/abs/0902.3802}
  {arXiv:0902.3802 [hep-ph]} \BibitemShut {NoStop}%
\bibitem [{\citenamefont {Gninenko}(2011)}]{Gninenko:2010pr}%
  \BibitemOpen
  \bibfield  {author} {\bibinfo {author} {\bibfnamefont {Sergei~N.}\
  \bibnamefont {Gninenko}},\ }\bibfield  {title} {\enquote {\bibinfo {title}
  {{A resolution of puzzles from the LSND, KARMEN, and MiniBooNE
  experiments}},}\ }\href {\doibase 10.1103/PhysRevD.83.015015} {\bibfield
  {journal} {\bibinfo  {journal} {Phys. Rev.}\ }\textbf {\bibinfo {volume}
  {D83}},\ \bibinfo {pages} {015015} (\bibinfo {year} {2011})},\ \Eprint
  {http://arxiv.org/abs/1009.5536} {arXiv:1009.5536 [hep-ph]} \BibitemShut
  {NoStop}%
\bibitem [{\citenamefont {McKeen}\ and\ \citenamefont
  {Pospelov}(2010)}]{McKeen:2010rx}%
  \BibitemOpen
  \bibfield  {author} {\bibinfo {author} {\bibfnamefont {David}\ \bibnamefont
  {McKeen}}\ and\ \bibinfo {author} {\bibfnamefont {Maxim}\ \bibnamefont
  {Pospelov}},\ }\bibfield  {title} {\enquote {\bibinfo {title} {{Muon Capture
  Constraints on Sterile Neutrino Properties}},}\ }\href {\doibase
  10.1103/PhysRevD.82.113018} {\bibfield  {journal} {\bibinfo  {journal} {Phys.
  Rev.}\ }\textbf {\bibinfo {volume} {D82}},\ \bibinfo {pages} {113018}
  (\bibinfo {year} {2010})},\ \Eprint {http://arxiv.org/abs/1011.3046}
  {arXiv:1011.3046 [hep-ph]} \BibitemShut {NoStop}%
\bibitem [{\citenamefont {Masip}\ \emph {et~al.}(2013)\citenamefont {Masip},
  \citenamefont {Masjuan},\ and\ \citenamefont {Meloni}}]{Masip:2012ke}%
  \BibitemOpen
  \bibfield  {author} {\bibinfo {author} {\bibfnamefont {Manuel}\ \bibnamefont
  {Masip}}, \bibinfo {author} {\bibfnamefont {Pere}\ \bibnamefont {Masjuan}}, \
  and\ \bibinfo {author} {\bibfnamefont {Davide}\ \bibnamefont {Meloni}},\
  }\bibfield  {title} {\enquote {\bibinfo {title} {{Heavy neutrino decays at
  MiniBooNE}},}\ }\href {\doibase 10.1007/JHEP01(2013)106} {\bibfield
  {journal} {\bibinfo  {journal} {JHEP}\ }\textbf {\bibinfo {volume} {01}},\
  \bibinfo {pages} {106} (\bibinfo {year} {2013})},\ \Eprint
  {http://arxiv.org/abs/1210.1519} {arXiv:1210.1519 [hep-ph]} \BibitemShut
  {NoStop}%
\bibitem [{\citenamefont {Masip}\ and\ \citenamefont
  {Masjuan}(2011)}]{Masip:2011qb}%
  \BibitemOpen
  \bibfield  {author} {\bibinfo {author} {\bibfnamefont {Manuel}\ \bibnamefont
  {Masip}}\ and\ \bibinfo {author} {\bibfnamefont {Pere}\ \bibnamefont
  {Masjuan}},\ }\bibfield  {title} {\enquote {\bibinfo {title} {{Heavy-neutrino
  decays at neutrino telescopes}},}\ }\href {\doibase
  10.1103/PhysRevD.83.091301} {\bibfield  {journal} {\bibinfo  {journal} {Phys.
  Rev.}\ }\textbf {\bibinfo {volume} {D83}},\ \bibinfo {pages} {091301}
  (\bibinfo {year} {2011})},\ \Eprint {http://arxiv.org/abs/1103.0689}
  {arXiv:1103.0689 [hep-ph]} \BibitemShut {NoStop}%
\bibitem [{\citenamefont {Gninenko}(2012)}]{Gninenko:2012rw}%
  \BibitemOpen
  \bibfield  {author} {\bibinfo {author} {\bibfnamefont {S.~N.}\ \bibnamefont
  {Gninenko}},\ }\bibfield  {title} {\enquote {\bibinfo {title} {{New limits on
  radiative sterile neutrino decays from a search for single photons in
  neutrino interactions}},}\ }\href {\doibase 10.1016/j.physletb.2012.02.071}
  {\bibfield  {journal} {\bibinfo  {journal} {Phys. Lett.}\ }\textbf {\bibinfo
  {volume} {B710}},\ \bibinfo {pages} {86--90} (\bibinfo {year} {2012})},\
  \Eprint {http://arxiv.org/abs/1201.5194} {arXiv:1201.5194 [hep-ph]}
  \BibitemShut {NoStop}%
\bibitem [{\citenamefont {Giunti}\ and\ \citenamefont
  {Studenikin}(2015)}]{Giunti:2014ixa}%
  \BibitemOpen
  \bibfield  {author} {\bibinfo {author} {\bibfnamefont {Carlo}\ \bibnamefont
  {Giunti}}\ and\ \bibinfo {author} {\bibfnamefont {Alexander}\ \bibnamefont
  {Studenikin}},\ }\bibfield  {title} {\enquote {\bibinfo {title} {{Neutrino
  electromagnetic interactions: a window to new physics}},}\ }\href {\doibase
  10.1103/RevModPhys.87.531} {\bibfield  {journal} {\bibinfo  {journal} {Rev.
  Mod. Phys.}\ }\textbf {\bibinfo {volume} {87}},\ \bibinfo {pages} {531}
  (\bibinfo {year} {2015})},\ \Eprint {http://arxiv.org/abs/1403.6344}
  {arXiv:1403.6344 [hep-ph]} \BibitemShut {NoStop}%
\bibitem [{\citenamefont {Aparici}\ \emph {et~al.}(2009)\citenamefont
  {Aparici}, \citenamefont {Kim}, \citenamefont {Santamaria},\ and\
  \citenamefont {Wudka}}]{Aparici:2009fh}%
  \BibitemOpen
  \bibfield  {author} {\bibinfo {author} {\bibfnamefont {Alberto}\ \bibnamefont
  {Aparici}}, \bibinfo {author} {\bibfnamefont {Kyungwook}\ \bibnamefont
  {Kim}}, \bibinfo {author} {\bibfnamefont {Arcadi}\ \bibnamefont
  {Santamaria}}, \ and\ \bibinfo {author} {\bibfnamefont {Jose}\ \bibnamefont
  {Wudka}},\ }\bibfield  {title} {\enquote {\bibinfo {title} {{Right-handed
  neutrino magnetic moments}},}\ }\href {\doibase 10.1103/PhysRevD.80.013010}
  {\bibfield  {journal} {\bibinfo  {journal} {Phys. Rev.}\ }\textbf {\bibinfo
  {volume} {D80}},\ \bibinfo {pages} {013010} (\bibinfo {year} {2009})},\
  \Eprint {http://arxiv.org/abs/0904.3244} {arXiv:0904.3244 [hep-ph]}
  \BibitemShut {NoStop}%
\bibitem [{\citenamefont {Aparici}(2013)}]{Aparici:2013xga}%
  \BibitemOpen
  \bibfield  {author} {\bibinfo {author} {\bibfnamefont {Alberto}\ \bibnamefont
  {Aparici}},\ }\emph {\bibinfo {title} {{Exotic properties of neutrinos using
  effective Lagrangians and specific models}}},\ \href
  {http://inspirehep.net/record/1267003/files/arXiv:1312.0554.pdf} {Ph.D.
  thesis},\ \bibinfo  {school} {Valencia U.} (\bibinfo {year} {2013}),\ \Eprint
  {http://arxiv.org/abs/1312.0554} {arXiv:1312.0554 [hep-ph]} \BibitemShut
  {NoStop}%
\bibitem [{\citenamefont {Caputo}\ \emph {et~al.}(2017)\citenamefont {Caputo},
  \citenamefont {Hernandez}, \citenamefont {Lopez-Pavon},\ and\ \citenamefont
  {Salvado}}]{Caputo:2017pit}%
  \BibitemOpen
  \bibfield  {author} {\bibinfo {author} {\bibfnamefont {A.}~\bibnamefont
  {Caputo}}, \bibinfo {author} {\bibfnamefont {P.}~\bibnamefont {Hernandez}},
  \bibinfo {author} {\bibfnamefont {J.}~\bibnamefont {Lopez-Pavon}}, \ and\
  \bibinfo {author} {\bibfnamefont {J.}~\bibnamefont {Salvado}},\ }\bibfield
  {title} {\enquote {\bibinfo {title} {{The seesaw portal in testable models of
  neutrino masses}},}\ }\href {\doibase 10.1007/JHEP06(2017)112} {\bibfield
  {journal} {\bibinfo  {journal} {JHEP}\ }\textbf {\bibinfo {volume} {06}},\
  \bibinfo {pages} {112} (\bibinfo {year} {2017})},\ \Eprint
  {http://arxiv.org/abs/1704.08721} {arXiv:1704.08721 [hep-ph]} \BibitemShut
  {NoStop}%
\bibitem [{\citenamefont {Abazajian}(2017)}]{Abazajian:2017tcc}%
  \BibitemOpen
  \bibfield  {author} {\bibinfo {author} {\bibfnamefont {Kevork~N.}\
  \bibnamefont {Abazajian}},\ }\bibfield  {title} {\enquote {\bibinfo {title}
  {{Sterile neutrinos in cosmology}},}\ }\href {\doibase
  10.1016/j.physrep.2017.10.003} {\bibfield  {journal} {\bibinfo  {journal}
  {Phys. Rept.}\ }\textbf {\bibinfo {volume} {711-712}},\ \bibinfo {pages}
  {1--28} (\bibinfo {year} {2017})},\ \Eprint {http://arxiv.org/abs/1705.01837}
  {arXiv:1705.01837 [hep-ph]} \BibitemShut {NoStop}%
\bibitem [{\citenamefont {Aguilar-Arevalo}\ \emph {et~al.}(2007)\citenamefont
  {Aguilar-Arevalo} \emph {et~al.}}]{AguilarArevalo:2007it}%
  \BibitemOpen
  \bibfield  {author} {\bibinfo {author} {\bibfnamefont {A.~A.}\ \bibnamefont
  {Aguilar-Arevalo}} \emph {et~al.} (\bibinfo {collaboration} {MiniBooNE}),\
  }\bibfield  {title} {\enquote {\bibinfo {title} {{A Search for electron
  neutrino appearance at the $\Delta m^{2} \sim 1$eV$^{2}$ scale}},}\ }\href
  {\doibase 10.1103/PhysRevLett.98.231801} {\bibfield  {journal} {\bibinfo
  {journal} {Phys. Rev. Lett.}\ }\textbf {\bibinfo {volume} {98}},\ \bibinfo
  {pages} {231801} (\bibinfo {year} {2007})},\ \Eprint
  {http://arxiv.org/abs/0704.1500} {arXiv:0704.1500 [hep-ex]} \BibitemShut
  {NoStop}%
\bibitem [{\citenamefont {Athanassopoulos}\ \emph {et~al.}(1996)\citenamefont
  {Athanassopoulos} \emph {et~al.}}]{Athanassopoulos:1996jb}%
  \BibitemOpen
  \bibfield  {author} {\bibinfo {author} {\bibfnamefont {C.}~\bibnamefont
  {Athanassopoulos}} \emph {et~al.} (\bibinfo {collaboration} {LSND}),\
  }\bibfield  {title} {\enquote {\bibinfo {title} {{Evidence for
  anti-muon-neutrino ---> anti-electron-neutrino oscillations from the LSND
  experiment at LAMPF}},}\ }\href {\doibase 10.1103/PhysRevLett.77.3082}
  {\bibfield  {journal} {\bibinfo  {journal} {Phys. Rev. Lett.}\ }\textbf
  {\bibinfo {volume} {77}},\ \bibinfo {pages} {3082--3085} (\bibinfo {year}
  {1996})},\ \Eprint {http://arxiv.org/abs/nucl-ex/9605003}
  {arXiv:nucl-ex/9605003 [nucl-ex]} \BibitemShut {NoStop}%
\bibitem [{\citenamefont {Hill}(2011)}]{Hill:2010zy}%
  \BibitemOpen
  \bibfield  {author} {\bibinfo {author} {\bibfnamefont {Richard~J.}\
  \bibnamefont {Hill}},\ }\bibfield  {title} {\enquote {\bibinfo {title} {{On
  the single photon background to $\nu_e$ appearance at MiniBooNE}},}\ }\href
  {\doibase 10.1103/PhysRevD.84.017501} {\bibfield  {journal} {\bibinfo
  {journal} {Phys. Rev.}\ }\textbf {\bibinfo {volume} {D84}},\ \bibinfo {pages}
  {017501} (\bibinfo {year} {2011})},\ \Eprint {http://arxiv.org/abs/1002.4215}
  {arXiv:1002.4215 [hep-ph]} \BibitemShut {NoStop}%
\bibitem [{\citenamefont {Abazajian}\ \emph {et~al.}(2012)\citenamefont
  {Abazajian} \emph {et~al.}}]{Abazajian:2012ys}%
  \BibitemOpen
  \bibfield  {author} {\bibinfo {author} {\bibfnamefont {K.~N.}\ \bibnamefont
  {Abazajian}} \emph {et~al.},\ }\bibfield  {title} {\enquote {\bibinfo {title}
  {{Light Sterile Neutrinos: A White Paper}},}\ }\href@noop {} {\  (\bibinfo
  {year} {2012})},\ \Eprint {http://arxiv.org/abs/1204.5379} {arXiv:1204.5379
  [hep-ph]} \BibitemShut {NoStop}%
\bibitem [{\citenamefont {Couchot}\ \emph {et~al.}(2017)\citenamefont
  {Couchot}, \citenamefont {Henrot-Versill\'{e}}, \citenamefont {Perdereau},
  \citenamefont {Plaszczynski}, \citenamefont {Rouill\'{e}~D'Orfeuil},
  \citenamefont {Spinelli},\ and\ \citenamefont {Tristram}}]{Couchot:2017pvz}%
  \BibitemOpen
  \bibfield  {author} {\bibinfo {author} {\bibfnamefont {F.}~\bibnamefont
  {Couchot}}, \bibinfo {author} {\bibfnamefont {S.}~\bibnamefont
  {Henrot-Versill\'{e}}}, \bibinfo {author} {\bibfnamefont {O.}~\bibnamefont
  {Perdereau}}, \bibinfo {author} {\bibfnamefont {S.}~\bibnamefont
  {Plaszczynski}}, \bibinfo {author} {\bibfnamefont {B.}~\bibnamefont
  {Rouill\'{e}~D'Orfeuil}}, \bibinfo {author} {\bibfnamefont {M.}~\bibnamefont
  {Spinelli}}, \ and\ \bibinfo {author} {\bibfnamefont {M.}~\bibnamefont
  {Tristram}},\ }\bibfield  {title} {\enquote {\bibinfo {title} {{Cosmological
  constraints on the neutrino mass including systematic uncertainties}},}\
  }\href {\doibase 10.1051/0004-6361/201730927} {\bibfield  {journal} {\bibinfo
   {journal} {Astron. Astrophys.}\ } (\bibinfo {year} {2017}),\
  10.1051/0004-6361/201730927},\ \bibinfo {note} {[Astron.
  Astrophys.606,A104(2017)]},\ \Eprint {http://arxiv.org/abs/1703.10829}
  {arXiv:1703.10829 [astro-ph.CO]} \BibitemShut {NoStop}%
\bibitem [{\citenamefont {Dasgupta}\ and\ \citenamefont
  {Kopp}(2014)}]{Dasgupta:2013zpn}%
  \BibitemOpen
  \bibfield  {author} {\bibinfo {author} {\bibfnamefont {Basudeb}\ \bibnamefont
  {Dasgupta}}\ and\ \bibinfo {author} {\bibfnamefont {Joachim}\ \bibnamefont
  {Kopp}},\ }\bibfield  {title} {\enquote {\bibinfo {title} {{Cosmologically
  Safe eV-Scale Sterile Neutrinos and Improved Dark Matter Structure}},}\
  }\href {\doibase 10.1103/PhysRevLett.112.031803} {\bibfield  {journal}
  {\bibinfo  {journal} {Phys. Rev. Lett.}\ }\textbf {\bibinfo {volume} {112}},\
  \bibinfo {pages} {031803} (\bibinfo {year} {2014})},\ \Eprint
  {http://arxiv.org/abs/1310.6337} {arXiv:1310.6337 [hep-ph]} \BibitemShut
  {NoStop}%
\bibitem [{\citenamefont {Hannestad}\ \emph {et~al.}(2014)\citenamefont
  {Hannestad}, \citenamefont {Hansen},\ and\ \citenamefont
  {Tram}}]{Hannestad:2013ana}%
  \BibitemOpen
  \bibfield  {author} {\bibinfo {author} {\bibfnamefont {Steen}\ \bibnamefont
  {Hannestad}}, \bibinfo {author} {\bibfnamefont {Rasmus~Sloth}\ \bibnamefont
  {Hansen}}, \ and\ \bibinfo {author} {\bibfnamefont {Thomas}\ \bibnamefont
  {Tram}},\ }\bibfield  {title} {\enquote {\bibinfo {title} {{How
  Self-Interactions can Reconcile Sterile Neutrinos with Cosmology}},}\ }\href
  {\doibase 10.1103/PhysRevLett.112.031802} {\bibfield  {journal} {\bibinfo
  {journal} {Phys. Rev. Lett.}\ }\textbf {\bibinfo {volume} {112}},\ \bibinfo
  {pages} {031802} (\bibinfo {year} {2014})},\ \Eprint
  {http://arxiv.org/abs/1310.5926} {arXiv:1310.5926 [astro-ph.CO]} \BibitemShut
  {NoStop}%
\bibitem [{\citenamefont {Antonello}\ \emph {et~al.}(2015)\citenamefont
  {Antonello} \emph {et~al.}}]{Antonello:2015lea}%
  \BibitemOpen
  \bibfield  {author} {\bibinfo {author} {\bibfnamefont {M.}~\bibnamefont
  {Antonello}} \emph {et~al.} (\bibinfo {collaboration} {LAr1-ND, ICARUS-WA104,
  MicroBooNE}),\ }\bibfield  {title} {\enquote {\bibinfo {title} {{A Proposal
  for a Three Detector Short-Baseline Neutrino Oscillation Program in the
  Fermilab Booster Neutrino Beam}},}\ }\href@noop {} {\  (\bibinfo {year}
  {2015})},\ \Eprint {http://arxiv.org/abs/1503.01520} {arXiv:1503.01520
  [physics.ins-det]} \BibitemShut {NoStop}%
\bibitem [{\citenamefont {Ballett}\ \emph {et~al.}(2017)\citenamefont
  {Ballett}, \citenamefont {Pascoli},\ and\ \citenamefont
  {Ross-Lonergan}}]{Ballett:2016opr}%
  \BibitemOpen
  \bibfield  {author} {\bibinfo {author} {\bibfnamefont {Peter}\ \bibnamefont
  {Ballett}}, \bibinfo {author} {\bibfnamefont {Silvia}\ \bibnamefont
  {Pascoli}}, \ and\ \bibinfo {author} {\bibfnamefont {Mark}\ \bibnamefont
  {Ross-Lonergan}},\ }\bibfield  {title} {\enquote {\bibinfo {title}
  {{MeV-scale sterile neutrino decays at the Fermilab Short-Baseline Neutrino
  program}},}\ }\href {\doibase 10.1007/JHEP04(2017)102} {\bibfield  {journal}
  {\bibinfo  {journal} {JHEP}\ }\textbf {\bibinfo {volume} {04}},\ \bibinfo
  {pages} {102} (\bibinfo {year} {2017})},\ \Eprint
  {http://arxiv.org/abs/1610.08512} {arXiv:1610.08512 [hep-ph]} \BibitemShut
  {NoStop}%
\bibitem [{\citenamefont {Pal}\ and\ \citenamefont
  {Wolfenstein}(1982)}]{Pal1982}%
  \BibitemOpen
  \bibfield  {author} {\bibinfo {author} {\bibfnamefont {Palash~B.}\
  \bibnamefont {Pal}}\ and\ \bibinfo {author} {\bibfnamefont {Lincoln}\
  \bibnamefont {Wolfenstein}},\ }\bibfield  {title} {\enquote {\bibinfo {title}
  {Radiative decays of massive neutrinos},}\ }\href {\doibase
  10.1103/PhysRevD.25.766} {\bibfield  {journal} {\bibinfo  {journal} {Phys.
  Rev. D}\ }\textbf {\bibinfo {volume} {25}},\ \bibinfo {pages} {766--773}
  (\bibinfo {year} {1982})}\BibitemShut {NoStop}%
\bibitem [{\citenamefont {Shrock}(1982)}]{Shrock1982}%
  \BibitemOpen
  \bibfield  {author} {\bibinfo {author} {\bibfnamefont {Robert~E.}\
  \bibnamefont {Shrock}},\ }\bibfield  {title} {\enquote {\bibinfo {title}
  {Electromagnetic properties and decays of dirac and majorana neutrinos in a
  general class of gauge theories},}\ }\href {\doibase
  https://doi.org/10.1016/0550-3213(82)90273-5} {\bibfield  {journal} {\bibinfo
   {journal} {Nuclear Physics B}\ }\textbf {\bibinfo {volume} {206}},\ \bibinfo
  {pages} {359 -- 379} (\bibinfo {year} {1982})}\BibitemShut {NoStop}%
\bibitem [{\citenamefont {Mohapatra}(1986)}]{Mohapatra:1986aw}%
  \BibitemOpen
  \bibfield  {author} {\bibinfo {author} {\bibfnamefont {R.~N.}\ \bibnamefont
  {Mohapatra}},\ }\bibfield  {title} {\enquote {\bibinfo {title} {{Mechanism
  for Understanding Small Neutrino Mass in Superstring Theories}},}\ }\href
  {\doibase 10.1103/PhysRevLett.56.561} {\bibfield  {journal} {\bibinfo
  {journal} {Phys. Rev. Lett.}\ }\textbf {\bibinfo {volume} {56}},\ \bibinfo
  {pages} {561--563} (\bibinfo {year} {1986})}\BibitemShut {NoStop}%
\bibitem [{\citenamefont {Mohapatra}\ and\ \citenamefont
  {Valle}(1986)}]{Mohapatra:1986bd}%
  \BibitemOpen
  \bibfield  {author} {\bibinfo {author} {\bibfnamefont {R.~N.}\ \bibnamefont
  {Mohapatra}}\ and\ \bibinfo {author} {\bibfnamefont {J.~W.~F.}\ \bibnamefont
  {Valle}},\ }\bibfield  {title} {\enquote {\bibinfo {title} {{Neutrino Mass
  and Baryon Number Nonconservation in Superstring Models}},}\ }\bibfield
  {booktitle} {\emph {\bibinfo {booktitle} {{Proceedings, 23RD International
  Conference on High Energy Physics, JULY 16-23, 1986, Berkeley, CA}}},\ }\href
  {\doibase 10.1103/PhysRevD.34.1642} {\bibfield  {journal} {\bibinfo
  {journal} {Phys. Rev.}\ }\textbf {\bibinfo {volume} {D34}},\ \bibinfo {pages}
  {1642} (\bibinfo {year} {1986})}\BibitemShut {NoStop}%
\bibitem [{\citenamefont {Grzadkowski}\ \emph {et~al.}(2010)\citenamefont
  {Grzadkowski}, \citenamefont {Iskrzynski}, \citenamefont {Misiak},\ and\
  \citenamefont {Rosiek}}]{Grzadkowski:2010es}%
  \BibitemOpen
  \bibfield  {author} {\bibinfo {author} {\bibfnamefont {B.}~\bibnamefont
  {Grzadkowski}}, \bibinfo {author} {\bibfnamefont {M.}~\bibnamefont
  {Iskrzynski}}, \bibinfo {author} {\bibfnamefont {M.}~\bibnamefont {Misiak}},
  \ and\ \bibinfo {author} {\bibfnamefont {J.}~\bibnamefont {Rosiek}},\
  }\bibfield  {title} {\enquote {\bibinfo {title} {{Dimension-Six Terms in the
  Standard Model Lagrangian}},}\ }\href {\doibase 10.1007/JHEP10(2010)085}
  {\bibfield  {journal} {\bibinfo  {journal} {JHEP}\ }\textbf {\bibinfo
  {volume} {10}},\ \bibinfo {pages} {085} (\bibinfo {year} {2010})},\ \Eprint
  {http://arxiv.org/abs/1008.4884} {arXiv:1008.4884 [hep-ph]} \BibitemShut
  {NoStop}%
\bibitem [{\citenamefont {Aguilar-Arevalo}\ \emph {et~al.}(2009)\citenamefont
  {Aguilar-Arevalo} \emph {et~al.}}]{PhysRevD.79.072002}%
  \BibitemOpen
  \bibfield  {author} {\bibinfo {author} {\bibfnamefont {A.~A.}\ \bibnamefont
  {Aguilar-Arevalo}} \emph {et~al.} (\bibinfo {collaboration} {MiniBooNE
  Collaboration}),\ }\bibfield  {title} {\enquote {\bibinfo {title} {Neutrino
  flux prediction at miniboone},}\ }\href {\doibase 10.1103/PhysRevD.79.072002}
  {\bibfield  {journal} {\bibinfo  {journal} {Phys. Rev. D}\ }\textbf {\bibinfo
  {volume} {79}},\ \bibinfo {pages} {072002} (\bibinfo {year}
  {2009})}\BibitemShut {NoStop}%
\bibitem [{\citenamefont {Auerbach}\ \emph {et~al.}(2001)\citenamefont
  {Auerbach} \emph {et~al.}}]{Auerbach:2001wg}%
  \BibitemOpen
  \bibfield  {author} {\bibinfo {author} {\bibfnamefont {L.~B.}\ \bibnamefont
  {Auerbach}} \emph {et~al.} (\bibinfo {collaboration} {LSND}),\ }\bibfield
  {title} {\enquote {\bibinfo {title} {{Measurement of electron - neutrino -
  electron elastic scattering}},}\ }\href {\doibase 10.1103/PhysRevD.63.112001}
  {\bibfield  {journal} {\bibinfo  {journal} {Phys. Rev.}\ }\textbf {\bibinfo
  {volume} {D63}},\ \bibinfo {pages} {112001} (\bibinfo {year} {2001})},\
  \Eprint {http://arxiv.org/abs/hep-ex/0101039} {arXiv:hep-ex/0101039 [hep-ex]}
  \BibitemShut {NoStop}%
\bibitem [{\citenamefont {Burman}\ \emph {et~al.}(1990)\citenamefont {Burman},
  \citenamefont {Potter},\ and\ \citenamefont {Smith}}]{BURMAN1990621}%
  \BibitemOpen
  \bibfield  {author} {\bibinfo {author} {\bibfnamefont {R.L.}\ \bibnamefont
  {Burman}}, \bibinfo {author} {\bibfnamefont {M.E.}\ \bibnamefont {Potter}}, \
  and\ \bibinfo {author} {\bibfnamefont {E.S.}\ \bibnamefont {Smith}},\
  }\bibfield  {title} {\enquote {\bibinfo {title} {Monte carlo simulation of
  neutrino production by medium-energy protons in a beam stop},}\ }\href
  {\doibase https://doi.org/10.1016/0168-9002(90)90012-U} {\bibfield  {journal}
  {\bibinfo  {journal} {Nuclear Instruments and Methods in Physics Research
  Section A: Accelerators, Spectrometers, Detectors and Associated Equipment}\
  }\textbf {\bibinfo {volume} {291}},\ \bibinfo {pages} {621 -- 633} (\bibinfo
  {year} {1990})}\BibitemShut {NoStop}%
\bibitem [{\citenamefont {Burman}\ and\ \citenamefont
  {Smith}(1989)}]{Burman_Smith_1989}%
  \BibitemOpen
  \bibfield  {author} {\bibinfo {author} {\bibfnamefont {R.L.}\ \bibnamefont
  {Burman}}\ and\ \bibinfo {author} {\bibfnamefont {E.S.}\ \bibnamefont
  {Smith}},\ }\bibfield  {title} {\enquote {\bibinfo {title} {Parameterization
  of pion production and reaction cross sections at lampf energies},}\ }\href
  {\doibase 10.2172/6167579} {\bibfield  {journal} {\bibinfo  {journal} {LAMPF
  Report}\ }\textbf {\bibinfo {volume} {LA-11502-MS}} (\bibinfo {year}
  {1989}),\ 10.2172/6167579}\BibitemShut {NoStop}%
\bibitem [{\citenamefont {Bernard}\ \emph {et~al.}(2001)\citenamefont
  {Bernard}, \citenamefont {Hemmert},\ and\ \citenamefont
  {Meissner}}]{Bernard:2000et}%
  \BibitemOpen
  \bibfield  {author} {\bibinfo {author} {\bibfnamefont {Veronique}\
  \bibnamefont {Bernard}}, \bibinfo {author} {\bibfnamefont {Thomas~R.}\
  \bibnamefont {Hemmert}}, \ and\ \bibinfo {author} {\bibfnamefont {Ulf-G.}\
  \bibnamefont {Meissner}},\ }\bibfield  {title} {\enquote {\bibinfo {title}
  {{Ordinary and radiative muon capture on the proton and the pseudoscalar
  form-factor of the nucleon}},}\ }\href {\doibase
  10.1016/S0375-9474(00)00520-0} {\bibfield  {journal} {\bibinfo  {journal}
  {Nucl. Phys.}\ }\textbf {\bibinfo {volume} {A686}},\ \bibinfo {pages}
  {290--316} (\bibinfo {year} {2001})},\ \Eprint
  {http://arxiv.org/abs/nucl-th/0001052} {arXiv:nucl-th/0001052 [nucl-th]}
  \BibitemShut {NoStop}%
\bibitem [{\citenamefont {Alvarez-Ruso}\ and\ \citenamefont
  {Saul-Sala}(2017)}]{Alvarez-Ruso:2017hdm}%
  \BibitemOpen
  \bibfield  {author} {\bibinfo {author} {\bibfnamefont {Luis}\ \bibnamefont
  {Alvarez-Ruso}}\ and\ \bibinfo {author} {\bibfnamefont {Eduardo}\
  \bibnamefont {Saul-Sala}},\ }\bibfield  {title} {\enquote {\bibinfo {title}
  {{Radiative decay of heavy neutrinos at MiniBooNE and MicroBooNE}},}\ }in\
  \href {http://inspirehep.net/record/1597406/files/arXiv:1705.00353.pdf}
  {\emph {\bibinfo {booktitle} {{Proceedings, Prospects in Neutrino Physics
  (NuPhys2016): London, UK, December 12-14, 2016}}}}\ (\bibinfo {year} {2017})\
  \Eprint {http://arxiv.org/abs/1705.00353} {arXiv:1705.00353 [hep-ph]}
  \BibitemShut {NoStop}%
\bibitem [{\citenamefont {Vannucci}(2014)}]{Vannucci:2014wna}%
  \BibitemOpen
  \bibfield  {author} {\bibinfo {author} {\bibfnamefont {F.}~\bibnamefont
  {Vannucci}},\ }\bibfield  {title} {\enquote {\bibinfo {title} {{The NOMAD
  Experiment at CERN}},}\ }\href {\doibase 10.1155/2014/129694} {\bibfield
  {journal} {\bibinfo  {journal} {Adv. High Energy Phys.}\ }\textbf {\bibinfo
  {volume} {2014}},\ \bibinfo {pages} {129694} (\bibinfo {year}
  {2014})}\BibitemShut {NoStop}%
\bibitem [{\citenamefont {Altegoer}\ \emph
  {et~al.}(1998{\natexlab{a}})\citenamefont {Altegoer} \emph
  {et~al.}}]{Altegoer1998a}%
  \BibitemOpen
  \bibfield  {author} {\bibinfo {author} {\bibfnamefont {J.}~\bibnamefont
  {Altegoer}} \emph {et~al.},\ }\bibfield  {title} {\enquote {\bibinfo {title}
  {The nomad experiment at the cern sps},}\ }\href {\doibase
  http://dx.doi.org/10.1016/S0168-9002(97)01079-6} {\bibfield  {journal}
  {\bibinfo  {journal} {Nuclear Instruments and Methods in Physics Research
  Section A: Accelerators, Spectrometers, Detectors and Associated Equipment}\
  }\textbf {\bibinfo {volume} {404}},\ \bibinfo {pages} {96 -- 128} (\bibinfo
  {year} {1998}{\natexlab{a}})}\BibitemShut {NoStop}%
\bibitem [{\citenamefont {Altegoer}\ \emph
  {et~al.}(1998{\natexlab{b}})\citenamefont {Altegoer} \emph
  {et~al.}}]{Altegoer1998b}%
  \BibitemOpen
  \bibfield  {author} {\bibinfo {author} {\bibfnamefont {J.}~\bibnamefont
  {Altegoer}} \emph {et~al.},\ }\bibfield  {title} {\enquote {\bibinfo {title}
  {Search for a new gauge boson in $\pi^0$ decays},}\ }\href {\doibase
  https://doi.org/10.1016/S0370-2693(98)00402-X} {\bibfield  {journal}
  {\bibinfo  {journal} {Physics Letters B}\ }\textbf {\bibinfo {volume}
  {428}},\ \bibinfo {pages} {197 -- 205} (\bibinfo {year}
  {1998}{\natexlab{b}})}\BibitemShut {NoStop}%
\bibitem [{\citenamefont {Gninenko}\ and\ \citenamefont
  {Krasnikov}(1999)}]{Gninenko1998a}%
  \BibitemOpen
  \bibfield  {author} {\bibinfo {author} {\bibfnamefont {S.~N.}\ \bibnamefont
  {Gninenko}}\ and\ \bibinfo {author} {\bibfnamefont {N.~V.}\ \bibnamefont
  {Krasnikov}},\ }\bibfield  {title} {\enquote {\bibinfo {title} {{Limits on
  the magnetic moment of sterile neutrino and two photon neutrino decay}},}\
  }\href {\doibase 10.1016/S0370-2693(99)00130-6} {\bibfield  {journal}
  {\bibinfo  {journal} {Phys. Lett.}\ }\textbf {\bibinfo {volume} {B450}},\
  \bibinfo {pages} {165--172} (\bibinfo {year} {1999})},\ \Eprint
  {http://arxiv.org/abs/hep-ph/9808370} {arXiv:hep-ph/9808370 [hep-ph]}
  \BibitemShut {NoStop}%
\bibitem [{\citenamefont {Gninenko}\ and\ \citenamefont
  {Krasnikov}(1998)}]{Gninenko1998b}%
  \BibitemOpen
  \bibfield  {author} {\bibinfo {author} {\bibfnamefont {S.N.}\ \bibnamefont
  {Gninenko}}\ and\ \bibinfo {author} {\bibfnamefont {N.V.}\ \bibnamefont
  {Krasnikov}},\ }\bibfield  {title} {\enquote {\bibinfo {title} {On search for
  a new light gauge boson from $\pi^0(\eta)\to\gamma x$ decays in neutrino
  experiments},}\ }\href {\doibase
  https://doi.org/10.1016/S0370-2693(98)00358-X} {\bibfield  {journal}
  {\bibinfo  {journal} {Physics Letters B}\ }\textbf {\bibinfo {volume}
  {427}},\ \bibinfo {pages} {307 -- 313} (\bibinfo {year} {1998})}\BibitemShut
  {NoStop}%
\bibitem [{\citenamefont {Anelli}\ \emph {et~al.}(2015)\citenamefont {Anelli}
  \emph {et~al.}}]{Anelli:2015pba}%
  \BibitemOpen
  \bibfield  {author} {\bibinfo {author} {\bibfnamefont {M.}~\bibnamefont
  {Anelli}} \emph {et~al.} (\bibinfo {collaboration} {SHiP}),\ }\bibfield
  {title} {\enquote {\bibinfo {title} {{A facility to Search for Hidden
  Particles (SHiP) at the CERN SPS}},}\ }\href@noop {} {\  (\bibinfo {year}
  {2015})},\ \Eprint {http://arxiv.org/abs/1504.04956} {arXiv:1504.04956
  [physics.ins-det]} \BibitemShut {NoStop}%
\bibitem [{\citenamefont {Alwall}\ \emph {et~al.}(2014)\citenamefont {Alwall},
  \citenamefont {Frederix}, \citenamefont {Frixione}, \citenamefont {Hirschi},
  \citenamefont {Maltoni}, \citenamefont {Mattelaer}, \citenamefont {Shao},
  \citenamefont {Stelzer}, \citenamefont {Torrielli},\ and\ \citenamefont
  {Zaro}}]{Alwall:2014hca}%
  \BibitemOpen
  \bibfield  {author} {\bibinfo {author} {\bibfnamefont {J.}~\bibnamefont
  {Alwall}}, \bibinfo {author} {\bibfnamefont {R.}~\bibnamefont {Frederix}},
  \bibinfo {author} {\bibfnamefont {S.}~\bibnamefont {Frixione}}, \bibinfo
  {author} {\bibfnamefont {V.}~\bibnamefont {Hirschi}}, \bibinfo {author}
  {\bibfnamefont {F.}~\bibnamefont {Maltoni}}, \bibinfo {author} {\bibfnamefont
  {O.}~\bibnamefont {Mattelaer}}, \bibinfo {author} {\bibfnamefont {H.~S.}\
  \bibnamefont {Shao}}, \bibinfo {author} {\bibfnamefont {T.}~\bibnamefont
  {Stelzer}}, \bibinfo {author} {\bibfnamefont {P.}~\bibnamefont {Torrielli}},
  \ and\ \bibinfo {author} {\bibfnamefont {M.}~\bibnamefont {Zaro}},\
  }\bibfield  {title} {\enquote {\bibinfo {title} {{The automated computation
  of tree-level and next-to-leading order differential cross sections, and
  their matching to parton shower simulations}},}\ }\href {\doibase
  10.1007/JHEP07(2014)079} {\bibfield  {journal} {\bibinfo  {journal} {JHEP}\
  }\textbf {\bibinfo {volume} {07}},\ \bibinfo {pages} {079} (\bibinfo {year}
  {2014})},\ \Eprint {http://arxiv.org/abs/1405.0301} {arXiv:1405.0301
  [hep-ph]} \BibitemShut {NoStop}%
\bibitem [{\citenamefont {Alloul}\ \emph {et~al.}(2014)\citenamefont {Alloul},
  \citenamefont {Christensen}, \citenamefont {Degrande}, \citenamefont {Duhr},\
  and\ \citenamefont {Fuks}}]{Alloul:2013bka}%
  \BibitemOpen
  \bibfield  {author} {\bibinfo {author} {\bibfnamefont {Adam}\ \bibnamefont
  {Alloul}}, \bibinfo {author} {\bibfnamefont {Neil~D.}\ \bibnamefont
  {Christensen}}, \bibinfo {author} {\bibfnamefont {Celine}\ \bibnamefont
  {Degrande}}, \bibinfo {author} {\bibfnamefont {Claude}\ \bibnamefont {Duhr}},
  \ and\ \bibinfo {author} {\bibfnamefont {Benjamin}\ \bibnamefont {Fuks}},\
  }\bibfield  {title} {\enquote {\bibinfo {title} {{FeynRules 2.0 - A complete
  toolbox for tree-level phenomenology}},}\ }\href {\doibase
  10.1016/j.cpc.2014.04.012} {\bibfield  {journal} {\bibinfo  {journal}
  {Comput. Phys. Commun.}\ }\textbf {\bibinfo {volume} {185}},\ \bibinfo
  {pages} {2250--2300} (\bibinfo {year} {2014})},\ \Eprint
  {http://arxiv.org/abs/1310.1921} {arXiv:1310.1921 [hep-ph]} \BibitemShut
  {NoStop}%
\bibitem [{\citenamefont {Christensen}\ and\ \citenamefont
  {Duhr}(2009)}]{Christensen:2008py}%
  \BibitemOpen
  \bibfield  {author} {\bibinfo {author} {\bibfnamefont {Neil~D.}\ \bibnamefont
  {Christensen}}\ and\ \bibinfo {author} {\bibfnamefont {Claude}\ \bibnamefont
  {Duhr}},\ }\bibfield  {title} {\enquote {\bibinfo {title} {{FeynRules -
  Feynman rules made easy}},}\ }\href {\doibase 10.1016/j.cpc.2009.02.018}
  {\bibfield  {journal} {\bibinfo  {journal} {Comput. Phys. Commun.}\ }\textbf
  {\bibinfo {volume} {180}},\ \bibinfo {pages} {1614--1641} (\bibinfo {year}
  {2009})},\ \Eprint {http://arxiv.org/abs/0806.4194} {arXiv:0806.4194
  [hep-ph]} \BibitemShut {NoStop}%
\bibitem [{\citenamefont {Adriani}\ \emph {et~al.}(1992)\citenamefont {Adriani}
  \emph {et~al.}}]{Adriani1992}%
  \BibitemOpen
  \bibfield  {author} {\bibinfo {author} {\bibfnamefont {O}~\bibnamefont
  {Adriani}} \emph {et~al.},\ }\bibfield  {title} {\enquote {\bibinfo {title}
  {{Search for anomalous production of single-photon events in $e^+e^-$
  annihilations at the $Z$ resonance}},}\ }\href {\doibase
  http://dx.doi.org/10.1016/0370-2693(92)91286-I} {\bibfield  {journal}
  {\bibinfo  {journal} {Phys. Lett. B}\ }\textbf {\bibinfo {volume} {297}},\
  \bibinfo {pages} {469--476} (\bibinfo {year} {1992})}\BibitemShut {NoStop}%
\bibitem [{\citenamefont {Akers}\ \emph {et~al.}(1995)\citenamefont {Akers}
  \emph {et~al.}}]{Akers1994}%
  \BibitemOpen
  \bibfield  {author} {\bibinfo {author} {\bibfnamefont {R.}~\bibnamefont
  {Akers}} \emph {et~al.} (\bibinfo {collaboration} {OPAL}),\ }\bibfield
  {title} {\enquote {\bibinfo {title} {{Measurement of single photon production
  in $e^+e^-$ collisions near the $Z^0$ resonance}},}\ }\href {\doibase
  10.1007/BF01571303} {\bibfield  {journal} {\bibinfo  {journal} {Z. Phys.}\
  }\textbf {\bibinfo {volume} {C65}},\ \bibinfo {pages} {47--66} (\bibinfo
  {year} {1995})}\BibitemShut {NoStop}%
\bibitem [{\citenamefont {Abreu}\ \emph {et~al.}(1997)\citenamefont {Abreu}
  \emph {et~al.}}]{Abreu1996}%
  \BibitemOpen
  \bibfield  {author} {\bibinfo {author} {\bibfnamefont {P.}~\bibnamefont
  {Abreu}} \emph {et~al.} (\bibinfo {collaboration} {DELPHI}),\ }\bibfield
  {title} {\enquote {\bibinfo {title} {{Search for new phenomena using single
  photon events in the DELPHI detector at LEP}},}\ }\href {\doibase
  10.1007/s002880050421} {\bibfield  {journal} {\bibinfo  {journal} {Z. Phys.}\
  }\textbf {\bibinfo {volume} {C74}},\ \bibinfo {pages} {577--586} (\bibinfo
  {year} {1997})}\BibitemShut {NoStop}%
\bibitem [{\citenamefont {Lopez}\ \emph {et~al.}(1997)\citenamefont {Lopez},
  \citenamefont {Nanopoulos},\ and\ \citenamefont {Zichichi}}]{Lopez1996}%
  \BibitemOpen
  \bibfield  {author} {\bibinfo {author} {\bibfnamefont {Jorge~L.}\
  \bibnamefont {Lopez}}, \bibinfo {author} {\bibfnamefont {Dimitri~V.}\
  \bibnamefont {Nanopoulos}}, \ and\ \bibinfo {author} {\bibfnamefont
  {A.}~\bibnamefont {Zichichi}},\ }\bibfield  {title} {\enquote {\bibinfo
  {title} {{Single photon signals at LEP in supersymmetric models with a light
  gravitino}},}\ }\href {\doibase 10.1103/PhysRevD.55.5813} {\bibfield
  {journal} {\bibinfo  {journal} {Phys. Rev.}\ }\textbf {\bibinfo {volume}
  {D55}},\ \bibinfo {pages} {5813--5825} (\bibinfo {year} {1997})},\ \Eprint
  {http://arxiv.org/abs/hep-ph/9611437} {arXiv:hep-ph/9611437 [hep-ph]}
  \BibitemShut {NoStop}%
\bibitem [{\citenamefont {Assmann}\ \emph {et~al.}(2002)\citenamefont
  {Assmann}, \citenamefont {Lamont},\ and\ \citenamefont
  {Myers}}]{Assmann2002}%
  \BibitemOpen
  \bibfield  {author} {\bibinfo {author} {\bibfnamefont {R.}~\bibnamefont
  {Assmann}}, \bibinfo {author} {\bibfnamefont {M.}~\bibnamefont {Lamont}}, \
  and\ \bibinfo {author} {\bibfnamefont {S.}~\bibnamefont {Myers}},\ }\bibfield
   {title} {\enquote {\bibinfo {title} {{A brief history of the LEP
  collider}},}\ }\bibfield  {booktitle} {\emph {\bibinfo {booktitle} {{The
  legacy of LEP and SLC. Proceedings, 7th Topical Seminar, Siena, Italy,
  October 8-11, 2001}}},\ }\href {\doibase 10.1016/S0920-5632(02)90005-8}
  {\bibfield  {journal} {\bibinfo  {journal} {Nucl. Phys. Proc. Suppl.}\
  }\textbf {\bibinfo {volume} {109B}},\ \bibinfo {pages} {17--31} (\bibinfo
  {year} {2002})}\BibitemShut {NoStop}%
\bibitem [{\citenamefont {Aaboud}\ \emph {et~al.}(2017)\citenamefont {Aaboud}
  \emph {et~al.}}]{Aaboud:2017dor}%
  \BibitemOpen
  \bibfield  {author} {\bibinfo {author} {\bibfnamefont {Morad}\ \bibnamefont
  {Aaboud}} \emph {et~al.} (\bibinfo {collaboration} {ATLAS}),\ }\bibfield
  {title} {\enquote {\bibinfo {title} {{Search for dark matter at $\sqrt{s}=13$
  TeV in final states containing an energetic photon and large missing
  transverse momentum with the ATLAS detector}},}\ }\href {\doibase
  10.1140/epjc/s10052-017-4965-8} {\bibfield  {journal} {\bibinfo  {journal}
  {Eur. Phys. J.}\ }\textbf {\bibinfo {volume} {C77}},\ \bibinfo {pages} {393}
  (\bibinfo {year} {2017})},\ \Eprint {http://arxiv.org/abs/1704.03848}
  {arXiv:1704.03848 [hep-ex]} \BibitemShut {NoStop}%
\bibitem [{\citenamefont {Aaboud}\ \emph {et~al.}(2016)\citenamefont {Aaboud}
  \emph {et~al.}}]{Aaboud:2016yuq}%
  \BibitemOpen
  \bibfield  {author} {\bibinfo {author} {\bibfnamefont {Morad}\ \bibnamefont
  {Aaboud}} \emph {et~al.} (\bibinfo {collaboration} {ATLAS}),\ }\bibfield
  {title} {\enquote {\bibinfo {title} {{Measurement of the photon
  identification efficiencies with the ATLAS detector using LHC Run-1 data}},}\
  }\href {\doibase 10.1140/epjc/s10052-016-4507-9} {\bibfield  {journal}
  {\bibinfo  {journal} {Eur. Phys. J.}\ }\textbf {\bibinfo {volume} {C76}},\
  \bibinfo {pages} {666} (\bibinfo {year} {2016})},\ \Eprint
  {http://arxiv.org/abs/1606.01813} {arXiv:1606.01813 [hep-ex]} \BibitemShut
  {NoStop}%
\bibitem [{\citenamefont {Khachatryan}\ \emph {et~al.}(2016)\citenamefont
  {Khachatryan} \emph {et~al.}}]{CMS:2015loa}%
  \BibitemOpen
  \bibfield  {author} {\bibinfo {author} {\bibfnamefont {Vardan}\ \bibnamefont
  {Khachatryan}} \emph {et~al.} (\bibinfo {collaboration} {CMS}),\ }\bibfield
  {title} {\enquote {\bibinfo {title} {{Search for supersymmetry in events with
  a photon, a lepton, and missing transverse momentum in pp collisions at
  $\sqrt s=$ 8 TeV}},}\ }\href {\doibase 10.1016/j.physletb.2016.03.039}
  {\bibfield  {journal} {\bibinfo  {journal} {Phys. Lett.}\ }\textbf {\bibinfo
  {volume} {B757}},\ \bibinfo {pages} {6--31} (\bibinfo {year} {2016})},\
  \Eprint {http://arxiv.org/abs/1508.01218} {arXiv:1508.01218 [hep-ex]}
  \BibitemShut {NoStop}%
\bibitem [{\citenamefont {Ruchayskiy}\ and\ \citenamefont
  {Ivashko}(2012)}]{Ruchayskiy:2012si}%
  \BibitemOpen
  \bibfield  {author} {\bibinfo {author} {\bibfnamefont {Oleg}\ \bibnamefont
  {Ruchayskiy}}\ and\ \bibinfo {author} {\bibfnamefont {Artem}\ \bibnamefont
  {Ivashko}},\ }\bibfield  {title} {\enquote {\bibinfo {title} {{Restrictions
  on the lifetime of sterile neutrinos from primordial nucleosynthesis}},}\
  }\href {\doibase 10.1088/1475-7516/2012/10/014} {\bibfield  {journal}
  {\bibinfo  {journal} {JCAP}\ }\textbf {\bibinfo {volume} {1210}},\ \bibinfo
  {pages} {014} (\bibinfo {year} {2012})},\ \Eprint
  {http://arxiv.org/abs/1202.2841} {arXiv:1202.2841 [hep-ph]} \BibitemShut
  {NoStop}%
\bibitem [{\citenamefont {Dodelson}\ and\ \citenamefont
  {Widrow}(1994)}]{Dodelson:1993je}%
  \BibitemOpen
  \bibfield  {author} {\bibinfo {author} {\bibfnamefont {Scott}\ \bibnamefont
  {Dodelson}}\ and\ \bibinfo {author} {\bibfnamefont {Lawrence~M.}\
  \bibnamefont {Widrow}},\ }\bibfield  {title} {\enquote {\bibinfo {title}
  {{Sterile-neutrinos as dark matter}},}\ }\href {\doibase
  10.1103/PhysRevLett.72.17} {\bibfield  {journal} {\bibinfo  {journal} {Phys.
  Rev. Lett.}\ }\textbf {\bibinfo {volume} {72}},\ \bibinfo {pages} {17--20}
  (\bibinfo {year} {1994})},\ \Eprint {http://arxiv.org/abs/hep-ph/9303287}
  {arXiv:hep-ph/9303287 [hep-ph]} \BibitemShut {NoStop}%
\bibitem [{\citenamefont {Fields}\ \emph {et~al.}(2014)\citenamefont {Fields},
  \citenamefont {Molaro},\ and\ \citenamefont {Sarkar}}]{Fields:2014uja}%
  \BibitemOpen
  \bibfield  {author} {\bibinfo {author} {\bibfnamefont {Brian~D.}\
  \bibnamefont {Fields}}, \bibinfo {author} {\bibfnamefont {Paolo}\
  \bibnamefont {Molaro}}, \ and\ \bibinfo {author} {\bibfnamefont {Subir}\
  \bibnamefont {Sarkar}},\ }\bibfield  {title} {\enquote {\bibinfo {title}
  {{Big-Bang Nucleosynthesis}},}\ }\href@noop {} {\bibfield  {journal}
  {\bibinfo  {journal} {Chin. Phys.}\ }\textbf {\bibinfo {volume} {C38}},\
  \bibinfo {pages} {339--344} (\bibinfo {year} {2014})},\ \Eprint
  {http://arxiv.org/abs/1412.1408} {arXiv:1412.1408 [astro-ph.CO]} \BibitemShut
  {NoStop}%
\bibitem [{\citenamefont {Raffelt}(1996)}]{Raffelt1996}%
  \BibitemOpen
  \bibfield  {author} {\bibinfo {author} {\bibfnamefont {G.~G.}\ \bibnamefont
  {Raffelt}},\ }\href@noop {} {\emph {\bibinfo {title} {{Stars as laboratories
  for fundamental physics}}}}\ (\bibinfo  {publisher} {University of Chicago
  Press},\ \bibinfo {year} {1996})\BibitemShut {NoStop}%
\bibitem [{\citenamefont {Dreiner}\ \emph
  {et~al.}(2003{\natexlab{a}})\citenamefont {Dreiner}, \citenamefont {Hanhart},
  \citenamefont {Langenfeld},\ and\ \citenamefont {Phillips}}]{Dreiner2003}%
  \BibitemOpen
  \bibfield  {author} {\bibinfo {author} {\bibfnamefont {H.~K.}\ \bibnamefont
  {Dreiner}}, \bibinfo {author} {\bibfnamefont {C.}~\bibnamefont {Hanhart}},
  \bibinfo {author} {\bibfnamefont {U.}~\bibnamefont {Langenfeld}}, \ and\
  \bibinfo {author} {\bibfnamefont {Daniel~R.}\ \bibnamefont {Phillips}},\
  }\bibfield  {title} {\enquote {\bibinfo {title} {{Supernovae and light
  neutralinos: SN1987A bounds on supersymmetry revisited}},}\ }\href {\doibase
  10.1103/PhysRevD.68.055004} {\bibfield  {journal} {\bibinfo  {journal} {Phys.
  Rev.}\ }\textbf {\bibinfo {volume} {D68}},\ \bibinfo {pages} {055004}
  (\bibinfo {year} {2003}{\natexlab{a}})},\ \Eprint
  {http://arxiv.org/abs/hep-ph/0304289} {arXiv:hep-ph/0304289 [hep-ph]}
  \BibitemShut {NoStop}%
\bibitem [{\citenamefont {Dreiner}\ \emph {et~al.}(2010)\citenamefont
  {Dreiner}, \citenamefont {Haber},\ and\ \citenamefont
  {Martin}}]{Dreiner:2008tw}%
  \BibitemOpen
  \bibfield  {author} {\bibinfo {author} {\bibfnamefont {Herbi~K.}\
  \bibnamefont {Dreiner}}, \bibinfo {author} {\bibfnamefont {Howard~E.}\
  \bibnamefont {Haber}}, \ and\ \bibinfo {author} {\bibfnamefont {Stephen~P.}\
  \bibnamefont {Martin}},\ }\bibfield  {title} {\enquote {\bibinfo {title}
  {{Two-component spinor techniques and Feynman rules for quantum field theory
  and supersymmetry}},}\ }\href {\doibase 10.1016/j.physrep.2010.05.002}
  {\bibfield  {journal} {\bibinfo  {journal} {Phys. Rept.}\ }\textbf {\bibinfo
  {volume} {494}},\ \bibinfo {pages} {1--196} (\bibinfo {year} {2010})},\
  \Eprint {http://arxiv.org/abs/0812.1594} {arXiv:0812.1594 [hep-ph]}
  \BibitemShut {NoStop}%
\bibitem [{\citenamefont {Dreiner}\ \emph {et~al.}(2014)\citenamefont
  {Dreiner}, \citenamefont {Fortin}, \citenamefont {Hanhart},\ and\
  \citenamefont {Ubaldi}}]{Dreiner:2013mua}%
  \BibitemOpen
  \bibfield  {author} {\bibinfo {author} {\bibfnamefont {Herbert~K.}\
  \bibnamefont {Dreiner}}, \bibinfo {author} {\bibfnamefont
  {Jean-Fran\c{c}ois}\ \bibnamefont {Fortin}}, \bibinfo {author} {\bibfnamefont
  {Christoph}\ \bibnamefont {Hanhart}}, \ and\ \bibinfo {author} {\bibfnamefont
  {Lorenzo}\ \bibnamefont {Ubaldi}},\ }\bibfield  {title} {\enquote {\bibinfo
  {title} {{Supernova constraints on MeV dark sectors from $e^+e^-$
  annihilations}},}\ }\href {\doibase 10.1103/PhysRevD.89.105015} {\bibfield
  {journal} {\bibinfo  {journal} {Phys. Rev.}\ }\textbf {\bibinfo {volume}
  {D89}},\ \bibinfo {pages} {105015} (\bibinfo {year} {2014})},\ \Eprint
  {http://arxiv.org/abs/1310.3826} {arXiv:1310.3826 [hep-ph]} \BibitemShut
  {NoStop}%
\bibitem [{\citenamefont {Fischer}\ \emph {et~al.}(2016)\citenamefont
  {Fischer}, \citenamefont {Chakraborty}, \citenamefont {Giannotti},
  \citenamefont {Mirizzi}, \citenamefont {Payez},\ and\ \citenamefont
  {Ringwald}}]{Fischer:2016cyd}%
  \BibitemOpen
  \bibfield  {author} {\bibinfo {author} {\bibfnamefont {Tobias}\ \bibnamefont
  {Fischer}}, \bibinfo {author} {\bibfnamefont {Sovan}\ \bibnamefont
  {Chakraborty}}, \bibinfo {author} {\bibfnamefont {Maurizio}\ \bibnamefont
  {Giannotti}}, \bibinfo {author} {\bibfnamefont {tlessandro}\ \bibnamefont
  {Mirizzi}}, \bibinfo {author} {\bibfnamefont {Alexandre}\ \bibnamefont
  {Payez}}, \ and\ \bibinfo {author} {\bibfnamefont {Andreas}\ \bibnamefont
  {Ringwald}},\ }\bibfield  {title} {\enquote {\bibinfo {title} {{Probing
  axions with the neutrino signal from the next galactic supernova}},}\ }\href
  {\doibase 10.1103/PhysRevD.94.085012} {\bibfield  {journal} {\bibinfo
  {journal} {Phys. Rev.}\ }\textbf {\bibinfo {volume} {D94}},\ \bibinfo {pages}
  {085012} (\bibinfo {year} {2016})},\ \Eprint
  {http://arxiv.org/abs/1605.08780} {arXiv:1605.08780 [astro-ph.HE]}
  \BibitemShut {NoStop}%
\bibitem [{\citenamefont {Chang}\ \emph {et~al.}(2017)\citenamefont {Chang},
  \citenamefont {Essig},\ and\ \citenamefont {McDermott}}]{Chang:2016ntp}%
  \BibitemOpen
  \bibfield  {author} {\bibinfo {author} {\bibfnamefont {Jae~Hyeok}\
  \bibnamefont {Chang}}, \bibinfo {author} {\bibfnamefont {Rouven}\
  \bibnamefont {Essig}}, \ and\ \bibinfo {author} {\bibfnamefont {Samuel~D.}\
  \bibnamefont {McDermott}},\ }\bibfield  {title} {\enquote {\bibinfo {title}
  {{Revisiting Supernova 1987A Constraints on Dark Photons}},}\ }\href
  {\doibase 10.1007/JHEP01(2017)107} {\bibfield  {journal} {\bibinfo  {journal}
  {JHEP}\ }\textbf {\bibinfo {volume} {01}},\ \bibinfo {pages} {107} (\bibinfo
  {year} {2017})},\ \Eprint {http://arxiv.org/abs/1611.03864} {arXiv:1611.03864
  [hep-ph]} \BibitemShut {NoStop}%
\bibitem [{\citenamefont {Hardy}\ and\ \citenamefont
  {Lasenby}(2017)}]{Hardy:2016kme}%
  \BibitemOpen
  \bibfield  {author} {\bibinfo {author} {\bibfnamefont {Edward}\ \bibnamefont
  {Hardy}}\ and\ \bibinfo {author} {\bibfnamefont {Robert}\ \bibnamefont
  {Lasenby}},\ }\bibfield  {title} {\enquote {\bibinfo {title} {{Stellar
  cooling bounds on new light particles: plasma mixing effects}},}\ }\href
  {\doibase 10.1007/JHEP02(2017)033} {\bibfield  {journal} {\bibinfo  {journal}
  {JHEP}\ }\textbf {\bibinfo {volume} {02}},\ \bibinfo {pages} {033} (\bibinfo
  {year} {2017})},\ \Eprint {http://arxiv.org/abs/1611.05852} {arXiv:1611.05852
  [hep-ph]} \BibitemShut {NoStop}%
\bibitem [{\citenamefont {Hirata}\ \emph {et~al.}(1987)\citenamefont {Hirata},
  \citenamefont {Kajita}, \citenamefont {Koshiba}, \citenamefont {Nakahata},
  \citenamefont {Oyama}, \citenamefont {Sato}, \citenamefont {Suzuki},
  \citenamefont {Takita}, \citenamefont {Totsuka}, \citenamefont {Kifune},
  \citenamefont {Suda}, \citenamefont {Takahashi}, \citenamefont {Tanimori},
  \citenamefont {Miyano}, \citenamefont {Yamada}, \citenamefont {Beier},
  \citenamefont {Feldscher}, \citenamefont {Kim}, \citenamefont {Mann},
  \citenamefont {Newcomer}, \citenamefont {Van}, \citenamefont {Zhang},\ and\
  \citenamefont {Cortez}}]{Hirata1987}%
  \BibitemOpen
  \bibfield  {author} {\bibinfo {author} {\bibfnamefont {K.}~\bibnamefont
  {Hirata}}, \bibinfo {author} {\bibfnamefont {T.}~\bibnamefont {Kajita}},
  \bibinfo {author} {\bibfnamefont {M.}~\bibnamefont {Koshiba}}, \bibinfo
  {author} {\bibfnamefont {M.}~\bibnamefont {Nakahata}}, \bibinfo {author}
  {\bibfnamefont {Y.}~\bibnamefont {Oyama}}, \bibinfo {author} {\bibfnamefont
  {N.}~\bibnamefont {Sato}}, \bibinfo {author} {\bibfnamefont {A.}~\bibnamefont
  {Suzuki}}, \bibinfo {author} {\bibfnamefont {M.}~\bibnamefont {Takita}},
  \bibinfo {author} {\bibfnamefont {Y.}~\bibnamefont {Totsuka}}, \bibinfo
  {author} {\bibfnamefont {T.}~\bibnamefont {Kifune}}, \bibinfo {author}
  {\bibfnamefont {T.}~\bibnamefont {Suda}}, \bibinfo {author} {\bibfnamefont
  {K.}~\bibnamefont {Takahashi}}, \bibinfo {author} {\bibfnamefont
  {T.}~\bibnamefont {Tanimori}}, \bibinfo {author} {\bibfnamefont
  {K.}~\bibnamefont {Miyano}}, \bibinfo {author} {\bibfnamefont
  {M.}~\bibnamefont {Yamada}}, \bibinfo {author} {\bibfnamefont {E.~W.}\
  \bibnamefont {Beier}}, \bibinfo {author} {\bibfnamefont {L.~R.}\ \bibnamefont
  {Feldscher}}, \bibinfo {author} {\bibfnamefont {S.~B.}\ \bibnamefont {Kim}},
  \bibinfo {author} {\bibfnamefont {A.~K.}\ \bibnamefont {Mann}}, \bibinfo
  {author} {\bibfnamefont {F.~M.}\ \bibnamefont {Newcomer}}, \bibinfo {author}
  {\bibfnamefont {R.}~\bibnamefont {Van}}, \bibinfo {author} {\bibfnamefont
  {W.}~\bibnamefont {Zhang}}, \ and\ \bibinfo {author} {\bibfnamefont {B.~G.}\
  \bibnamefont {Cortez}},\ }\bibfield  {title} {\enquote {\bibinfo {title}
  {Observation of a neutrino burst from the supernova sn1987a},}\ }\href
  {\doibase 10.1103/PhysRevLett.58.1490} {\bibfield  {journal} {\bibinfo
  {journal} {Phys. Rev. Lett.}\ }\textbf {\bibinfo {volume} {58}},\ \bibinfo
  {pages} {1490--1493} (\bibinfo {year} {1987})}\BibitemShut {NoStop}%
\bibitem [{\citenamefont {Alekseev}\ \emph {et~al.}(1988)\citenamefont
  {Alekseev}, \citenamefont {Alekseeva}, \citenamefont {Krivosheina},\ and\
  \citenamefont {Volchenko}}]{Alekseev:1988gp}%
  \BibitemOpen
  \bibfield  {author} {\bibinfo {author} {\bibfnamefont {E.~N.}\ \bibnamefont
  {Alekseev}}, \bibinfo {author} {\bibfnamefont {L.~N.}\ \bibnamefont
  {Alekseeva}}, \bibinfo {author} {\bibfnamefont {I.~V.}\ \bibnamefont
  {Krivosheina}}, \ and\ \bibinfo {author} {\bibfnamefont {V.~I.}\ \bibnamefont
  {Volchenko}},\ }\bibfield  {title} {\enquote {\bibinfo {title} {{Detection of
  the Neutrino Signal From {SN1987A} in the {LMC} Using the Inr Baksan
  Underground Scintillation Telescope}},}\ }\href {\doibase
  10.1016/0370-2693(88)91651-6} {\bibfield  {journal} {\bibinfo  {journal}
  {Phys. Lett.}\ }\textbf {\bibinfo {volume} {B205}},\ \bibinfo {pages}
  {209--214} (\bibinfo {year} {1988})}\BibitemShut {NoStop}%
\bibitem [{\citenamefont {Bionta}\ \emph {et~al.}(1987)\citenamefont {Bionta}
  \emph {et~al.}}]{Bionta:1987qt}%
  \BibitemOpen
  \bibfield  {author} {\bibinfo {author} {\bibfnamefont {R.~M.}\ \bibnamefont
  {Bionta}} \emph {et~al.},\ }\bibfield  {title} {\enquote {\bibinfo {title}
  {{Observation of a Neutrino Burst in Coincidence with Supernova SN 1987a in
  the Large Magellanic Cloud}},}\ }\href {\doibase 10.1103/PhysRevLett.58.1494}
  {\bibfield  {journal} {\bibinfo  {journal} {Phys. Rev. Lett.}\ }\textbf
  {\bibinfo {volume} {58}},\ \bibinfo {pages} {1494} (\bibinfo {year}
  {1987})}\BibitemShut {NoStop}%
\bibitem [{\citenamefont {Gondolo}\ and\ \citenamefont
  {Gelmini}(1991)}]{Gondolo:1990dk}%
  \BibitemOpen
  \bibfield  {author} {\bibinfo {author} {\bibfnamefont {Paolo}\ \bibnamefont
  {Gondolo}}\ and\ \bibinfo {author} {\bibfnamefont {Graciela}\ \bibnamefont
  {Gelmini}},\ }\bibfield  {title} {\enquote {\bibinfo {title} {{Cosmic
  abundances of stable particles: Improved analysis}},}\ }\href {\doibase
  10.1016/0550-3213(91)90438-4} {\bibfield  {journal} {\bibinfo  {journal}
  {Nucl. Phys.}\ }\textbf {\bibinfo {volume} {B360}},\ \bibinfo {pages}
  {145--179} (\bibinfo {year} {1991})}\BibitemShut {NoStop}%
\bibitem [{\citenamefont {Cannoni}(2014)}]{Cannoni:2013bza}%
  \BibitemOpen
  \bibfield  {author} {\bibinfo {author} {\bibfnamefont {M.}~\bibnamefont
  {Cannoni}},\ }\bibfield  {title} {\enquote {\bibinfo {title} {{Relativistic
  $<\sigma v_\text{rel}>$ in the calculation of relics abundances: a closer
  look}},}\ }\href {\doibase 10.1103/PhysRevD.89.103533} {\bibfield  {journal}
  {\bibinfo  {journal} {Phys. Rev.}\ }\textbf {\bibinfo {volume} {D89}},\
  \bibinfo {pages} {103533} (\bibinfo {year} {2014})},\ \Eprint
  {http://arxiv.org/abs/1311.4494} {arXiv:1311.4494 [astro-ph.CO]} \BibitemShut
  {NoStop}%
\bibitem [{\citenamefont {Dreiner}\ \emph
  {et~al.}(2003{\natexlab{b}})\citenamefont {Dreiner}, \citenamefont {Hanhart},
  \citenamefont {Langenfeld},\ and\ \citenamefont {Phillips}}]{Dreiner:2003wh}%
  \BibitemOpen
  \bibfield  {author} {\bibinfo {author} {\bibfnamefont {H.~K.}\ \bibnamefont
  {Dreiner}}, \bibinfo {author} {\bibfnamefont {C.}~\bibnamefont {Hanhart}},
  \bibinfo {author} {\bibfnamefont {U.}~\bibnamefont {Langenfeld}}, \ and\
  \bibinfo {author} {\bibfnamefont {Daniel~R.}\ \bibnamefont {Phillips}},\
  }\bibfield  {title} {\enquote {\bibinfo {title} {{Supernovae and light
  neutralinos: SN1987A bounds on supersymmetry revisited}},}\ }\href {\doibase
  10.1103/PhysRevD.68.055004} {\bibfield  {journal} {\bibinfo  {journal} {Phys.
  Rev.}\ }\textbf {\bibinfo {volume} {D68}},\ \bibinfo {pages} {055004}
  (\bibinfo {year} {2003}{\natexlab{b}})},\ \Eprint
  {http://arxiv.org/abs/hep-ph/0304289} {arXiv:hep-ph/0304289 [hep-ph]}
  \BibitemShut {NoStop}%
\bibitem [{\citenamefont {Fradette}\ and\ \citenamefont
  {Pospelov}(2017)}]{Fradette:2017sdd}%
  \BibitemOpen
  \bibfield  {author} {\bibinfo {author} {\bibfnamefont {Anthony}\ \bibnamefont
  {Fradette}}\ and\ \bibinfo {author} {\bibfnamefont {Maxim}\ \bibnamefont
  {Pospelov}},\ }\bibfield  {title} {\enquote {\bibinfo {title} {{BBN for the
  LHC: constraints on lifetimes of the Higgs portal scalars}},}\ }\href
  {\doibase 10.1103/PhysRevD.96.075033} {\bibfield  {journal} {\bibinfo
  {journal} {Phys. Rev.}\ }\textbf {\bibinfo {volume} {D96}},\ \bibinfo {pages}
  {075033} (\bibinfo {year} {2017})},\ \Eprint
  {http://arxiv.org/abs/1706.01920} {arXiv:1706.01920 [hep-ph]} \BibitemShut
  {NoStop}%
\bibitem [{\citenamefont {{Colgate}}\ and\ \citenamefont
  {{White}}(1966)}]{1966ApJ143626C}%
  \BibitemOpen
  \bibfield  {author} {\bibinfo {author} {\bibfnamefont {S.~A.}\ \bibnamefont
  {{Colgate}}}\ and\ \bibinfo {author} {\bibfnamefont {R.~H.}\ \bibnamefont
  {{White}}},\ }\bibfield  {title} {\enquote {\bibinfo {title} {{The
  Hydrodynamic Behavior of Supernovae Explosions}},}\ }\href {\doibase
  10.1086/148549} {\bibfield  {journal} {\bibinfo  {journal} {\apj}\ }\textbf
  {\bibinfo {volume} {143}},\ \bibinfo {pages} {626} (\bibinfo {year}
  {1966})}\BibitemShut {NoStop}%
\bibitem [{\citenamefont {Rampp}\ and\ \citenamefont
  {Janka}(2000)}]{Rampp:2000ws}%
  \BibitemOpen
  \bibfield  {author} {\bibinfo {author} {\bibfnamefont {Markus}\ \bibnamefont
  {Rampp}}\ and\ \bibinfo {author} {\bibfnamefont {H.~Thomas}\ \bibnamefont
  {Janka}},\ }\bibfield  {title} {\enquote {\bibinfo {title} {{Spherically
  symmetric simulation with Boltzmann neutrino transport of core collapse and
  post bounce evolution of a 15 solar mass star}},}\ }\href {\doibase
  10.1086/312837} {\bibfield  {journal} {\bibinfo  {journal} {Astrophys. J.}\
  }\textbf {\bibinfo {volume} {539}},\ \bibinfo {pages} {L33--L36} (\bibinfo
  {year} {2000})},\ \Eprint {http://arxiv.org/abs/astro-ph/0005438}
  {arXiv:astro-ph/0005438 [astro-ph]} \BibitemShut {NoStop}%
\bibitem [{\citenamefont {Liebendoerfer}\ \emph {et~al.}(2001)\citenamefont
  {Liebendoerfer}, \citenamefont {Mezzacappa}, \citenamefont {Thielemann},
  \citenamefont {Messer}, \citenamefont {Hix},\ and\ \citenamefont
  {Bruenn}}]{Liebendoerfer:2000cq}%
  \BibitemOpen
  \bibfield  {author} {\bibinfo {author} {\bibfnamefont {Matthias}\
  \bibnamefont {Liebendoerfer}}, \bibinfo {author} {\bibfnamefont {Anthony}\
  \bibnamefont {Mezzacappa}}, \bibinfo {author} {\bibfnamefont
  {Friederich-Karl}\ \bibnamefont {Thielemann}}, \bibinfo {author}
  {\bibfnamefont {O.~E.~Bronson}\ \bibnamefont {Messer}}, \bibinfo {author}
  {\bibfnamefont {W.~Raphael}\ \bibnamefont {Hix}}, \ and\ \bibinfo {author}
  {\bibfnamefont {Stephen~W.}\ \bibnamefont {Bruenn}},\ }\bibfield  {title}
  {\enquote {\bibinfo {title} {{Probing the gravitational well: no supernova
  explosion in spherical symmetry with general relativistic boltzmann neutrino
  transport}},}\ }\href {\doibase 10.1103/PhysRevD.63.103004} {\bibfield
  {journal} {\bibinfo  {journal} {Phys. Rev.}\ }\textbf {\bibinfo {volume}
  {D63}},\ \bibinfo {pages} {103004} (\bibinfo {year} {2001})},\ \Eprint
  {http://arxiv.org/abs/astro-ph/0006418} {arXiv:astro-ph/0006418 [astro-ph]}
  \BibitemShut {NoStop}%
\bibitem [{\citenamefont {Müller}\ \emph {et~al.}(2017)\citenamefont
  {Müller}, \citenamefont {Melson}, \citenamefont {Heger},\ and\ \citenamefont
  {Janka}}]{Muller:2017hht}%
  \BibitemOpen
  \bibfield  {author} {\bibinfo {author} {\bibfnamefont {B.}~\bibnamefont
  {Müller}}, \bibinfo {author} {\bibfnamefont {T.}~\bibnamefont {Melson}},
  \bibinfo {author} {\bibfnamefont {A.}~\bibnamefont {Heger}}, \ and\ \bibinfo
  {author} {\bibfnamefont {H.~Th.}\ \bibnamefont {Janka}},\ }\bibfield  {title}
  {\enquote {\bibinfo {title} {{Supernova simulations from a 3D progenitor
  model - Impact of perturbations and evolution of explosion properties}},}\
  }\href {\doibase 10.1093/mnras/stx1962} {\bibfield  {journal} {\bibinfo
  {journal} {Mon. Not. Roy. Astron. Soc.}\ }\textbf {\bibinfo {volume} {472}},\
  \bibinfo {pages} {491--513} (\bibinfo {year} {2017})},\ \Eprint
  {http://arxiv.org/abs/1705.00620} {arXiv:1705.00620 [astro-ph.SR]}
  \BibitemShut {NoStop}%
\bibitem [{\citenamefont {Zatsepin}\ and\ \citenamefont
  {Smirnov}(1978)}]{Zatsepin:1978ac}%
  \BibitemOpen
  \bibfield  {author} {\bibinfo {author} {\bibfnamefont {G.~T.}\ \bibnamefont
  {Zatsepin}}\ and\ \bibinfo {author} {\bibfnamefont {A.~{\relax Yu}.}\
  \bibnamefont {Smirnov}},\ }\bibfield  {title} {\enquote {\bibinfo {title}
  {{Is the Radiative Decay of a Neutral Lepton a Stripping Mechanism of
  Supernova Shells?}}}\ }\href@noop {} {\bibfield  {journal} {\bibinfo
  {journal} {Pisma Zh. Eksp. Teor. Fiz.}\ }\textbf {\bibinfo {volume} {28}},\
  \bibinfo {pages} {379--381} (\bibinfo {year} {1978})}\BibitemShut {NoStop}%
\bibitem [{\citenamefont {Fuller}\ \emph {et~al.}(2009)\citenamefont {Fuller},
  \citenamefont {Kusenko},\ and\ \citenamefont {Petraki}}]{Fuller:2009zz}%
  \BibitemOpen
  \bibfield  {author} {\bibinfo {author} {\bibfnamefont {George~M.}\
  \bibnamefont {Fuller}}, \bibinfo {author} {\bibfnamefont {Alexander}\
  \bibnamefont {Kusenko}}, \ and\ \bibinfo {author} {\bibfnamefont {Kalliopi}\
  \bibnamefont {Petraki}},\ }\bibfield  {title} {\enquote {\bibinfo {title}
  {{Heavy sterile neutrinos and supernova explosions}},}\ }\href {\doibase
  10.1016/j.physletb.2008.11.016} {\bibfield  {journal} {\bibinfo  {journal}
  {Phys. Lett.}\ }\textbf {\bibinfo {volume} {B670}},\ \bibinfo {pages}
  {281--284} (\bibinfo {year} {2009})},\ \Eprint
  {http://arxiv.org/abs/0806.4273} {arXiv:0806.4273 [astro-ph]} \BibitemShut
  {NoStop}%
\bibitem [{\citenamefont {Chou}\ \emph {et~al.}(2017)\citenamefont {Chou},
  \citenamefont {Curtin},\ and\ \citenamefont {Lubatti}}]{Chou:2016lxi}%
  \BibitemOpen
  \bibfield  {author} {\bibinfo {author} {\bibfnamefont {John~Paul}\
  \bibnamefont {Chou}}, \bibinfo {author} {\bibfnamefont {David}\ \bibnamefont
  {Curtin}}, \ and\ \bibinfo {author} {\bibfnamefont {H.~J.}\ \bibnamefont
  {Lubatti}},\ }\bibfield  {title} {\enquote {\bibinfo {title} {{New Detectors
  to Explore the Lifetime Frontier}},}\ }\href {\doibase
  10.1016/j.physletb.2017.01.043} {\bibfield  {journal} {\bibinfo  {journal}
  {Phys. Lett.}\ }\textbf {\bibinfo {volume} {B767}},\ \bibinfo {pages}
  {29--36} (\bibinfo {year} {2017})},\ \Eprint
  {http://arxiv.org/abs/1606.06298} {arXiv:1606.06298 [hep-ph]} \BibitemShut
  {NoStop}%
\bibitem [{\citenamefont {Gligorov}\ \emph {et~al.}(2018)\citenamefont
  {Gligorov}, \citenamefont {Knapen}, \citenamefont {Papucci},\ and\
  \citenamefont {Robinson}}]{Gligorov:2017nwh}%
  \BibitemOpen
  \bibfield  {author} {\bibinfo {author} {\bibfnamefont {Vladimir~V.}\
  \bibnamefont {Gligorov}}, \bibinfo {author} {\bibfnamefont {Simon}\
  \bibnamefont {Knapen}}, \bibinfo {author} {\bibfnamefont {Michele}\
  \bibnamefont {Papucci}}, \ and\ \bibinfo {author} {\bibfnamefont {Dean~J.}\
  \bibnamefont {Robinson}},\ }\bibfield  {title} {\enquote {\bibinfo {title}
  {{Searching for Long-lived Particles: A Compact Detector for Exotics at
  LHCb}},}\ }\href {\doibase 10.1103/PhysRevD.97.015023} {\bibfield  {journal}
  {\bibinfo  {journal} {Phys. Rev.}\ }\textbf {\bibinfo {volume} {D97}},\
  \bibinfo {pages} {015023} (\bibinfo {year} {2018})},\ \Eprint
  {http://arxiv.org/abs/1708.09395} {arXiv:1708.09395 [hep-ph]} \BibitemShut
  {NoStop}%
\bibitem [{\citenamefont {Feng}\ \emph {et~al.}(2018)\citenamefont {Feng},
  \citenamefont {Galon}, \citenamefont {Kling},\ and\ \citenamefont
  {Trojanowski}}]{Feng:2017uoz}%
  \BibitemOpen
  \bibfield  {author} {\bibinfo {author} {\bibfnamefont {Jonathan}\
  \bibnamefont {Feng}}, \bibinfo {author} {\bibfnamefont {Iftah}\ \bibnamefont
  {Galon}}, \bibinfo {author} {\bibfnamefont {Felix}\ \bibnamefont {Kling}}, \
  and\ \bibinfo {author} {\bibfnamefont {Sebastian}\ \bibnamefont
  {Trojanowski}},\ }\bibfield  {title} {\enquote {\bibinfo {title} {{ForwArd
  Search ExpeRiment at the LHC}},}\ }\href {\doibase
  10.1103/PhysRevD.97.035001} {\bibfield  {journal} {\bibinfo  {journal} {Phys.
  Rev.}\ }\textbf {\bibinfo {volume} {D97}},\ \bibinfo {pages} {035001}
  (\bibinfo {year} {2018})},\ \Eprint {http://arxiv.org/abs/1708.09389}
  {arXiv:1708.09389 [hep-ph]} \BibitemShut {NoStop}%
\bibitem [{\citenamefont {Kling}\ and\ \citenamefont
  {Trojanowski}(2018)}]{Kling:2018wct}%
  \BibitemOpen
  \bibfield  {author} {\bibinfo {author} {\bibfnamefont {Felix}\ \bibnamefont
  {Kling}}\ and\ \bibinfo {author} {\bibfnamefont {Sebastian}\ \bibnamefont
  {Trojanowski}},\ }\bibfield  {title} {\enquote {\bibinfo {title} {{Heavy
  Neutral Leptons at FASER}},}\ }\href@noop {} {\  (\bibinfo {year} {2018})},\
  \Eprint {http://arxiv.org/abs/1801.08947} {arXiv:1801.08947 [hep-ph]}
  \BibitemShut {NoStop}%
\bibitem [{\citenamefont {Magill}\ and\ \citenamefont
  {Plestid}(2017{\natexlab{a}})}]{Magill2016}%
  \BibitemOpen
  \bibfield  {author} {\bibinfo {author} {\bibfnamefont {Gabriel}\ \bibnamefont
  {Magill}}\ and\ \bibinfo {author} {\bibfnamefont {Ryan}\ \bibnamefont
  {Plestid}},\ }\bibfield  {title} {\enquote {\bibinfo {title} {Neutrino
  trident production at the intensity frontier},}\ }\href {\doibase
  10.1103/PhysRevD.95.073004} {\bibfield  {journal} {\bibinfo  {journal} {Phys.
  Rev.}\ }\textbf {\bibinfo {volume} {D95}},\ \bibinfo {pages} {073004}
  (\bibinfo {year} {2017}{\natexlab{a}})},\ \Eprint
  {http://arxiv.org/abs/1612.05642} {arXiv:1612.05642 [hep-ph]} \BibitemShut
  {NoStop}%
\bibitem [{\citenamefont {Jentschura}\ and\ \citenamefont
  {Serbo}(2009)}]{Jentschura2009}%
  \BibitemOpen
  \bibfield  {author} {\bibinfo {author} {\bibfnamefont {U.~D.}\ \bibnamefont
  {Jentschura}}\ and\ \bibinfo {author} {\bibfnamefont {V.~G.}\ \bibnamefont
  {Serbo}},\ }\bibfield  {title} {\enquote {\bibinfo {title} {{Nuclear form
  factor, validity of the equivalent photon approximation and Coulomb
  corrections to muon pair production in photon-nucleus and nucleus-nucleus
  collisions}},}\ }\href {\doibase 10.1140/epjc/s10052-009-1147-3} {\bibfield
  {journal} {\bibinfo  {journal} {Eur. Phys. J.}\ }\textbf {\bibinfo {volume}
  {C64}},\ \bibinfo {pages} {309--317} (\bibinfo {year} {2009})},\ \Eprint
  {http://arxiv.org/abs/0908.3853} {arXiv:0908.3853 [hep-ph]} \BibitemShut
  {NoStop}%
\bibitem [{\citenamefont {Perdrisat}\ \emph {et~al.}(2007)\citenamefont
  {Perdrisat}, \citenamefont {Punjabi},\ and\ \citenamefont
  {Vanderhaeghen}}]{Perdrisat2006}%
  \BibitemOpen
  \bibfield  {author} {\bibinfo {author} {\bibfnamefont {C.~F.}\ \bibnamefont
  {Perdrisat}}, \bibinfo {author} {\bibfnamefont {V.}~\bibnamefont {Punjabi}},
  \ and\ \bibinfo {author} {\bibfnamefont {M.}~\bibnamefont {Vanderhaeghen}},\
  }\bibfield  {title} {\enquote {\bibinfo {title} {{Nucleon Electromagnetic
  Form Factors}},}\ }\href {\doibase 10.1016/j.ppnp.2007.05.001} {\bibfield
  {journal} {\bibinfo  {journal} {Prog. Part. Nucl. Phys.}\ }\textbf {\bibinfo
  {volume} {59}},\ \bibinfo {pages} {694--764} (\bibinfo {year} {2007})},\
  \Eprint {http://arxiv.org/abs/hep-ph/0612014} {arXiv:hep-ph/0612014 [hep-ph]}
  \BibitemShut {NoStop}%
\bibitem [{\citenamefont {Beck}\ and\ \citenamefont
  {Holstein}(2001)}]{Beck:2001dz}%
  \BibitemOpen
  \bibfield  {author} {\bibinfo {author} {\bibfnamefont {Douglas~H.}\
  \bibnamefont {Beck}}\ and\ \bibinfo {author} {\bibfnamefont {Barry~R.}\
  \bibnamefont {Holstein}},\ }\bibfield  {title} {\enquote {\bibinfo {title}
  {{Nucleon structure and parity violating electron scattering}},}\ }\href
  {\doibase 10.1142/S0218301301000381} {\bibfield  {journal} {\bibinfo
  {journal} {Int. J. Mod. Phys.}\ }\textbf {\bibinfo {volume} {E10}},\ \bibinfo
  {pages} {1--41} (\bibinfo {year} {2001})},\ \Eprint
  {http://arxiv.org/abs/hep-ph/0102053} {arXiv:hep-ph/0102053 [hep-ph]}
  \BibitemShut {NoStop}%
\bibitem [{\citenamefont {deNiverville}\ \emph {et~al.}(2017)\citenamefont
  {deNiverville}, \citenamefont {Chen}, \citenamefont {Pospelov},\ and\
  \citenamefont {Ritz}}]{deNiverville:2016rqh}%
  \BibitemOpen
  \bibfield  {author} {\bibinfo {author} {\bibfnamefont {Patrick}\ \bibnamefont
  {deNiverville}}, \bibinfo {author} {\bibfnamefont {Chien-Yi}\ \bibnamefont
  {Chen}}, \bibinfo {author} {\bibfnamefont {Maxim}\ \bibnamefont {Pospelov}},
  \ and\ \bibinfo {author} {\bibfnamefont {Adam}\ \bibnamefont {Ritz}},\
  }\bibfield  {title} {\enquote {\bibinfo {title} {{Light dark matter in
  neutrino beams: production modelling and scattering signatures at MiniBooNE,
  T2K and SHiP}},}\ }\href {\doibase 10.1103/PhysRevD.95.035006} {\bibfield
  {journal} {\bibinfo  {journal} {Phys. Rev.}\ }\textbf {\bibinfo {volume}
  {D95}},\ \bibinfo {pages} {035006} (\bibinfo {year} {2017})},\ \Eprint
  {http://arxiv.org/abs/1609.01770} {arXiv:1609.01770 [hep-ph]} \BibitemShut
  {NoStop}%
\bibitem [{\citenamefont {Bonesini}\ \emph {et~al.}(2001)\citenamefont
  {Bonesini}, \citenamefont {Marchionni}, \citenamefont {Pietropaolo},\ and\
  \citenamefont {Tabarelli~de Fatis}}]{BMPT:Bonesini:2001iz}%
  \BibitemOpen
  \bibfield  {author} {\bibinfo {author} {\bibfnamefont {M.}~\bibnamefont
  {Bonesini}}, \bibinfo {author} {\bibfnamefont {A.}~\bibnamefont
  {Marchionni}}, \bibinfo {author} {\bibfnamefont {F.}~\bibnamefont
  {Pietropaolo}}, \ and\ \bibinfo {author} {\bibfnamefont {T.}~\bibnamefont
  {Tabarelli~de Fatis}},\ }\bibfield  {title} {\enquote {\bibinfo {title} {{On
  Particle production for high-energy neutrino beams}},}\ }\href {\doibase
  10.1007/s100520100656} {\bibfield  {journal} {\bibinfo  {journal} {Eur. Phys.
  J.}\ }\textbf {\bibinfo {volume} {C20}},\ \bibinfo {pages} {13--27} (\bibinfo
  {year} {2001})},\ \Eprint {http://arxiv.org/abs/hep-ph/0101163}
  {arXiv:hep-ph/0101163 [hep-ph]} \BibitemShut {NoStop}%
\bibitem [{\citenamefont {Mariani}\ \emph {et~al.}(2011)\citenamefont
  {Mariani}, \citenamefont {Cheng}, \citenamefont {Conrad},\ and\ \citenamefont
  {Shaevitz}}]{Mariani:2011zd}%
  \BibitemOpen
  \bibfield  {author} {\bibinfo {author} {\bibfnamefont {C.}~\bibnamefont
  {Mariani}}, \bibinfo {author} {\bibfnamefont {G.}~\bibnamefont {Cheng}},
  \bibinfo {author} {\bibfnamefont {J.~M.}\ \bibnamefont {Conrad}}, \ and\
  \bibinfo {author} {\bibfnamefont {M.~H.}\ \bibnamefont {Shaevitz}},\
  }\bibfield  {title} {\enquote {\bibinfo {title} {{Improved Parameterization
  of $K^+$ Production in p-Be Collisions at Low Energy Using Feynman
  Scaling}},}\ }\href {\doibase 10.1103/PhysRevD.84.114021} {\bibfield
  {journal} {\bibinfo  {journal} {Phys. Rev.}\ }\textbf {\bibinfo {volume}
  {D84}},\ \bibinfo {pages} {114021} (\bibinfo {year} {2011})},\ \Eprint
  {http://arxiv.org/abs/1110.0417} {arXiv:1110.0417 [hep-ex]} \BibitemShut
  {NoStop}%
\bibitem [{\citenamefont {Dicus}\ and\ \citenamefont
  {Repko}(1997)}]{Dicus1997}%
  \BibitemOpen
  \bibfield  {author} {\bibinfo {author} {\bibfnamefont {Duane~A.}\
  \bibnamefont {Dicus}}\ and\ \bibinfo {author} {\bibfnamefont {Wayne~W.}\
  \bibnamefont {Repko}},\ }\bibfield  {title} {\enquote {\bibinfo {title}
  {{Photon-neutrino interactions}},}\ }\href {\doibase
  10.1103/PhysRevLett.79.569} {\bibfield  {journal} {\bibinfo  {journal} {Phys.
  Rev. Lett.}\ }\textbf {\bibinfo {volume} {79}},\ \bibinfo {pages} {569--571}
  (\bibinfo {year} {1997})},\ \Eprint {http://arxiv.org/abs/hep-ph/9703210}
  {arXiv:hep-ph/9703210 [hep-ph]} \BibitemShut {NoStop}%
\bibitem [{\citenamefont {Abbasabadi}\ \emph {et~al.}(2001)\citenamefont
  {Abbasabadi}, \citenamefont {Devoto},\ and\ \citenamefont
  {Repko}}]{Abbasabadi2000}%
  \BibitemOpen
  \bibfield  {author} {\bibinfo {author} {\bibfnamefont {Ali}\ \bibnamefont
  {Abbasabadi}}, \bibinfo {author} {\bibfnamefont {Alberto}\ \bibnamefont
  {Devoto}}, \ and\ \bibinfo {author} {\bibfnamefont {Wayne~W.}\ \bibnamefont
  {Repko}},\ }\bibfield  {title} {\enquote {\bibinfo {title} {{High-energy
  photon neutrino elastic scattering}},}\ }\href {\doibase
  10.1103/PhysRevD.63.093001} {\bibfield  {journal} {\bibinfo  {journal} {Phys.
  Rev.}\ }\textbf {\bibinfo {volume} {D63}},\ \bibinfo {pages} {093001}
  (\bibinfo {year} {2001})},\ \Eprint {http://arxiv.org/abs/hep-ph/0012257}
  {arXiv:hep-ph/0012257 [hep-ph]} \BibitemShut {NoStop}%
\bibitem [{\citenamefont {Belusevic}\ and\ \citenamefont
  {Smith}(1988)}]{Belusevic1988}%
  \BibitemOpen
  \bibfield  {author} {\bibinfo {author} {\bibfnamefont {R.}~\bibnamefont
  {Belusevic}}\ and\ \bibinfo {author} {\bibfnamefont {J.}~\bibnamefont
  {Smith}},\ }\bibfield  {title} {\enquote {\bibinfo {title} {W-z interference
  in \ensuremath{\nu}-nucleus scattering},}\ }\href {\doibase
  10.1103/PhysRevD.37.2419} {\bibfield  {journal} {\bibinfo  {journal} {Phys.
  Rev. D}\ }\textbf {\bibinfo {volume} {37}},\ \bibinfo {pages} {2419--2422}
  (\bibinfo {year} {1988})}\BibitemShut {NoStop}%
\bibitem [{\citenamefont {Budnev}\ \emph {et~al.}(1975)\citenamefont {Budnev},
  \citenamefont {Ginzburg}, \citenamefont {Meledin},\ and\ \citenamefont
  {Serbo}}]{Budnev1975}%
  \BibitemOpen
  \bibfield  {author} {\bibinfo {author} {\bibfnamefont {V.~M.}\ \bibnamefont
  {Budnev}}, \bibinfo {author} {\bibfnamefont {I.~F.}\ \bibnamefont
  {Ginzburg}}, \bibinfo {author} {\bibfnamefont {G.~V.}\ \bibnamefont
  {Meledin}}, \ and\ \bibinfo {author} {\bibfnamefont {V.~G.}\ \bibnamefont
  {Serbo}},\ }\bibfield  {title} {\enquote {\bibinfo {title} {{The Two photon
  particle production mechanism. Physical problems. Applications. Equivalent
  photon approximation}},}\ }\href {\doibase 10.1016/0370-1573(75)90009-5}
  {\bibfield  {journal} {\bibinfo  {journal} {Phys. Rept.}\ }\textbf {\bibinfo
  {volume} {15}},\ \bibinfo {pages} {181--281} (\bibinfo {year}
  {1975})}\BibitemShut {NoStop}%
\bibitem [{\citenamefont {Magill}\ and\ \citenamefont
  {Plestid}(2017{\natexlab{b}})}]{Magill:2017mps}%
  \BibitemOpen
  \bibfield  {author} {\bibinfo {author} {\bibfnamefont {Gabriel}\ \bibnamefont
  {Magill}}\ and\ \bibinfo {author} {\bibfnamefont {Ryan}\ \bibnamefont
  {Plestid}},\ }\bibfield  {title} {\enquote {\bibinfo {title} {{Probing new
  charged scalars with neutrino trident production}},}\ }\href@noop {} {\
  (\bibinfo {year} {2017}{\natexlab{b}})},\ \Eprint
  {http://arxiv.org/abs/1710.08431} {arXiv:1710.08431 [hep-ph]} \BibitemShut
  {NoStop}%
\bibitem [{\citenamefont {Gell-Mann}(1961)}]{Gell-Mann1961}%
  \BibitemOpen
  \bibfield  {author} {\bibinfo {author} {\bibfnamefont {Murray}\ \bibnamefont
  {Gell-Mann}},\ }\bibfield  {title} {\enquote {\bibinfo {title} {The reaction
  $\ensuremath{\gamma}+\ensuremath{\gamma}\ensuremath{\rightarrow}\ensuremath{\nu}+\overline{\ensuremath{\nu}}$},}\
  }\href {\doibase 10.1103/PhysRevLett.6.70} {\bibfield  {journal} {\bibinfo
  {journal} {Phys. Rev. Lett.}\ }\textbf {\bibinfo {volume} {6}},\ \bibinfo
  {pages} {70--71} (\bibinfo {year} {1961})}\BibitemShut {NoStop}%
\bibitem [{\citenamefont {Yang}(1950)}]{Yang1950}%
  \BibitemOpen
  \bibfield  {author} {\bibinfo {author} {\bibfnamefont {C.~N.}\ \bibnamefont
  {Yang}},\ }\bibfield  {title} {\enquote {\bibinfo {title} {Selection rules
  for the dematerialization of a particle into two photons},}\ }\href {\doibase
  10.1103/PhysRev.77.242} {\bibfield  {journal} {\bibinfo  {journal} {Phys.
  Rev.}\ }\textbf {\bibinfo {volume} {77}},\ \bibinfo {pages} {242--245}
  (\bibinfo {year} {1950})}\BibitemShut {NoStop}%
\bibitem [{\citenamefont {Amsler}\ \emph {et~al.}(2008 and 2009 partial update
  for the 2010 edition)\citenamefont {Amsler} \emph {et~al.}}]{Amsler2009}%
  \BibitemOpen
  \bibfield  {author} {\bibinfo {author} {\bibfnamefont {C.}~\bibnamefont
  {Amsler}} \emph {et~al.} (\bibinfo {collaboration} {Particle Data Group}),\
  }\href@noop {} {\bibfield  {journal} {\bibinfo  {journal} {Physics Letters}\
  }\textbf {\bibinfo {volume} {B667}},\ \bibinfo {pages} {1} (\bibinfo {year}
  {2008 and 2009 partial update for the 2010 edition})}\BibitemShut {NoStop}%
\end{thebibliography}%

\end{document}